\documentclass[aps,prb,superscriptaddress,twocolumn]{revtex4-2}

\usepackage{setspace}
\usepackage{lmodern}

\usepackage[latin9]{inputenc}
\usepackage{amsmath}
\usepackage{amssymb}
\usepackage{graphicx}
\usepackage{epstopdf}
\usepackage{color}
\usepackage{enumerate}
\usepackage{ulem}
\usepackage{bm} 
\usepackage{physics}

\usepackage{array}
\usepackage[labelfont=bf]{caption}
\usepackage{multirow}
\usepackage{floatpag}
\usepackage[misc]{ifsym}
\usepackage{sistyle}
\usepackage{upgreek}

\begin{document}

\title{Evidence for chiral supercurrent in quantum Hall Josephson junctions}

\author{Hadrien Vignaud}
\altaffiliation{These authors contributed equally to this work.}
\affiliation{Univ. Grenoble Alpes, CNRS, Grenoble INP, Institut N\'{e}el, 38000 Grenoble, France}
\author{David Perconte}
\altaffiliation{These authors contributed equally to this work.}
\affiliation{Univ. Grenoble Alpes, CNRS, Grenoble INP, Institut N\'{e}el, 38000 Grenoble, France}
\author{Wenmin Yang}
\author{Bilal Kousar}
\author{Edouard Wagner}
\author{Fr\'{e}d\'{e}ric Gay}
\affiliation{Univ. Grenoble Alpes, CNRS, Grenoble INP, Institut N\'{e}el, 38000 Grenoble, France}
\author{Kenji Watanabe}
\affiliation{Research Center for Electronic and Optical Materials, National Institute for Materials Science, 1-1 Namiki, Tsukuba 305-0044, Japan}
\author{Takashi Taniguchi}
\affiliation{Research Center for Materials Nanoarchitectonics, National Institute for Materials Science,  1-1 Namiki, Tsukuba 305-0044, Japan}
\author{Herv\'e Courtois}
\affiliation{Univ. Grenoble Alpes, CNRS, Grenoble INP, Institut N\'{e}el, 38000 Grenoble, France}
\author{Zheng Han}
\affiliation{State Key Laboratory of Quantum Optics and Quantum Optics Devices, Institute of Opto-Electronics, Shanxi University, Taiyuan 030006, P. R. China}
\affiliation{Collaborative Innovation Center of Extreme Optics, Shanxi University, Taiyuan 030006, P. R. China}
\affiliation{Univ. Grenoble Alpes, CNRS, Grenoble INP, Institut N\'{e}el, 38000 Grenoble, France}
\author{Hermann Sellier}
\affiliation{Univ. Grenoble Alpes, CNRS, Grenoble INP, Institut N\'{e}el, 38000 Grenoble, France}
\author{Benjamin Sac\'{e}p\'{e}}
\email{benjamin.sacepe@neel.cnrs.fr}
\altaffiliation[Present address: ]{Google Quantum AI, Mountain View, CA, USA}
\affiliation{Univ. Grenoble Alpes, CNRS, Grenoble INP, Institut N\'{e}el, 38000 Grenoble, France}

\begin{abstract}
Hybridizing superconductivity with the quantum Hall (QH) effects has major potential for designing novel circuits capable of inducing and manipulating non-Abelian states for topological quantum computation\cite{Stern2013,Alicea2015,Alicea16}.
However, despite recent experimental progress towards this hybridization\cite{Rickhaus12,Komatsu12,Shalom2015,Wan2015,Amet2016,Lee2017,Park17,Seredinski19,Zhao2020a,Wang2021,Gul2022,Hatefipour22,Zhao22}, concrete evidence for a chiral QH Josephson junction\cite{Ma1993,Zyuzin94} --the elemental building block for coherent superconducting-QH circuits-- is still lacking. Its expected signature is an unusual chiral supercurrent flowing in QH edge channels, which oscillates with a specific $2\phi_0$ magnetic flux periodicity\cite{Ma1993,Zyuzin94,Stone2011,VanOstaay2011,Alavirad2018} ($\phi_0=h/2e$ is the superconducting flux quantum, $h$ the Planck constant and $e$ the electron charge). Here, we show that ultra-narrow Josephson junctions defined in encapsulated graphene nanoribbons exhibit such a chiral supercurrent, visible up to 8 teslas, and carried by the spin-degenerate edge channel of the QH plateau of resistance $h/2e^2\simeq 12.9$ k$\Omega$. We observe reproducible $2\phi_0$-periodic oscillation of the supercurrent, which emerges at constant filling factor when the area of the loop formed by the QH edge channel is constant, within a magnetic-length correction that we resolve in the data. Furthermore, by varying the junction geometry, we show that reducing the superconductor/normal interface length is pivotal to obtain a measurable supercurrent on QH plateaus, in agreement with theories predicting dephasing along the superconducting interface\cite{Alavirad2018,Manesco22,Kurilovich22,Tang22}. Our findings mark a critical milestone along the path to explore correlated and fractional QH-based superconducting devices that should host non-Abelian Majorana and parafermion zero modes\cite{Qi2010,Clarke2012,Lindner2012,Vaezi13,Clarke2014,Mong2014,Beenakker14,SanJose15,Finocchiaro18,Snizhko18,Nielsen22}.
\end{abstract}

\maketitle 

The correlated and topological orders in the multifaceted QH effects have significant potential as a resource for inducing topological superconductivity in superconducting QH hybrid circuits\cite{Qi2010,Clarke2012,Lindner2012,Vaezi13,Clarke2014,Mong2014,Beenakker14,SanJose15,Finocchiaro18,Snizhko18,Nielsen22}. 
Several blueprints for the experimental realization of Majorana zero modes or their fractionalized generalization, the parafermions and Fibonacci anyons, have been drawn up on the basis of hybridizing spin-polarized\cite{Clarke2012,SanJose15,Finocchiaro18,Galambos22} or fractional QH states\cite{Clarke2012,Alicea2015,Alicea16}. Owing to their non-local nature and non-commutative braiding properties, these non-Abelian zero modes are expected to be the basis for fault-tolerant topological quantum computation\cite{Nayak2008,Stern2013,Mong2014,Alicea16}. 

\begin{figure*}[ht]
\includegraphics[width=12cm]{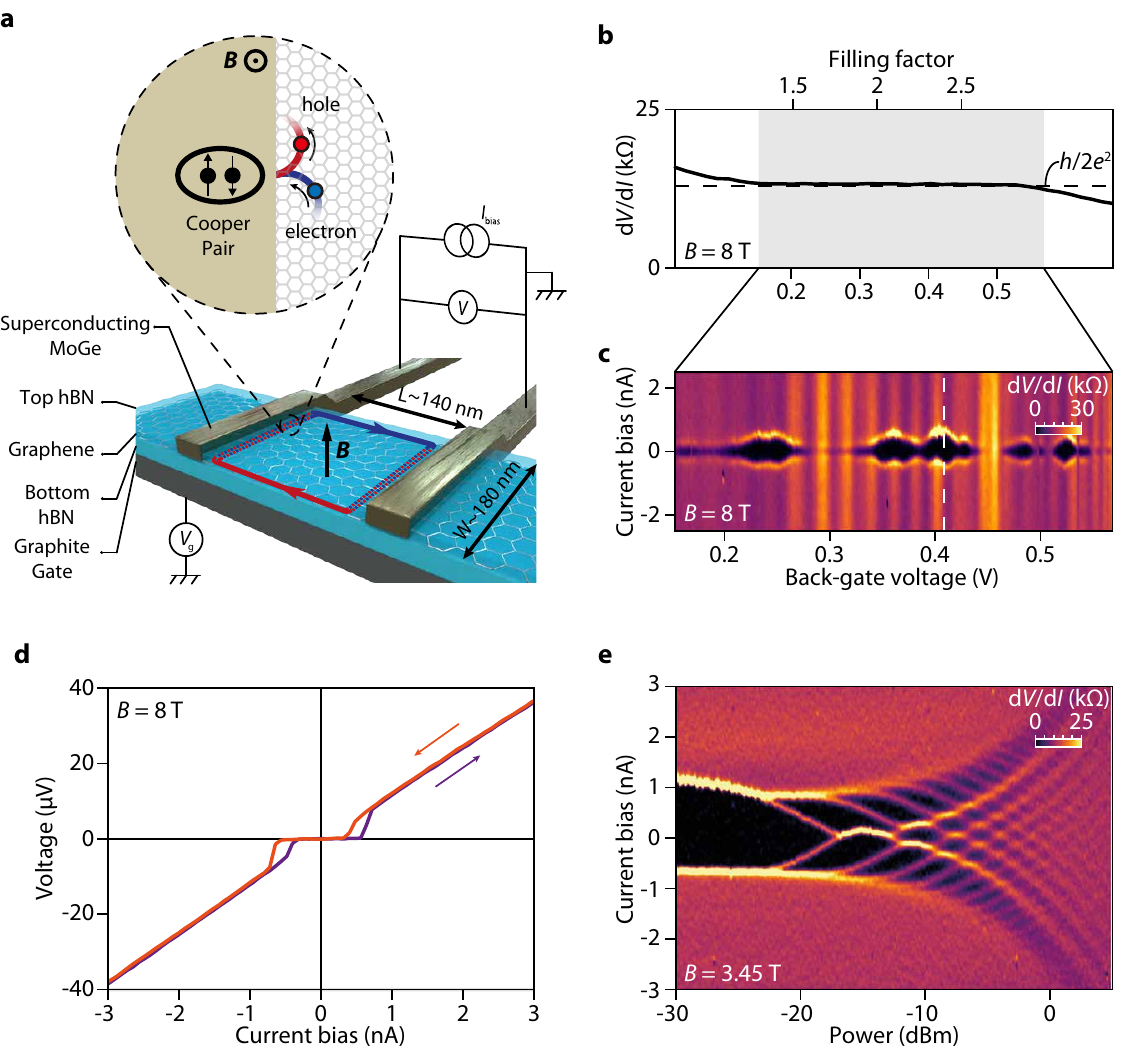}
\centering
\caption{\textbf{Josephson effect through quantum Hall edge channels.} \textbf{a}, Schematics of the Josephson junction consisting of a graphene nanoribbon encapsulated in hexagonal boron nitride (hBN) resting atop a graphite back-gate electrode. Edge contacts are made of superconducting MoGe electrodes. Under magnetic field, $B$, charge carriers of the QH edge channel undergo successive Andreev reflections along the superconducting electrode, which convert incident quasi-electrons (blue channel) into quasi-holes (red channel) and vice-versa, as illustrated with semi-classical cyclotron trajectories in the insert. This electron-hole mixture forms a CAES (dashed blue-red channel) along the interface. 
\textbf{b}, Differential resistance $\text{d}V/\text{d}I$ of device HV88-B measured at 8 T with a dc current bias of 1.2 $\upmu \rm{A}$ and plotted as a function of back-gate voltage, showing a well quantized resistance plateau at $\nu = 2$. 
\textbf{c}, $\text{d}V/\text{d}I$ as a function of back-gate voltage and dc current bias over the plateau region (grey area in \textbf{b}). Superconducting pockets (in black) alternate with finite resistance regions. 
\textbf{d}, Superconducting $I-V$ curves at $V_{\rm g} = 0.41$ V (white line in \textbf{c}). The hysteresis is characterized by separate switching and retrapping currents.
\textbf{e}, Shapiro map of differential resistance measured in device HV88-C as a function of RF power ($f = 1$ GHz) and dc current bias at $\nu = 2.15$ and $B = 3.45$ T.}
\label{Fig1}
\end{figure*}

However, as appealing as this approach may be, coupling superconductivity and QH effect remains an outstanding experimental challenge.
The main dilemma arises from the perpendicular magnetic field, $B$, required for the QH effect in a two-dimensional electron gas (2DEG). While magnetic field generates and strengthens the QH effect by increasing energy gaps in the Landau level spectrum, it adversely affects superconductivity until breakdown at the upper critical field. Furthermore, QH effect profoundly modifies the charge transfer process at the superconductor/QH interface --the Andreev process converting an incident electron into a retro-reflected hole. There, electrons and Andreev-reflected holes have the same chirality, that is, co-propagating forward and forming chiral Andreev edge states (CAES) along the interface\cite{Takagaki98,Hoppe2000}. 
In a Josephson junction connecting two superconducting electrodes through a 2DEG in the QH regime, the CAES and QH edge states form a chiral loop that connects both electrodes (see Fig. \ref{Fig1}a). 
This yields a \textit{chiral supercurrent} that is non-locally split between the two edges of the 2DEG, and subjected to Aharonov-Bohm quantum interference with a $2\phi_0$ flux periodicity, twice that of conventional Josephson junctions\cite{Ma1993,Zyuzin94,Stone2011,VanOstaay2011,Alavirad2018}.

Experimentally, employing graphene in contact with high upper critical field superconductors has led to recent advances in superconducting-QH hybrid devices\cite{Rickhaus12,Komatsu12,Shalom2015,Wan2015,Amet2016,Lee2017,Park17,Seredinski19,Zhao2020a,Wang2021,Gul2022,Hatefipour22,Zhao22}, with evidence of CAES along superconducting interfaces\cite{Zhao2020a,Hatefipour22} and crossed Andreev conversion through narrow contacts\cite{Lee2017,Gul2022}. In 2016, a supercurrent in graphene Josephson junctions in the QH regime was reported\cite{Amet2016}, visible at relatively low magnetic field ($<2$T) and high filling factors ($\geq 6$). Yet, the junction exhibited a standard $\phi_0$-periodic oscillation, in contradiction with the chiral supercurrent prediction. Further work\cite{Seredinski19} suggested that additional conduction channels induced by charge accumulation along etched graphene edges may yield two trivial Josephson junctions in parallel, one per edge, explaining the SQUID-like $\phi_0$-oscillation.

Here, we report on the observation of a chiral supercurrent carried by a single, spin-degenerate QH edge channel in a graphene QH-Josephson junction at bulk filling factor $\nu=2$. Our approach builds on theoretical predictions pointing out that the supercurrent amplitude is inversely proportional to the 2DEG perimeter\cite{Ma1993,Zyuzin94,Stone2011,VanOstaay2011}, and that long superconductor-QH interfaces are detrimental to coherence\cite{Alavirad2018,Manesco22,Kurilovich22,Tang22}. We therefore purposely designed ultra-narrow graphene Josephson junctions, with contact width of 125 to 330 nm (see Extended Data Table~\ref{TabI}), an order of magnitude smaller than any previous work\cite{Rickhaus12,Shalom2015,Amet2016,Seredinski19}, which allows us to unveil the chiral supercurrent.   

We fabricated QH-Josephson junctions with exfoliated graphene nanoribbons encapsulated between hexagonal boron-nitride flakes\cite{Wang13}. The resulting heterostructures lay atop exfoliated graphite flakes that serve as back-gate electrode to tune the charge carrier density with a voltage $V_{\rm g}$ (Fig. \ref{Fig1}a). We chose MoGe as superconducting contact material, with a high upper critical field ($H_{c2}\simeq 12.5$ T, see SI) and a contact resistance of $460 \pm 21$ $\Omega.\upmu$m averaged from 13 devices. To ensure a good quantization of the QH plateau\cite{Abanin2008, Williams2009}, we designed junctions with aspect ratio $L/W \sim 1$ , where $W$ is the width of the superconductor-graphene contact and $L$ the distance between the two contacts (see Extended Data Table \ref{TabI}). All measurements are performed at a temperature of $0.01$ K.

\bigskip
\textbf{Supercurrent on $\nu = 2$ quantum Hall plateau}
\bigskip

An essential benchmark for demonstrating superconducting transport through QH edge channels is the conjunction of a well-defined supercurrent and a QH plateau in the normal state. This is validated in Fig. \ref{Fig1}b and c that display the two-terminal resistance as a function of back-gate voltage at $8$ T for a junction with $L= 140$ nm and $W=180$ nm. 
Figure \ref{Fig1}b displays the normal state resistance measured at a high current bias of $1.2\,\upmu$A, which shows a well quantized $h/2e^2$ plateau, indicating edge transport without bulk backscattering. 
Concomitantly, the differential resistance map at low current bias in Fig. \ref{Fig1}c reveals pockets of supercurrent with zero resistance state color coded in black. Individual current-voltage characteristics (Fig. \ref{Fig1}d) show a well-defined, hysteretic supercurrent with zero voltage drop below a switching current of $0.56$ nA. 

To get insight into the current-phase relation of the junction, we performed Shapiro steps measurements under irradiation by a microwave tone of frequency $f=1$ GHz. Fig. \ref{Fig1}e displays a typical map of $\text{d}V/\text{d}I$ as a function of current bias and microwave power, measured on the $h/2e^2$ QH plateau. As the microwave power is ramped up, Shapiro voltage steps develop and are visible as zero $\text{d}V/\text{d}I$ areas color coded in black. The height of these steps is $2.0\, \upmu \rm{V}$ (see Extended Data Fig. \ref{ExtDataFig2}) in agreement with the theoretical expectation $hf/(2e) = 2.1 \, \upmu \rm{V}$. This observation was reproduced at different frequencies (see SI), and indicates a conventional $2\pi$-periodic current-phase relation, as expected for the spin-degenerate QH edge channel at filling factor 2 (ref.\cite{Ma1993,Stone2011,VanOstaay2011}).

\bigskip
\textbf{Quantum interference of chiral supercurrent}
\bigskip

\begin{figure*}[!ht]
\includegraphics[width=1\linewidth]{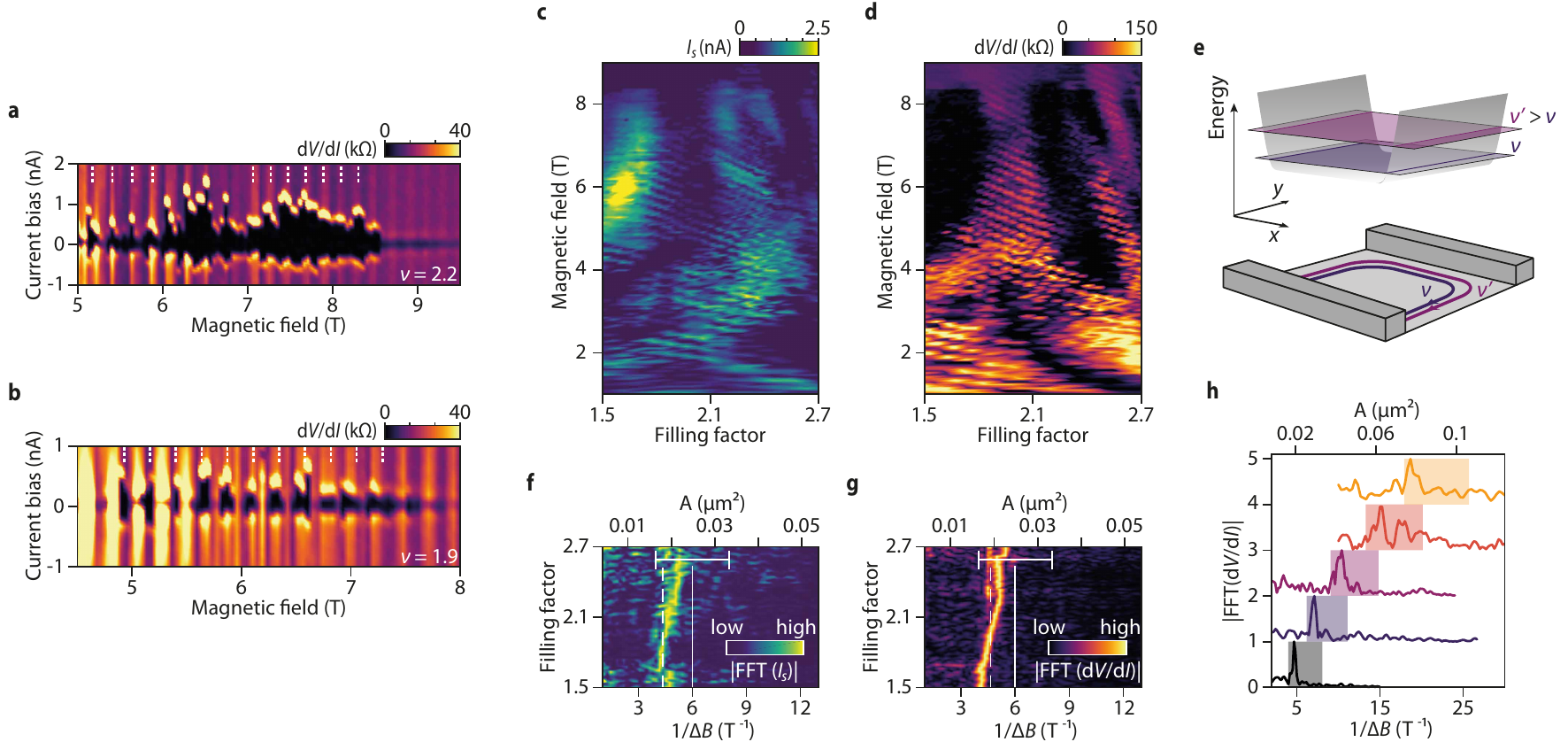}
\centering
\caption{\textbf{$2\phi_0$-periodic chiral supercurrent oscillations.} 
\textbf{a}, \textbf{b}, Differential resistance $\text{d}V/\text{d}I$ at constant filling factors $\nu = 2.2$ in (\textbf{a}) and $\nu = 1.9$ in (\textbf{b}) as a function of magnetic field and dc current bias in device HV88-B. 
The two groups of white lines indicate the expected period for $2\phi_0$-periodic oscillations involving 70\% and 80\% of the graphene area on the left and right combs, respectively, in \textbf{a} and 70\% of the graphene area in \textbf{b}.
\textbf{c}, \textbf{d}, Maps of the switching current and of the zero bias differential resistance as a function of magnetic field and filling factor, respectively. 
\textbf{f}, \textbf{g}, Fourier transforms of the magnetic field dependent oscillations of the supercurrent map in (\textbf{c}) and the differential resistance taken at a finite bias of 4 nA from the set of $I-V$ curves used in (\textbf{d}), respectively, as a function of frequency $1/\Delta B$, equivalent area $A$ for $2\phi_0$-periodic oscillations, and filling factor. Magnetic field ranges used are 2T to 7.5T in (\textbf{f}), and 4T to 10T in (\textbf{g}). For both Fourier transforms, a slowly varying background has been subtracted.    
The white solid lines indicate the frequency corresponding to the $2\phi_0$-periodic oscillations for the geometric area, whereas the white dashed lines take into account a correction of the area $(L-2l_B) \times (W-2l_B)$ that accounts for a better estimate of the positioning of the QH edge states at $l_B$ (taken at the average value of 4.5T in (\textbf{f}) and 7 T in (\textbf{g})) from the physical edge\cite{Coissard22b}. The white error bars indicate the experimental uncertainty on the geometry evaluation (see Methods and Extended Data Fig. \ref{ExtDataFig1}). 
\textbf{e}, Schematics of the Landau levels and of the position of their edge states in the device for two different filling factors at a fixed magnetic field. 
\textbf{h}, Fourier transforms of the differential resistance versus magnetic field at constant filling factor for junctions with different sizes (devices B to F from bottom to top, sample HV88). The color-coded rectangles indicate the frequencies for $2\phi_0$ oscillation based on the geometric area. Their width corresponds to the uncertainty in the geometric area.}
\label{Fig2}
\end{figure*}

We now turn to the central result of this work, that is, demonstration of the distinct $2\phi_0$ flux periodicity of the chiral supercurrent. Supercurrent oscillations are readily seen in Fig. \ref{Fig2}a and b, which display two typical $\text{d}V/\text{d}I$ maps as a function of magnetic field and current bias, at filling factor $\nu = 1.9$ and $2.2$, respectively (see SI for the evaluation of $\nu$). 
At constant filling factor (non-constant back-gate voltage), superconducting pockets are observed over a large $B$-range showing periodic oscillations.
To evaluate the periodicity of the supercurrent oscillations, we superpose on Fig. \ref{Fig2}a and b a comb of white dashed lines equally spaced by $2\phi_0/A$, that is, twice the superconducting flux quantum divided by the effective graphene area. Clearly, maxima of the supercurrent oscillation match the $2\phi_0$ flux periodicity, the hallmark of the chiral supercurrent. Slight shifts of the oscillation are sometimes seen, such as for the last two white dashed lines above 7 T in Fig. \ref{Fig2}b.

The key to this finding lies in analyzing the supercurrent oscillation at constant filling factor instead of fixed back-gate voltage. Indeed, for the latter,  the charge carrier density is constant and the filling factor decreases with $B$, leading to an inward displacement of the QH edge channels relative to the physical edge of the graphene flake and hence a decrease in the effective area. For our ultra-narrow junctions, such an area variation is sufficient to alter the flux periodicity (see SI Fig. S9). Conversely, at constant filling factor, that is, at constant Fermi level position in the cyclotron gap, the edge channel remains at a nearly $B$-independent spatial position, enabling us to unveil the $2\phi_0$ flux periodicity.

To substantiate our findings we present in Fig. \ref{Fig2}c a full mapping of the supercurrent (see Methods) over the $h/2e^2$ QH plateau. 
The $2\phi_0$-periodic supercurrent oscillations form fringes with negative slope, extending all over the plateau, on top of a slowly varying supercurrent background. 
Figure \ref{Fig2}f shows a Fourier transform of these fringes performed at constant filling factor. 
The peak at $1/\Delta B$, which indicates the period $\Delta B$ of the supercurrent oscillations, is close but slightly smaller than the value expected for $2\phi_0$-periodic oscillations related to the geometric graphene area (see white solid line and error bar related to the uncertainty in the nanoribbon width).
This slightly smaller effective area can be accounted for by taking into account the distance between the actual edge state and the physical graphene edge, which is of the order of the magnetic length $l_{\rm B}=\sqrt{\hbar/eB}$ (see white dashed line).
The close match between the periodicity and the corresponding geometric graphene area indicates that the QH edge channel is located in the immediate vicinity of the crystal edge, in agreement with recent scanning tunneling spectroscopy results\cite{Coissard22b}.

In addition, one can notice a shift of the peak to higher frequency upon increasing $\nu $. 
This shift reflects the increase in area with $\nu $ due to the slight displacement of the QH edge channel towards the graphene edge, as sketched in Fig. \ref{Fig2}e, therefore providing a signature of the Landau level dispersion at the edge. From Fig. \ref{Fig2}f, we obtain an area variation of $4.5\times 10^{-3}$ $\upmu$m$^2$, which corresponds to an edge channel displacement of $7$ nm, i.e. of the order of the magnetic length.

Importantly, the supercurrent oscillations are accompanied by resistance oscillations of the resistive state, as seen in Figs. \ref{Fig1}c, \ref{Fig2}a and \ref{Fig2}b. 
The $\text{d}V/\text{d}I$ map at zero bias in Fig. \ref{Fig2}d shows similar fringes with negative slope as for the critical current in Fig. 2c (both maps are extracted from the same set of data).
The Fourier transform of the resistance oscillations (Fig. \ref{Fig2}g) also reveals a $2\phi_0$ flux periodicity with the same frequency shift with $\nu $ as the supercurrent. 
We ascribe these resistive state oscillations to a consequence of the Aharonov-Bohm phase picked by the charge carriers circulating around the QH channel loop, while undergoing chiral Andreev reflections along the superconducting interfaces. 
Note that, contrary to usual Josephson junctions, maxima of supercurrent do not necessarily coincide with minima of resistive state (see e.g. Fig. \ref{Fig2}a and b). 
This is not surprising as the phase accumulated by CAES differs from that accumulated in the normal state by an additional phase shift picked up upon Andreev reflections.

\begin{figure}[ht!]
\includegraphics[width=1\columnwidth]{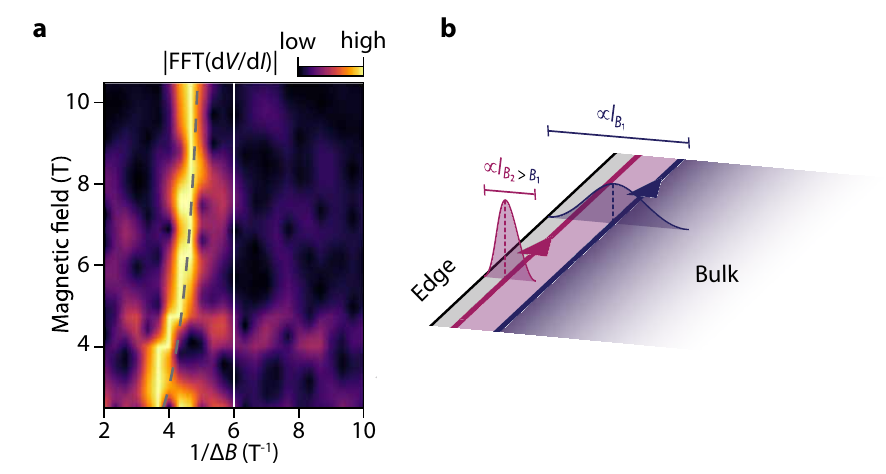}
\caption{\textbf{Edge state wavefunction shrinkage.} \textbf{a}, Fourier transform of the differential resistance at filling factor $\nu = 1.9$ and finite current bias of 3 nA of the same data as in Fig. \ref{Fig2}d, computed in a magnetic field window of $3$ T sliding along the magnetic field axis, and plotted as a function of inverse period $1/\Delta B$ and magnetic field. It reflects the $B$-dependence of the oscillation period visible in Fig. \ref{Fig2}d. A slowly varying background has been subtracted. The white solid line indicates the frequency corresponding to the $2\phi_0$-periodic oscillations for the geometric area $W\times L$. The dashed line shows the expected frequency with a correction of the area as $(L-2l_B) \times (W-2l_B)$. 
\textbf{b}, Schematics of the edge state position at the edge of the graphene flake for two different values of magnetic field. The width of the gaussian wavefunction of the zeroth Landau level is proportional to the magnetic length $l_B$, which decreases with $B$. As a result, the mean position of the edge channel shifts towards the physical edge upon increasing $B$, leading to an increase of the effective area. 
The area variations set by the Landau level wavefunction shrinkage illustrated in (\textbf{b}) are in excellent agreement with the observed frequency shift in (\textbf{a}).}
	\label{Fig3}
\end{figure}

Inspecting Figs. \ref{Fig2}c and d more closely, we see a decrease in the oscillation period with magnetic field, which translates to an increase in effective area. This variation can be accounted for by a decrease in the width of the Landau level wavefunction that scales as $l_{\rm B}$, which in turn results in the displacement of the QH edge channels towards the graphene edge (see Fig. \ref{Fig3}b). Here, unlike Figs. \ref{Fig2}f, g and e, the area variation is not related to the edge dispersion of the Landau level, but to the wavefunction shrinkage. An approximate correction of the flux area by the magnetic length $(L-2l_B)\times (W-2l_B)$, assuming a displacement of $l_B$ relative to the graphene edges, is shown as a grey dashed line in Fig. \ref{Fig3}a. This correction is superimposed in Fig. \ref{Fig3}a on the Fourier transform of Fig. \ref{Fig2}d, which is computed here in a $B$-window sliding along the $B$-axis at fixed filling factor, and fits remarkably well the observed shift. Such a fine analysis of the area variation due to the wavefunction shrinkage with magnetic field confirms that the $2\phi_0$ flux periodicity is observed over the whole range of magnetic field.

We obtained equivalent results for six other junctions for both critical current and resistive state oscillations (see SI). As a summary, we present in Fig. \ref{Fig2}h the Fourier transforms of the oscillation for five junctions of different sizes, which show an excellent agreement with the expected $2\phi_0$ flux periodicity indicated by colored rectangles. Such a systematic set of data therefore provides a conclusive demonstration of the $2\phi_0$ flux periodicity of the chiral supercurrent. 

\bigskip
\textbf{Velocity renormalization at superconducting interface}
\bigskip

\begin{figure*}[!ht]
\includegraphics[width=12cm]{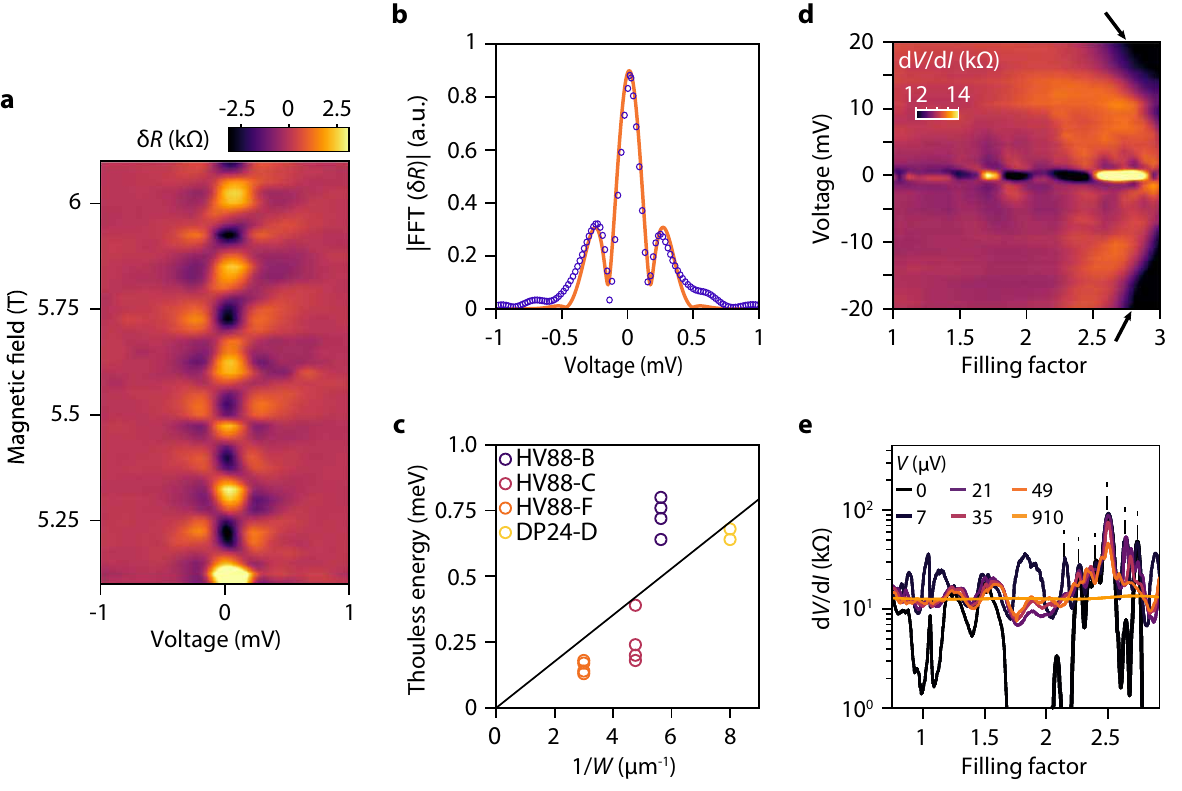}
\centering
\caption{\textbf{Aharonov-Bohm interferometry with chiral Andreev edge states.} 
\textbf{a}, Checkerboard pattern of the differential resistance oscillation (after background subtraction) as a function of the measured voltage and magnetic field, at a fixed filling factor for device HV88-B. 
\textbf{b}, Fourier transform amplitude of (\textbf{a}) as a function of measured voltage, showing a lobe structure. The solid line is a fit using the theory of QH Fabry-P\'{e}rot interferometers of Ref.\cite{Deprez21} with a bias asymmetry parameter $x=0.02$. The maxima of the two side lobes are separated by the Thouless energy, which is plotted in (\textbf{c}) for various devices as a function of the inverse of the superconducting interface width $1/W$. Different points for a given sample correspond to measurements at different filling factors. The black line is a linear fit with the slope yielding a renormalized edge velocity of $2.1 \pm 0.5 \times 10^4$ m/s. 
\textbf{d}, Differential resistance map of device HV88-C at $3$ T as a function of filling factor and measured dc voltage across the junction. The QH breakdown at the edge of the plateau is visible at high voltage, as indicated by the two black arrows. 
\textbf{e}, Linecuts measured at the same filling factor and magnetic field as in (\textbf{d}) but with higher resolution and lower bias, showing superconducting pockets as those seen in Fig. \ref{Fig2}c. The oscillations disappear on a scale of the order of the Thouless energy, while the large-scale fluctuations subsist up to about 1.4 m$\rm{V}$. Note that the $2\phi_0$-periodic oscillations are visible at low voltage bias as indicated by the dashed black lines.}
\label{Fig4}
\end{figure*}

We now analyse the energy dependence of the resistance oscillations. Figure \ref{Fig4}a shows $\text{d}V/\text{d}I$ as a function of the measured voltage across the junction and magnetic field at constant filling factor $\nu=2.6$. The $2\phi_0$-periodic oscillations undergo a phase shift of $\pi$ at finite bias $V=155\,\upmu$V, forming a checkerboard pattern resembling that of Fabry-P\'{e}rot QH interferometers\cite{Deprez21,Ronen2021}. While latter contain a loop of QH edge channels partially transmitted through quantum point contacts, here the configuration is different since the quantum point contacts are replaced by superconducting interfaces, which are partially reflecting the QH edge states through Andreev processes with finite transparency. The checkerboard pattern thus results from the phase shift $4\pi\delta \epsilon /E_{\rm Th}$ due to the finite energy $\delta \epsilon $ of the injected electrons, where $E_{\rm Th}=h/\tau$ is the ballistic Thouless energy related to the traveling time $\tau $ along half of the loop perimeter. 
Following Ref.\cite{Deprez21}, we plot the Fourier transform of the checkerboard pattern in Fig. \ref{Fig4}b and extract $E_{\rm Th} = 640\,\upmu$eV. 
Importantly, this value is smaller than what is expected for a QH edge state velocity $v_{\rm QH} = 1.4\times 10^5$ m$/$s measured in QH interferometers \cite{Deprez21}, which would give $E_{\rm Th}=1.8$ meV for the loop with half-perimeter $L+W = 318$ nm. This estimate is instead consistent with the data at 14T above the critical field of the electrode shown in Extended Data Figure 5, in which we measured $E_{\rm Th}= 1.6$ meV. 

This discrepancy can be understood if we consider the CAES velocity renormalization along the superconducting interface\cite{Hoppe2000,Alavirad2018}. Due to the time delay of the Andreev process, that is, the extra time of the order of $h/\Delta$ that an impinging electron spends in the electrode with a superconducting gap $\Delta $, the edge velocity $v_{\rm SC}$ along the superconducting interface drops significantly. As a result, $\tau =(L/v_{\rm QH}+W/v_{\rm SC})$ and the Thouless energy is dominated by the interface width $W$, $E_{\rm Th} \sim hv_{\rm SC}/W$, when $v_{\rm SC}\ll v_{\rm QH}$. From the analysis of four junctions of different sizes (see other checkerboards in Extended Data Fig.~\ref{ExtDataFig3}), taken between 4.5 and 6 T, the evolution of $E_{\rm Th}$ with $1/W$ shown in Fig. \ref{Fig3}c gives an estimate of $v_{\rm SC} \sim 2.1\pm 0.5 \times 10^4$ m$/$s, which is much smaller than $v_{\rm QH}$. 
Theoretically, the edge state velocity renormalization for a non ideal interface can be derived from the Andreev bound states spectrum\cite{Hoppe2000} (see Methods). From Eq. (\ref{eqn_vsc}) in Methods, we obtain $v_{\rm SC} \sim 6\times 10^4$ m$/$s, whose order of magnitude is in good agreement with the experimental estimate. Interestingly, such a renormalization leads to a suppression of $v_{\rm SC}$ when $\Delta $ vanishes at the upper critical field of the electrodes. This effect may be delicate to observe due to a crossover at $E_{\rm Th} \gtrsim \Delta $ to a regime where normal electrons take over the interference pattern.

The key parameter that limits the supercurrent in QH Josephson junctions at finite temperature is therefore the width of the superconducting interface, which severely reduces the Thouless energy via edge state velocity renormalization\cite{Hoppe2000,Alavirad2018}. 
This could provide an explanation for measurable supercurrent limited to small magnetic fields and high filling factors in the large junctions of Ref.\cite{Amet2016}. 
In contrast, in the graphene nanoribbons studied here, a well-defined critical current withstands up to 8 T on the $h/2e^2$ QH plateau. 
Furthermore, contrary to conventional Josephson junctions, increasing interface transparency further suppresses supercurrent by reducing $v_{\rm SC}$ (see Eq. (\ref{eqn_vsc}) in Methods) and consequently the Thouless energy. 
Counter-intuitively, low transparency contacts may thus be a way to enhance critical current. 

For consistency check, we studied wider junctions with $W= 2.3$ to $2.4\,\upmu$m and $L=100$ to $300$ nm on sample HV88 with similar contact transparencies (see HV88-G and HV88-H in SI Table S1). In these devices, the supercurrent is absent on QH plateaus and mostly visible at the transition between plateaus, where transport takes place through the bulk of the junction (see Extended Data Fig.~\ref{ExtDataFig4}). Note that other mechanisms involving decoherence due to disorder at the interface\cite{Manesco22,Kurilovich22} or vortices\cite{Kurilovich2022b,Tang22} could be equally relevant and lead to the same conclusion, although more delicate to assess experimentally\cite{Zhao22}.

Inspecting the current-voltage characteristics at higher bias, we further observe signatures of CAES interference along the superconducting interface, which emerge as filling factor dependent resistance fluctuations\cite{Chtchelkatchev2007,Batov09}. Figure \ref{Fig4}d and e display the differential resistance fluctuations related to the background fluctuations of the switching current and resistance seen in Figs. \ref{Fig2}b and d. Those are strongly suppressed above a voltage bias of about $1.4$ mV, which is close to $2 \Delta/e$ as usually expected in standard hybrid Josephson junctions\cite{Blonder82}. The fluctuation pattern echoes recent conductance fluctuations observed downstream superconducting interfaces\cite{Zhao2020a}, ascribed to CAES interference involving intervalley-scattering\cite{Manesco2021}.

Interestingly, Fig. \ref{Fig4}d reveals the breakdown of the $h/2e^2$ QH plateau with a drop in resistance at high current bias near the transition to the next QH plateau (see black arrows). Although at $\nu=2$ this breakdown occurs here at relatively large bias with respect to superconducting features, the situation may be detrimental for other QH plateaus with small energy gaps as those of the fractional QH states. For these correlated states, the need to reduce the superconducting interface width to obtain a robust supercurrent should therefore meet another constraint related to edge transport, since QH breakdown critical current is significantly weakened in narrow structures. 

Finally, we shall recall the spurious charging effects in Fabry-P\'{e}rot QH interferometers that have competed with and obscured the Aharonov-Bohm effect for some time\cite{halperin11}. It is therefore appropriate to examine the role of the Coulomb interaction in our small devices. As shown in Extended Data Figure \ref{ExtDataFig5}, we observed Coulomb blockade signatures at the edges of the QH plateau and in the transition to the next plateau, in the regime where a compressible island should form in the bulk. However, the Coulomb blockade quickly disappears as soon as the QH plateau is entered to give way to Aharonov-Bohm interference. Furthermore, the slope of the interference fringes in the $\nu-B$ plane is negative, which is in agreement with Aharonov-Bohm dominated oscillations\cite{halperin11}. The distinct separation between Coulomb blockade signatures near the plateau transition and Aharonov-Bohm interference on the plateau, along with the negative fringes slope and correct area hence rule out charging effects in the $2\phi_0$-periodic oscillations.

\section*{Methods}

\subsection*{Sample fabrication and parameters}

The graphene and hBN flakes used to fabricate the heterostructures were exfoliated and selected according to their surface cleanliness as determined by atomic force microscopy (AFM). AFM images were further used to locate the nanoribbons for subsequent lithography steps, and to measure the nanoribbon width at different positions with an accuracy of about 30 nm, as shown in Extended Data Fig. 1. Graphene/hBN heterostructures were assembled with the van der Waals pick-up technique\cite{Wang13} using a polycarbonate polymer. The heterostructures were deposited on a thin layer of graphite serving as a back-gate electrode. Large scale wiring (Ti/Au bilayer) and superconducting MoGe electrodes were patterned by two consecutive fabrication sequences including electron-beam lithography, metal deposition and liftoff. MoGe electrodes were deposited by DC-magnetron sputtering of a $\rm{Mo}_{50}\rm{Ge}_{50}$ target, after etching of the stack with a CHF$_3$/O$_2$ plasma directly through the resist pattern used to define the contacts. 
The critical temperature of MoGe tested on a separate film is 5.9 K corresponding to a superconducting gap of $\Delta = 895\,\upmu$eV. The upper critical field of the MoGe electrodes of the sample varies between 12.5 to 13 T (see Fig. S1).
The geometrical parameters of the two samples studied in this work, HV88 and DP24, are reported in Extended Data Table \ref{TabI}. The top hBN, bottom hBN and graphite thicknesses are 12.7, 23.2 and 7.7 nm for HV88, and 25, 36 and 20 nm for DP24.

\begin{figure*}[!ht]
\centering
\includegraphics[width=16cm]{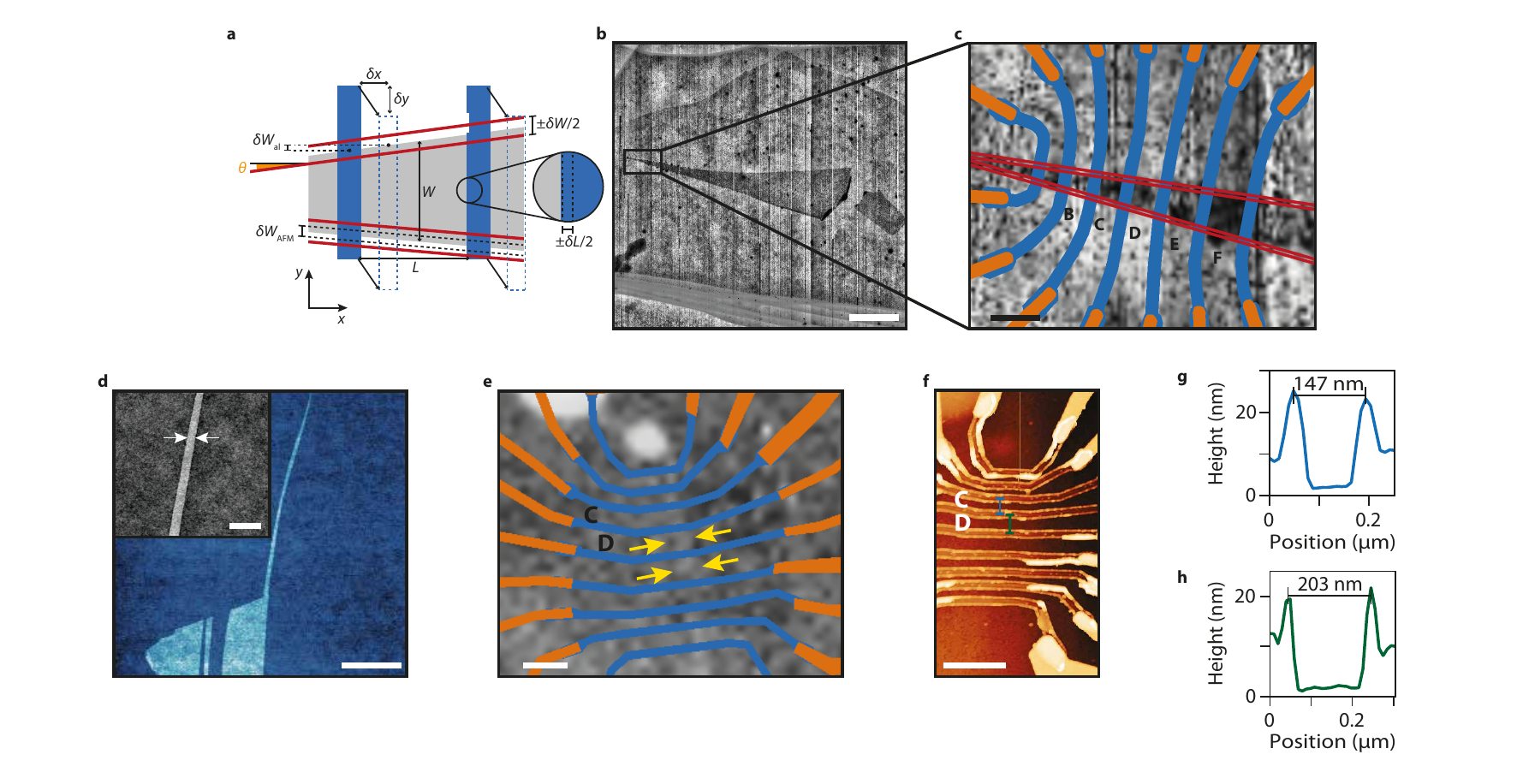}
\caption{\textbf{hBN-encapsulated graphene nanoribbon samples.} 
\textbf{a}, Schematic describing the uncertainty in the junction width $\delta W$ (red lines) and length $\delta L$ (vertical black dashed lines in the zoom). Horizontal black dashed lines indicate the uncertainty in the graphene width $\delta W_{\rm{AFM}}$ from AFM imaging. Blue rectangles represent MoGe electrodes correctly aligned, dashed blue lines indicate the MoGe electrode position with a misalignment $(\delta x, \delta y)$, which results in an uncertainty $\delta W_{\rm{al}}$ on the graphene edges position. 
\textbf{b}, AFM picture for sample HV88, the encapsulated graphene is readily visible with contrast enhancement. 
\textbf{c}, Zoom on the graphene narrow part in (\textbf{b}), with the lithography pattern overlayed (Ti/Au lines in orange and MoGe electrodes in blue). Red lines indicate the total width uncertainty $\pm \delta W /2$ around the graphene edges. 
\textbf{d}, AFM picture of the graphene nanoribbon of sample DP24 before hBN encapsulation. Inset: higher resolution AFM picture. The arrows indicate the nanoribbon width of 125 nm. 
\textbf{e}, AFM picture of sample DP24 after hBN encapsulation. The yellow arrows indicate the nanoribbon position. The lithography pattern is overlayed on the picture.
The scale bars are 5 $\upmu$m (\textbf{b}), 400 nm (\textbf{c}), 2 $\upmu$m (\textbf{d}), 500 nm (\textbf{d}) inset, 500 nm (\textbf{e}).
\textbf{f}, AFM picture of sample DP24 contact fabrication. The scale bar is 1 $\upmu$m.
\textbf{g}, \textbf{h}, Height profiles along the blue and green lines in the AFM image in \textbf{f}. The excess thickness on the contact edge results from the liftoff process of the MoGe. It provides a good estimate of the position of the contact to graphene.}
\label{ExtDataFig1}
\end{figure*}

\begin{figure*}[!ht]
\centering
\includegraphics[width=7cm]{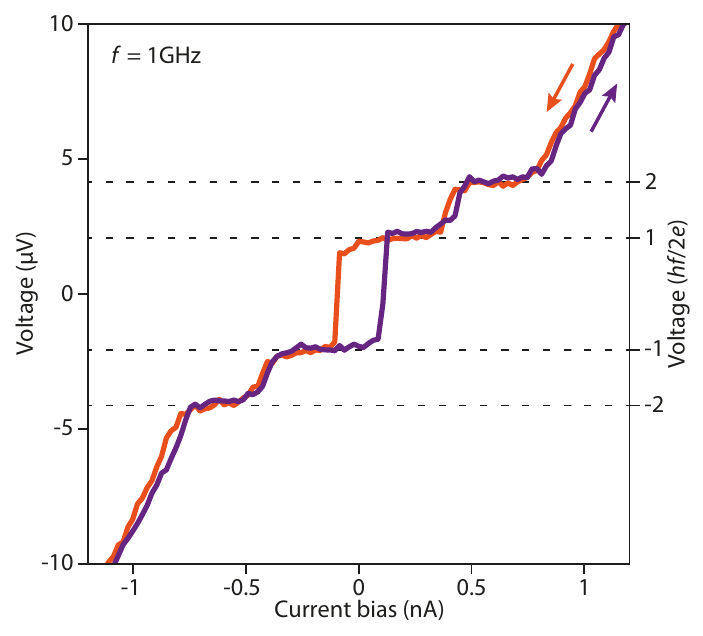}
\caption{\textbf{Shapiro steps of the chiral supercurrent.} $I-V$ characteristics measured at 3.45 T on sample HV88-C, under microwave radiation of frequency $f=1$ GHz. The $I-V$ curves are extracted from the data of Fig. \ref{Fig1}e at a microwave power of -13.8 dBm. The color-coded arrows indicate the current sweep direction.} 
\label{ExtDataFig2}
\end{figure*}

\begin{figure*}[!ht]
\centering
\includegraphics[width=12cm]{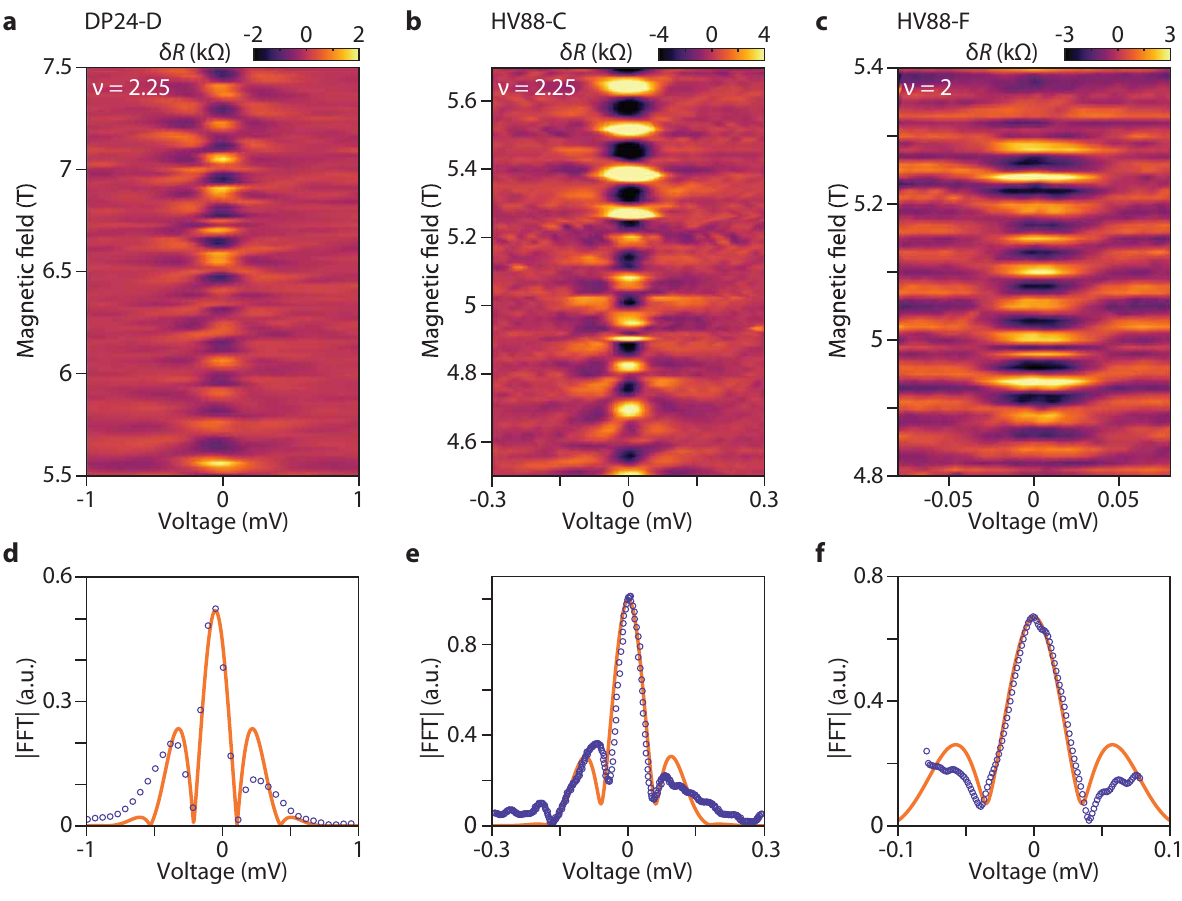}
\caption{\textbf{Cherkerboard patterns.} 
\textbf{a-c}, Differential resistance modulation with magnetic field and measured voltage showing a characteristic checkerboard pattern. 
\textbf{d-f}, Fourier transform amplitude at the checkerboard oscillation frequency as a function of the measured voltage. The orange curve is a fit to the data following the same procedure as in Ref.\cite{Deprez21}. 
\textbf{a,d}, Junction DP24-D with fit parameters $E_{\rm{Th}} = 640 \, \upmu \rm{V}$ and $x = 0$. 
\textbf{b,e}, Junction HV88-C with fit parameters $E_{\rm{Th}} = 240 \, \upmu \rm{V}$ and $x = 0.02$. 
\textbf{c,f}, Junction HV88-F with fit parameters $E_{\rm{Th}} = 140 \, \upmu \rm{V}$ and $x = 0.02$.}
\label{ExtDataFig3}
\end{figure*}

\begin{figure*}[!ht]
\centering
\includegraphics[width=12cm]{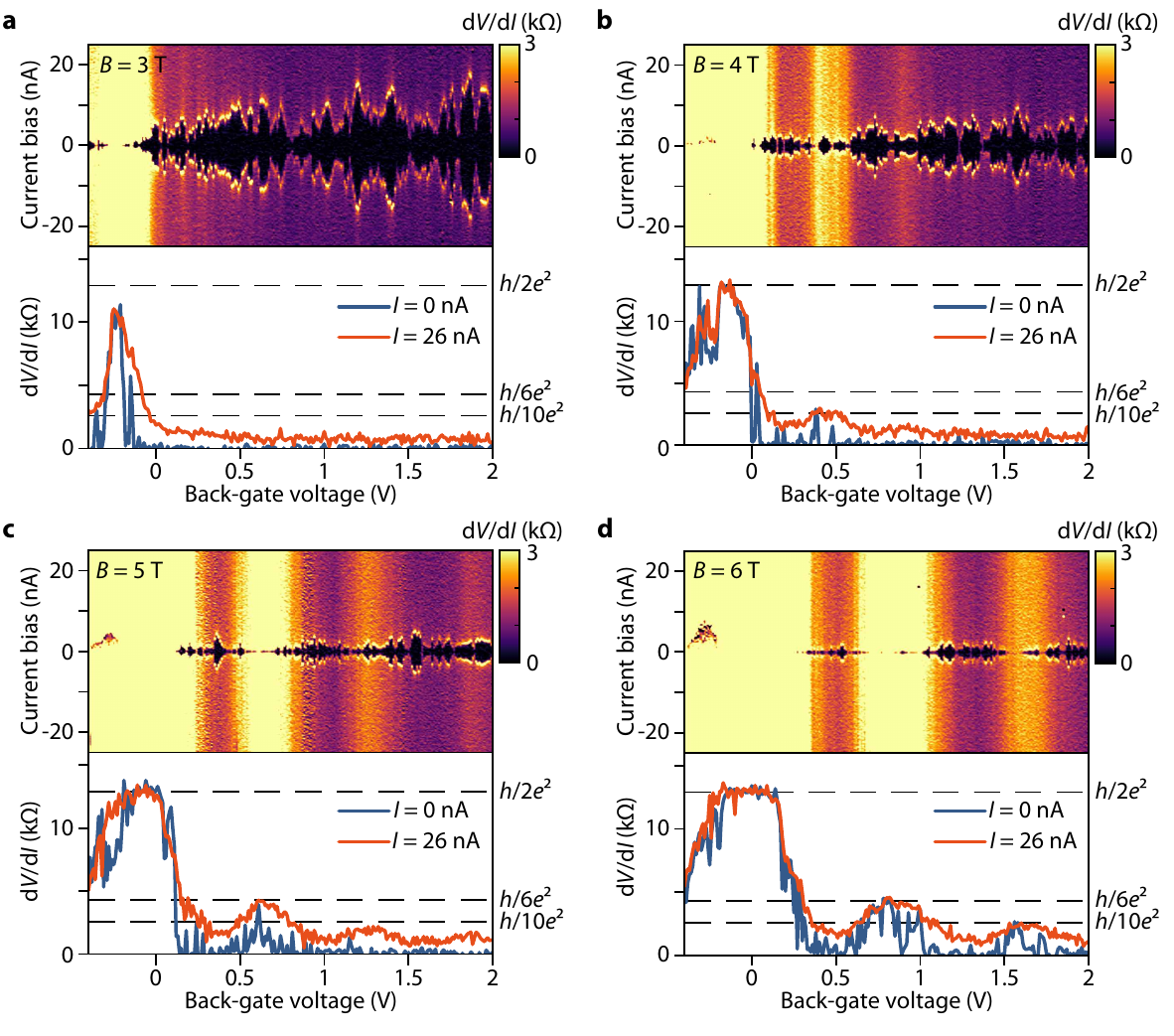}
\caption{\textbf{Wide graphene Josephson junction.} 
 \textbf{a}-\textbf{d}, Top panels show differential resistance maps as a function of back-gate voltage and dc current bias of sample HV88-G ($W = 2.3\,\upmu$m, $L=107$ nm). Bottom panels are linecuts of the differential resistance at dc current biases of 0 nA and 26 nA, which show the emergence of supercurrent pockets (zero resistance reached by the blue curve), and the corresponding resistive state (yellow curve). The magnetic field is indicated in each top panel. Supercurrent is visible only when the resistance is not quantized, that is, at filling factors of QH plateaus that are not developed, or in between plateaus. When a QH plateau emerges, as for instance the $h/2e^2$ plateau for $B\geq4$ T or the $h/6e^2$ plateau for $B\geq6$ T, the supercurrent vanishes. This weakness is even more marked in sample HV88-H shown in SI, which is twice longer. Note that the oscillatory behavior of the resistance (see red curve at 26 nA) is characteristic of QH devices in two-terminal configuration with $L\ll W$, see Ref.\cite{Abanin2008, Williams2009}.} 
\label{ExtDataFig4}
\end{figure*}

Each junction area $A$ is estimated from the superposition of the lithographic design with the AFM image of the heterostructure as illustrated in Extended Data Fig. \ref{ExtDataFig1}, and confirmed with the AFM image of the device after contact fabrication (see Extended Data Fig. \ref{ExtDataFig1}f-h). The precision on the flake width from the AFM image $\delta W_{\rm{AFM}}$ is estimated to be $\pm 30 \ \rm{nm}$. The junction width uncertainty $\delta W$ is dominated by this $\delta W_{\rm{AFM}}$ with negligible contribution from lithographic alignment uncertainty ($\delta W_{\rm{al}}$). Indeed, a lithographic misalignment $\delta y$ along the junction width direction, estimated to be smaller than 100 nm, has no effect on the junction width since the electrodes are parallel over more than 300 nanometers as shown in Extended Data Fig. \ref{ExtDataFig1}a. The alignment precision of the superconducting electrodes with respect to the flake in the $x$ direction is estimated to be about $\delta x = \pm 130$ nm. The resulting width uncertainty is given by $\delta W_{\rm{al}} = \delta x \sin{\theta}$ with $\theta$ the nanoribbon opening angle as defined in Extended Data Fig. \ref{ExtDataFig1}a. With $\theta_{\rm{HV88}} = 7.5^\circ$ and $\theta_{\rm{DP24}} = 0.6^\circ$, the width uncertainty from alignment $\delta W_{\rm{al}}$ is about $\pm 17 \ \rm{nm}$ for HV88 and $\pm 1.4 \ \rm{nm}$ for DP24. The total uncertainty on the junction width $\delta W$ is thus given by $\delta W = \sqrt{\delta W_{\rm{AFM}}^2+\delta W_{\rm{al}}^2}$, hence $\pm 35 \ \rm{nm} $ for HV88 and $\pm 30 \ \rm{nm} $ for DP24. The uncertainty on the junction length $\delta L$ originating from lithographic development and etching is estimated from dose test measurements to be about $\pm 20 \ \rm{nm}$. The resulting uncertainty in the junction area $\delta A = L\Delta W + W\delta L$ is reported for all junctions in Extended Data Table \ref{TabI}.

\subsection{Measurements}

Measurements were performed in a dilution refrigerator with a base temperature of $0.01$ K and equipped with a 14 T superconducting solenoid magnet. The sub nA switching current measurements (see Fig. \ref{Fig1}) were resolved thanks to a heavily filtered wiring comprising feedthrough $\pi$-filters (Tusonix \#4209-053) at room temperature, custom-designed all stainless steel pico-coax cables (0.05 mm inner conductor diameter, 0.5 mm outer diameter), copper powder filters at the mixing chamber plate, and 47 nF (NPO dielectric\cite{teyssandier10}) capacitor to ground on the sample holder. $I-V$ characteristics were measured in a pseudo-four probe configuration as illustrated in Fig. \ref{Fig1}a with an acquisition card (National Instruments NI-6346). The voltage drop across the junction was measured with a differential FET amplifier (DLPVA-100-F-D from FEMTO Messtechnik GmbH). Oversampling at 30 to 150 kHz per point with the acquisition card, along with very low level of low-frequency noise obtained by ground loop mitigation, enabled us to obtain well defined $I-V$ curves with an acquisition time of $1$ to $5$ s per curve. This fast $I-V$ acquisition enabled us to measure more than 30 000 $I-V$ curves in the $V_{\rm g}-B$ plane to entirely map out the QH plateau and to observe the critical current and resistance oscillations. Differential resistance data were obtained by numerically differentiating $I-V$ characteristics. Only the resistance data of Fig. S13c and d were obtained by standard lockin amplifier technique. Shapiro maps were obtained with a microwave radiation source feeding a coaxial cable with the open end positioned about one millimeter above the devices. 

\subsection{Edge state velocity renormalization}

The energy spectrum of the Andreev bound states along a superconducting interface has been calculated by Hoppe et. al.\cite{Hoppe2000}. For a non-ideal interface, it reads:
\begin{equation}
\label{eqn_e_position}
E = \Delta \frac{(2N+1)\pi \pm \arccos{\Gamma_0}-2X k_{\rm{F}}}{q+\pi \Delta/(\hbar \omega_{\rm{c}})},
\end{equation}
with $X$ the distance from the interface, $N$ the Landau level index, and $\Gamma_0$ is an oscillatory function of the filling factor. $q = 2s/(1+s^2+w^2)$ with $s = v_{\rm{N}}/v_{\rm{S}}$ the Fermi velocity mismatch between graphene and the superconductor and $w = k_{\rm{F}}U_0/E_{\rm{F}}$ is the scattering at the interface corresponding to 2Z in the Blonder-Tinkham-Klapwijk model\cite{Zulicke01,Blonder82}. 
While $s$ can be evaluated to be 1.4 (Fermi velocity is $v_{\rm{S}} = 5.2 \times 10^5\,\rm{m/s}$ in MoGe\cite{Kim2018}, and $v_{\rm{N}} = 1 \times 10^6\,\rm{m/s}$ in graphene), there is no experimental way to assess $U_0$. However, using the expression $\mathcal{T}=\frac{v_{\rm{N}} v_{\rm{S}}}{(v_{\rm{N}} + v_{\rm{S}})^2/4 + U_0^2}$ of the transparency for a non-perfect superconductor-metal interface of Ref.\cite{Mazin95}, we can identify $q = \mathcal{T}/(2-\mathcal{T})$ with $\mathcal{T} = 4s/((1+s)^2+w^2)$.
Then, the edge velocity is given by $v_{\rm{SC}} = - \frac{l_B^2}{\hbar} \frac{\partial E}{\partial X}$, which, with Eq. (\ref{eqn_e_position}), leads to:
\begin{equation}
\label{eqn_vsc}
v_{\rm{SC}} = \frac{2v_{\rm{N}}}{\pi+q\frac{\hbar\omega_{\rm{c}}}{\Delta}}.
\end{equation}
The contact transparency $\mathcal{T} = 0.5$ (see SI table S1) gives $q = 0.3$. With a superconducting critical temperature $T_{\rm{c}} =5.9$ K for MoGe (see SI) corresponding a superconducting gap $\Delta_0 = 1.76 k_{\rm{B}} T_{\rm{c}} = 895 \ \upmu \rm{eV}$, we estimate $\Delta = \Delta_0 \sqrt{1-(H/H_{\rm{c2}})^2} = 820 \, \upmu \rm{eV}$ at 5 T, with $H_{\rm{c2}} = 12.5$ T (see SI). With $\hbar \omega_{\rm{c}} = 81 \rm{meV}$ at 5 T, we obtain $v_{\rm{SC}} = 6 \times 10^4 \ \rm{m/s}$.

\subsection{Coulomb blockade at the plateau transition}

\begin{figure*}[!ht]
\centering
\includegraphics[width=12cm]{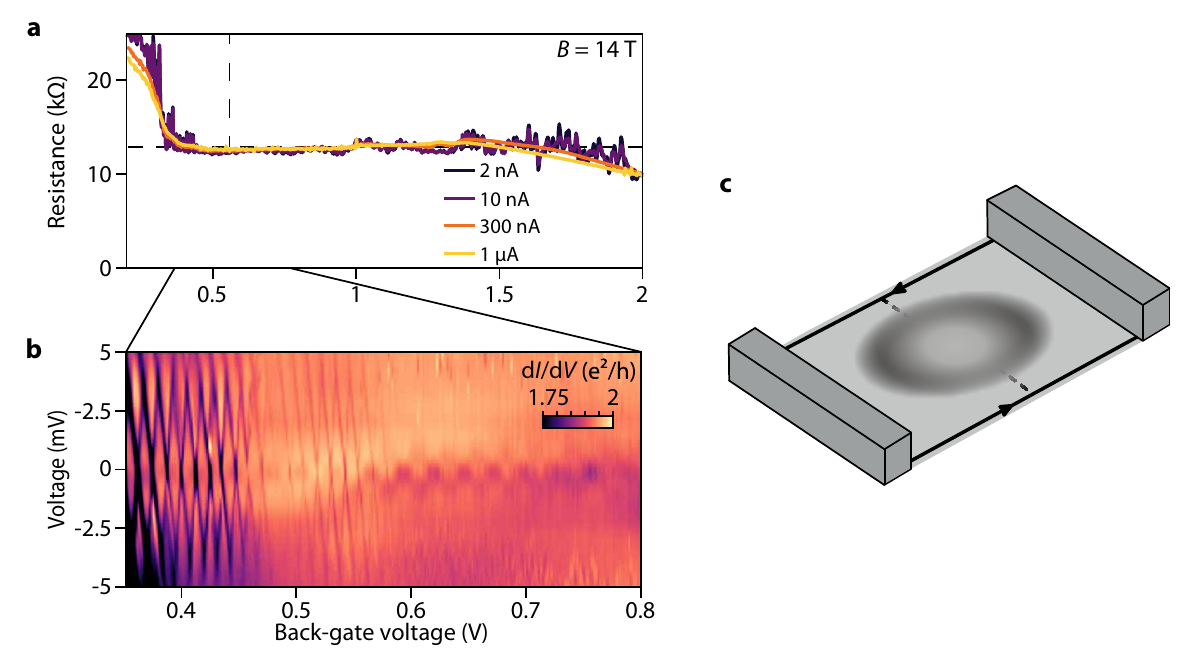}	
\caption{\textbf{Coulomb blockade on the QH plateau edge.} 
	\textbf{a}, Differential resistance as a function of back-gate voltage for different dc current biases at $B=14$ T on sample DP24-D. At low current, Coulomb diamonds appear as high frequency and high amplitude oscillations visible at the transition to the $\nu=1$ plateau. $2\phi_0$-quantum interference gives rise to oscillations with larger period and a very small amplitude on the plateau. The vertical dash line indicates the limit of Coulomb diamonds seen in (\textbf{b}). The horizontal dashed line indicates the  resistance value of $h/2e^2$. 
	\textbf{b}, Differential conductance as a function of back-gate voltage and applied voltage across the junction. Measurements are performed in voltage bias configuration. The Coulomb diamonds disappearance coincides with the start of the checkerboard pattern of the Aharonov-Bohm interference. 
	\textbf{c}, Schematics of the scattering process through a compressible island in the bulk\cite{halperin11}, with the back-gate voltage corresponding to the edge of the plateau, leading to the Coulomb diamonds in (\textbf{b}).}
	\label{ExtDataFig5}
\end{figure*}

The Coulomb blockade, an ubiquitous phenomenon in small mesoscopic systems, is found to occur at the transition between the plateaus. To illustrate this effect, we focus in Extended Data Fig. \ref{ExtDataFig5} on the left side of the $h/2e^2$ plateau upon approaching the transition to $h/e^2$ plateau, at $B=14$ T, that is, above the critical field of the superconducting electrodes. A clear transition from the Fabry-P\'{e}rot checkerboard pattern, for $V_{\rm g}> 0.6 $ V, to Coulomb diamonds, for $V_{\rm g}< 0.6 $ V, occurs (see Extended Data Fig. \ref{ExtDataFig5}b). Note that we have continuously observed $h/e$-periodic resistance oscillations beyond the critical field of the electrode, but with much lower visibility and higher Thouless energy. Contrary to the orthodox picture, the Coulomb blockade leads to quantized conductance inside the diamond and a decrease in the conductance at resonance (see diamonds in Extended Data Fig. \ref{ExtDataFig5}b delimited by darker conductance lines). Physically, as the back-gate voltage is decreased near the plateau transition, charge carriers are progressively removed from the zeroth Landau level, resulting in the emergence of a compressible island in the graphene bulk (see Extended Data Fig. \ref{ExtDataFig5}c). As a result, this island, when tunnel-coupled to the QH edge channels on both sides of the sample, induces back-scattering at resonance, hence decreasing the QH conductance.

From the Coulomb diamonds in Extended Data Fig. \ref{ExtDataFig5}b, we extract a charging energy of $1.8$ meV as well as an estimate of the diameter of the compressible island of $\sim 140$ nm, assuming it to be circular, which is a reasonable size for the dimension of the junction ($L \times W = 200 \times 125$ nm$^2$). This, together with the remarkable regularity of the diamonds, confirm the picture of a single and large compressible island in the graphene bulk upon approaching the transition between plateaus.

\section*{Acknowledgments}
We thank D. Basko, C. Beenakker, M. Feigel'Man, L. Glazman, M. Houzet, V. Kurilovich, J. Meyer, Y. Nazarov, K. Snizhko and A. Stern for valuable discussions. We thank F. Blondelle for technical support on the experimental apparatus. Samples were prepared at the Nanofab facility of the N\'eel Institute. 
This work has received funding from the European Union's Horizon 2020 research and innovation program under the ERC grant \textit{SUPERGRAPH} No. 866365. It also benefited from a French government grant managed by the ANR agency under the 'France 2030 plan', with reference ANR-22-PETQ-0003. B.S., H.S. and W.Y. acknowledge support from the QuantERA II Program that has received funding from the European Union's Horizon 2020 research and innovation program under Grant Agreement No 101017733. K.W. and T.T. acknowledge support from the JSPS KAKENHI (Grant Numbers 20H00354, 21H05233 and 23H02052) and World Premier International Research Center Initiative (WPI), MEXT, Japan.

\newpage 


\bibliography{Biblio-QHJJ,biblio-SI}

\begin{thebibliography}{64}%
\makeatletter
\providecommand \@ifxundefined [1]{%
 \@ifx{#1\undefined}
}%
\providecommand \@ifnum [1]{%
 \ifnum #1\expandafter \@firstoftwo
 \else \expandafter \@secondoftwo
 \fi
}%
\providecommand \@ifx [1]{%
 \ifx #1\expandafter \@firstoftwo
 \else \expandafter \@secondoftwo
 \fi
}%
\providecommand \natexlab [1]{#1}%
\providecommand \enquote  [1]{``#1''}%
\providecommand \bibnamefont  [1]{#1}%
\providecommand \bibfnamefont [1]{#1}%
\providecommand \citenamefont [1]{#1}%
\providecommand \href@noop [0]{\@secondoftwo}%
\providecommand \href [0]{\begingroup \@sanitize@url \@href}%
\providecommand \@href[1]{\@@startlink{#1}\@@href}%
\providecommand \@@href[1]{\endgroup#1\@@endlink}%
\providecommand \@sanitize@url [0]{\catcode `\\12\catcode `\$12\catcode
  `\&12\catcode `\#12\catcode `\^12\catcode `\_12\catcode `\%12\relax}%
\providecommand \@@startlink[1]{}%
\providecommand \@@endlink[0]{}%
\providecommand \url  [0]{\begingroup\@sanitize@url \@url }%
\providecommand \@url [1]{\endgroup\@href {#1}{\urlprefix }}%
\providecommand \urlprefix  [0]{URL }%
\providecommand \Eprint [0]{\href }%
\providecommand \doibase [0]{https://doi.org/}%
\providecommand \selectlanguage [0]{\@gobble}%
\providecommand \bibinfo  [0]{\@secondoftwo}%
\providecommand \bibfield  [0]{\@secondoftwo}%
\providecommand \translation [1]{[#1]}%
\providecommand \BibitemOpen [0]{}%
\providecommand \bibitemStop [0]{}%
\providecommand \bibitemNoStop [0]{.\EOS\space}%
\providecommand \EOS [0]{\spacefactor3000\relax}%
\providecommand \BibitemShut  [1]{\csname bibitem#1\endcsname}%
\let\auto@bib@innerbib\@empty
\bibitem [{\citenamefont {Stern}\ and\ \citenamefont
  {Lindner}(2013)}]{Stern2013}%
  \BibitemOpen
  \bibfield  {author} {\bibinfo {author} {\bibfnamefont {A.}~\bibnamefont
  {Stern}}\ and\ \bibinfo {author} {\bibfnamefont {N.~H.}\ \bibnamefont
  {Lindner}},\ }\bibfield  {title} {\bibinfo {title} {{Topological quantum
  computation--from basic concepts to first experiments}},\ }\href
  {https://doi.org/10.1126/science.1231473} {\bibfield  {journal} {\bibinfo
  {journal} {Science}\ }\textbf {\bibinfo {volume} {339}},\ \bibinfo {pages}
  {1179} (\bibinfo {year} {2013})}\BibitemShut {NoStop}%
\bibitem [{\citenamefont {Alicea}\ and\ \citenamefont
  {Stern}(2015)}]{Alicea2015}%
  \BibitemOpen
  \bibfield  {author} {\bibinfo {author} {\bibfnamefont {J.}~\bibnamefont
  {Alicea}}\ and\ \bibinfo {author} {\bibfnamefont {A.}~\bibnamefont {Stern}},\
  }\bibfield  {title} {\bibinfo {title} {{Designer non-Abelian anyon platforms:
  from Majorana to Fibonacci}},\ }\href
  {https://doi.org/10.1088/0031-8949/2015/T164/014006} {\bibfield  {journal}
  {\bibinfo  {journal} {Physica Scripta}\ }\textbf {\bibinfo {volume} {2015}},\
  \bibinfo {pages} {014006} (\bibinfo {year} {2015})}\BibitemShut {NoStop}%
\bibitem [{\citenamefont {Alicea}\ and\ \citenamefont
  {Fendley}(2016)}]{Alicea16}%
  \BibitemOpen
  \bibfield  {author} {\bibinfo {author} {\bibfnamefont {J.}~\bibnamefont
  {Alicea}}\ and\ \bibinfo {author} {\bibfnamefont {P.}~\bibnamefont
  {Fendley}},\ }\bibfield  {title} {\bibinfo {title} {Topological phases with
  parafermions: theory and blueprints},\ }\href
  {https://doi.org/10.1146/annurev-conmatphys-031115-011336} {\bibfield
  {journal} {\bibinfo  {journal} {Annual Review of Condensed Matter Physics}\
  }\textbf {\bibinfo {volume} {7}},\ \bibinfo {pages} {119} (\bibinfo {year}
  {2016})}\BibitemShut {NoStop}%
\bibitem [{\citenamefont {Rickhaus}\ \emph {et~al.}(2012)\citenamefont
  {Rickhaus}, \citenamefont {Weiss}, \citenamefont {Marot},\ and\ \citenamefont
  {Schonenberger}}]{Rickhaus12}%
  \BibitemOpen
  \bibfield  {author} {\bibinfo {author} {\bibfnamefont {P.}~\bibnamefont
  {Rickhaus}}, \bibinfo {author} {\bibfnamefont {M.}~\bibnamefont {Weiss}},
  \bibinfo {author} {\bibfnamefont {L.}~\bibnamefont {Marot}},\ and\ \bibinfo
  {author} {\bibfnamefont {C.}~\bibnamefont {Schonenberger}},\ }\bibfield
  {title} {\bibinfo {title} {{Quantum Hall effect in graphene with
  superconducting electrodes}},\ }\href {https://doi.org/10.1021/nl204415s}
  {\bibfield  {journal} {\bibinfo  {journal} {Nano letters}\ }\textbf {\bibinfo
  {volume} {12}},\ \bibinfo {pages} {1942} (\bibinfo {year}
  {2012})}\BibitemShut {NoStop}%
\bibitem [{\citenamefont {Komatsu}\ \emph {et~al.}(2012)\citenamefont
  {Komatsu}, \citenamefont {Li}, \citenamefont {Autier-Laurent}, \citenamefont
  {Bouchiat},\ and\ \citenamefont {Gu\'eron}}]{Komatsu12}%
  \BibitemOpen
  \bibfield  {author} {\bibinfo {author} {\bibfnamefont {K.}~\bibnamefont
  {Komatsu}}, \bibinfo {author} {\bibfnamefont {C.}~\bibnamefont {Li}},
  \bibinfo {author} {\bibfnamefont {S.}~\bibnamefont {Autier-Laurent}},
  \bibinfo {author} {\bibfnamefont {H.}~\bibnamefont {Bouchiat}},\ and\
  \bibinfo {author} {\bibfnamefont {S.}~\bibnamefont {Gu\'eron}},\ }\bibfield
  {title} {\bibinfo {title} {{Superconducting proximity effect in long
  superconductor/graphene/superconductor junctions: From specular Andreev
  reflection at zero field to the quantum Hall regime}},\ }\href
  {https://doi.org/10.1103/PhysRevB.86.115412} {\bibfield  {journal} {\bibinfo
  {journal} {Phys. Rev. B}\ }\textbf {\bibinfo {volume} {86}},\ \bibinfo
  {pages} {115412} (\bibinfo {year} {2012})}\BibitemShut {NoStop}%
\bibitem [{\citenamefont {{Ben Shalom}}\ \emph {et~al.}(2016)\citenamefont
  {{Ben Shalom}}, \citenamefont {Zhu}, \citenamefont {Fal'ko}, \citenamefont
  {Mishchenko}, \citenamefont {Kretinin}, \citenamefont {Novoselov},
  \citenamefont {Woods}, \citenamefont {Watanabe}, \citenamefont {Taniguchi},
  \citenamefont {Geim},\ and\ \citenamefont {Prance}}]{Shalom2015}%
  \BibitemOpen
  \bibfield  {author} {\bibinfo {author} {\bibfnamefont {M.}~\bibnamefont {{Ben
  Shalom}}}, \bibinfo {author} {\bibfnamefont {M.~J.}\ \bibnamefont {Zhu}},
  \bibinfo {author} {\bibfnamefont {V.~I.}\ \bibnamefont {Fal'ko}}, \bibinfo
  {author} {\bibfnamefont {A.}~\bibnamefont {Mishchenko}}, \bibinfo {author}
  {\bibfnamefont {A.~V.}\ \bibnamefont {Kretinin}}, \bibinfo {author}
  {\bibfnamefont {K.~S.}\ \bibnamefont {Novoselov}}, \bibinfo {author}
  {\bibfnamefont {C.~R.}\ \bibnamefont {Woods}}, \bibinfo {author}
  {\bibfnamefont {K.}~\bibnamefont {Watanabe}}, \bibinfo {author}
  {\bibfnamefont {T.}~\bibnamefont {Taniguchi}}, \bibinfo {author}
  {\bibfnamefont {A.~K.}\ \bibnamefont {Geim}},\ and\ \bibinfo {author}
  {\bibfnamefont {J.~R.}\ \bibnamefont {Prance}},\ }\bibfield  {title}
  {\bibinfo {title} {{Quantum oscillations of the critical current and
  high-field superconducting proximity in ballistic graphene}},\ }\href
  {https://doi.org/10.1038/nphys3592} {\bibfield  {journal} {\bibinfo
  {journal} {Nature Physics}\ }\textbf {\bibinfo {volume} {12}},\ \bibinfo
  {pages} {318} (\bibinfo {year} {2016})}\BibitemShut {NoStop}%
\bibitem [{\citenamefont {Wan}\ \emph {et~al.}(2015)\citenamefont {Wan},
  \citenamefont {Kazakov}, \citenamefont {Manfra}, \citenamefont {Pfeiffer},
  \citenamefont {West},\ and\ \citenamefont {Rokhinson}}]{Wan2015}%
  \BibitemOpen
  \bibfield  {author} {\bibinfo {author} {\bibfnamefont {Z.}~\bibnamefont
  {Wan}}, \bibinfo {author} {\bibfnamefont {A.}~\bibnamefont {Kazakov}},
  \bibinfo {author} {\bibfnamefont {M.~J.}\ \bibnamefont {Manfra}}, \bibinfo
  {author} {\bibfnamefont {L.~N.}\ \bibnamefont {Pfeiffer}}, \bibinfo {author}
  {\bibfnamefont {K.~W.}\ \bibnamefont {West}},\ and\ \bibinfo {author}
  {\bibfnamefont {L.~P.}\ \bibnamefont {Rokhinson}},\ }\bibfield  {title}
  {\bibinfo {title} {{Induced superconductivity in high-mobility
  two-dimensional electron gas in gallium arsenide heterostructures}},\ }\href
  {https://doi.org/10.1038/ncomms8426} {\bibfield  {journal} {\bibinfo
  {journal} {Nature Communications}\ }\textbf {\bibinfo {volume} {6}},\
  \bibinfo {pages} {7426} (\bibinfo {year} {2015})}\BibitemShut {NoStop}%
\bibitem [{\citenamefont {Amet}\ \emph {et~al.}(2016)\citenamefont {Amet},
  \citenamefont {Ke}, \citenamefont {Borzenets}, \citenamefont {Wang},
  \citenamefont {Watanabe}, \citenamefont {Taniguchi}, \citenamefont {Deacon},
  \citenamefont {Yamamoto}, \citenamefont {Bomze}, \citenamefont {Tarucha},\
  and\ \citenamefont {Finkelstein}}]{Amet2016}%
  \BibitemOpen
  \bibfield  {author} {\bibinfo {author} {\bibfnamefont {F.}~\bibnamefont
  {Amet}}, \bibinfo {author} {\bibfnamefont {C.~T.}\ \bibnamefont {Ke}},
  \bibinfo {author} {\bibfnamefont {I.~V.}\ \bibnamefont {Borzenets}}, \bibinfo
  {author} {\bibfnamefont {J.}~\bibnamefont {Wang}}, \bibinfo {author}
  {\bibfnamefont {K.}~\bibnamefont {Watanabe}}, \bibinfo {author}
  {\bibfnamefont {T.}~\bibnamefont {Taniguchi}}, \bibinfo {author}
  {\bibfnamefont {R.~S.}\ \bibnamefont {Deacon}}, \bibinfo {author}
  {\bibfnamefont {M.}~\bibnamefont {Yamamoto}}, \bibinfo {author}
  {\bibfnamefont {Y.}~\bibnamefont {Bomze}}, \bibinfo {author} {\bibfnamefont
  {S.}~\bibnamefont {Tarucha}},\ and\ \bibinfo {author} {\bibfnamefont
  {G.}~\bibnamefont {Finkelstein}},\ }\bibfield  {title} {\bibinfo {title}
  {{Supercurrent in the quantum Hall regime}},\ }\href
  {https://doi.org/10.1126/science.aad6203} {\bibfield  {journal} {\bibinfo
  {journal} {Science}\ }\textbf {\bibinfo {volume} {352}},\ \bibinfo {pages}
  {966} (\bibinfo {year} {2016})}\BibitemShut {NoStop}%
\bibitem [{\citenamefont {Lee}\ \emph {et~al.}(2017)\citenamefont {Lee},
  \citenamefont {Huang}, \citenamefont {Efetov}, \citenamefont {Wei},
  \citenamefont {Hart}, \citenamefont {Taniguchi}, \citenamefont {Watanabe},
  \citenamefont {Yacoby},\ and\ \citenamefont {Kim}}]{Lee2017}%
  \BibitemOpen
  \bibfield  {author} {\bibinfo {author} {\bibfnamefont {G.-H.}\ \bibnamefont
  {Lee}}, \bibinfo {author} {\bibfnamefont {K.-F.}\ \bibnamefont {Huang}},
  \bibinfo {author} {\bibfnamefont {D.~K.}\ \bibnamefont {Efetov}}, \bibinfo
  {author} {\bibfnamefont {D.~S.}\ \bibnamefont {Wei}}, \bibinfo {author}
  {\bibfnamefont {S.}~\bibnamefont {Hart}}, \bibinfo {author} {\bibfnamefont
  {T.}~\bibnamefont {Taniguchi}}, \bibinfo {author} {\bibfnamefont
  {K.}~\bibnamefont {Watanabe}}, \bibinfo {author} {\bibfnamefont
  {A.}~\bibnamefont {Yacoby}},\ and\ \bibinfo {author} {\bibfnamefont
  {P.}~\bibnamefont {Kim}},\ }\bibfield  {title} {\bibinfo {title} {{Inducing
  superconducting correlation in quantum Hall edge states}},\ }\href
  {https://doi.org/10.1038/nphys4084} {\bibfield  {journal} {\bibinfo
  {journal} {Nature Physics}\ }\textbf {\bibinfo {volume} {13}},\ \bibinfo
  {pages} {693} (\bibinfo {year} {2017})}\BibitemShut {NoStop}%
\bibitem [{\citenamefont {Park}\ \emph {et~al.}(2017)\citenamefont {Park},
  \citenamefont {Kim}, \citenamefont {Watanabe}, \citenamefont {Taniguchi},\
  and\ \citenamefont {Lee}}]{Park17}%
  \BibitemOpen
  \bibfield  {author} {\bibinfo {author} {\bibfnamefont {G.-H.}\ \bibnamefont
  {Park}}, \bibinfo {author} {\bibfnamefont {M.}~\bibnamefont {Kim}}, \bibinfo
  {author} {\bibfnamefont {K.}~\bibnamefont {Watanabe}}, \bibinfo {author}
  {\bibfnamefont {T.}~\bibnamefont {Taniguchi}},\ and\ \bibinfo {author}
  {\bibfnamefont {H.-J.}\ \bibnamefont {Lee}},\ }\bibfield  {title} {\bibinfo
  {title} {{Propagation of superconducting coherence via chiral quantum-Hall
  edge channels}},\ }\href {https://doi.org/10.1038/s41598-017-11209-w}
  {\bibfield  {journal} {\bibinfo  {journal} {Scientific Reports}\ }\textbf
  {\bibinfo {volume} {7}},\ \bibinfo {pages} {1} (\bibinfo {year}
  {2017})}\BibitemShut {NoStop}%
\bibitem [{\citenamefont {Seredinski}\ \emph {et~al.}(2019)\citenamefont
  {Seredinski}, \citenamefont {Draelos}, \citenamefont {Arnault}, \citenamefont
  {Wei}, \citenamefont {Li}, \citenamefont {Fleming}, \citenamefont {Watanabe},
  \citenamefont {Taniguchi}, \citenamefont {Amet},\ and\ \citenamefont
  {Finkelstein}}]{Seredinski19}%
  \BibitemOpen
  \bibfield  {author} {\bibinfo {author} {\bibfnamefont {A.}~\bibnamefont
  {Seredinski}}, \bibinfo {author} {\bibfnamefont {A.~W.}\ \bibnamefont
  {Draelos}}, \bibinfo {author} {\bibfnamefont {E.~G.}\ \bibnamefont
  {Arnault}}, \bibinfo {author} {\bibfnamefont {M.-T.}\ \bibnamefont {Wei}},
  \bibinfo {author} {\bibfnamefont {H.}~\bibnamefont {Li}}, \bibinfo {author}
  {\bibfnamefont {T.}~\bibnamefont {Fleming}}, \bibinfo {author} {\bibfnamefont
  {K.}~\bibnamefont {Watanabe}}, \bibinfo {author} {\bibfnamefont
  {T.}~\bibnamefont {Taniguchi}}, \bibinfo {author} {\bibfnamefont
  {F.}~\bibnamefont {Amet}},\ and\ \bibinfo {author} {\bibfnamefont
  {G.}~\bibnamefont {Finkelstein}},\ }\bibfield  {title} {\bibinfo {title}
  {{Quantum Hall-based superconducting interference device}},\ }\href
  {https://doi.org/10.1126/sciadv.aaw8693} {\bibfield  {journal} {\bibinfo
  {journal} {Science Advances}\ }\textbf {\bibinfo {volume} {5}},\ \bibinfo
  {pages} {eaaw8693} (\bibinfo {year} {2019})}\BibitemShut {NoStop}%
\bibitem [{\citenamefont {Zhao}\ \emph {et~al.}(2020)\citenamefont {Zhao},
  \citenamefont {Arnault}, \citenamefont {Bondarev}, \citenamefont
  {Seredinski}, \citenamefont {Larson}, \citenamefont {Draelos}, \citenamefont
  {Li}, \citenamefont {Watanabe}, \citenamefont {Taniguchi}, \citenamefont
  {Amet}, \citenamefont {Baranger},\ and\ \citenamefont
  {Finkelstein}}]{Zhao2020a}%
  \BibitemOpen
  \bibfield  {author} {\bibinfo {author} {\bibfnamefont {L.}~\bibnamefont
  {Zhao}}, \bibinfo {author} {\bibfnamefont {E.~G.}\ \bibnamefont {Arnault}},
  \bibinfo {author} {\bibfnamefont {A.}~\bibnamefont {Bondarev}}, \bibinfo
  {author} {\bibfnamefont {A.}~\bibnamefont {Seredinski}}, \bibinfo {author}
  {\bibfnamefont {T.~F.~Q.}\ \bibnamefont {Larson}}, \bibinfo {author}
  {\bibfnamefont {A.~W.}\ \bibnamefont {Draelos}}, \bibinfo {author}
  {\bibfnamefont {H.}~\bibnamefont {Li}}, \bibinfo {author} {\bibfnamefont
  {K.}~\bibnamefont {Watanabe}}, \bibinfo {author} {\bibfnamefont
  {T.}~\bibnamefont {Taniguchi}}, \bibinfo {author} {\bibfnamefont
  {F.}~\bibnamefont {Amet}}, \bibinfo {author} {\bibfnamefont {H.~U.}\
  \bibnamefont {Baranger}},\ and\ \bibinfo {author} {\bibfnamefont
  {G.}~\bibnamefont {Finkelstein}},\ }\bibfield  {title} {\bibinfo {title}
  {{Interference of chiral Andreev edge states}},\ }\href
  {https://doi.org/10.1038/s41567-020-0898-5} {\bibfield  {journal} {\bibinfo
  {journal} {Nature Physics}\ }\textbf {\bibinfo {volume} {16}},\ \bibinfo
  {pages} {862} (\bibinfo {year} {2020})}\BibitemShut {NoStop}%
\bibitem [{\citenamefont {Wang}\ \emph {et~al.}(2021)\citenamefont {Wang},
  \citenamefont {Telford}, \citenamefont {Benyamini}, \citenamefont
  {Jesudasan}, \citenamefont {Raychaudhuri}, \citenamefont {Watanabe},
  \citenamefont {Taniguchi}, \citenamefont {Hone}, \citenamefont {Dean},\ and\
  \citenamefont {Pasupathy}}]{Wang2021}%
  \BibitemOpen
  \bibfield  {author} {\bibinfo {author} {\bibfnamefont {D.}~\bibnamefont
  {Wang}}, \bibinfo {author} {\bibfnamefont {E.~J.}\ \bibnamefont {Telford}},
  \bibinfo {author} {\bibfnamefont {A.}~\bibnamefont {Benyamini}}, \bibinfo
  {author} {\bibfnamefont {J.}~\bibnamefont {Jesudasan}}, \bibinfo {author}
  {\bibfnamefont {P.}~\bibnamefont {Raychaudhuri}}, \bibinfo {author}
  {\bibfnamefont {K.}~\bibnamefont {Watanabe}}, \bibinfo {author}
  {\bibfnamefont {T.}~\bibnamefont {Taniguchi}}, \bibinfo {author}
  {\bibfnamefont {J.}~\bibnamefont {Hone}}, \bibinfo {author} {\bibfnamefont
  {C.~R.}\ \bibnamefont {Dean}},\ and\ \bibinfo {author} {\bibfnamefont
  {A.~N.}\ \bibnamefont {Pasupathy}},\ }\bibfield  {title} {\bibinfo {title}
  {{Andreev Reflections in NbN/Graphene Junctions under Large Magnetic
  Fields}},\ }\href {https://doi.org/10.1021/acs.nanolett.1c02020} {\bibfield
  {journal} {\bibinfo  {journal} {Nano Letters}\ }\textbf {\bibinfo {volume}
  {21}},\ \bibinfo {pages} {8229} (\bibinfo {year} {2021})}\BibitemShut
  {NoStop}%
\bibitem [{\citenamefont {G{\"{u}}l}\ \emph {et~al.}(2022)\citenamefont
  {G{\"{u}}l}, \citenamefont {Ronen}, \citenamefont {Lee}, \citenamefont
  {Shapourian}, \citenamefont {Zauberman}, \citenamefont {H.}, \citenamefont
  {Watanabe}, \citenamefont {Taniguchi}, \citenamefont {Vishwanath},
  \citenamefont {Yacoby},\ and\ \citenamefont {Kim}}]{Gul2022}%
  \BibitemOpen
  \bibfield  {author} {\bibinfo {author} {\bibfnamefont {{\"{O}}.}~\bibnamefont
  {G{\"{u}}l}}, \bibinfo {author} {\bibfnamefont {Y.}~\bibnamefont {Ronen}},
  \bibinfo {author} {\bibfnamefont {S.~Y.}\ \bibnamefont {Lee}}, \bibinfo
  {author} {\bibfnamefont {H.}~\bibnamefont {Shapourian}}, \bibinfo {author}
  {\bibfnamefont {J.}~\bibnamefont {Zauberman}}, \bibinfo {author}
  {\bibfnamefont {L.~Y.}\ \bibnamefont {H.}}, \bibinfo {author} {\bibfnamefont
  {K.}~\bibnamefont {Watanabe}}, \bibinfo {author} {\bibfnamefont
  {T.}~\bibnamefont {Taniguchi}}, \bibinfo {author} {\bibfnamefont
  {A.}~\bibnamefont {Vishwanath}}, \bibinfo {author} {\bibfnamefont
  {A.}~\bibnamefont {Yacoby}},\ and\ \bibinfo {author} {\bibfnamefont
  {P.}~\bibnamefont {Kim}},\ }\bibfield  {title} {\bibinfo {title} {{Andreev
  reflection in the fractional quantum Hall state}},\ }\href
  {https://doi.org/10.1103/PhysRevX.12.021057} {\bibfield  {journal} {\bibinfo
  {journal} {Physical Review X}\ }\textbf {\bibinfo {volume} {12}},\ \bibinfo
  {pages} {021057} (\bibinfo {year} {2022})}\BibitemShut {NoStop}%
\bibitem [{\citenamefont {Hatefipour}\ \emph {et~al.}(2022)\citenamefont
  {Hatefipour}, \citenamefont {Cuozzo}, \citenamefont {Kanter}, \citenamefont
  {Strickland}, \citenamefont {Allemang}, \citenamefont {Lu}, \citenamefont
  {Rossi},\ and\ \citenamefont {Shabani}}]{Hatefipour22}%
  \BibitemOpen
  \bibfield  {author} {\bibinfo {author} {\bibfnamefont {M.}~\bibnamefont
  {Hatefipour}}, \bibinfo {author} {\bibfnamefont {J.~J.}\ \bibnamefont
  {Cuozzo}}, \bibinfo {author} {\bibfnamefont {J.}~\bibnamefont {Kanter}},
  \bibinfo {author} {\bibfnamefont {W.~M.}\ \bibnamefont {Strickland}},
  \bibinfo {author} {\bibfnamefont {C.~R.}\ \bibnamefont {Allemang}}, \bibinfo
  {author} {\bibfnamefont {T.-M.}\ \bibnamefont {Lu}}, \bibinfo {author}
  {\bibfnamefont {E.}~\bibnamefont {Rossi}},\ and\ \bibinfo {author}
  {\bibfnamefont {J.}~\bibnamefont {Shabani}},\ }\bibfield  {title} {\bibinfo
  {title} {{Induced Superconducting Pairing in Integer Quantum Hall Edge
  States}},\ }\href {https://doi.org/10.1021/acs.nanolett.2c01413} {\bibfield
  {journal} {\bibinfo  {journal} {Nano Letters}\ }\textbf {\bibinfo {volume}
  {22}},\ \bibinfo {pages} {6173} (\bibinfo {year} {2022})}\BibitemShut
  {NoStop}%
\bibitem [{\citenamefont {Zhao}\ \emph {et~al.}(2022)\citenamefont {Zhao},
  \citenamefont {Iftikhar}, \citenamefont {Larson}, \citenamefont {Arnault},
  \citenamefont {Watanabe}, \citenamefont {Taniguchi}, \citenamefont {Amet},\
  and\ \citenamefont {Finkelstein}}]{Zhao22}%
  \BibitemOpen
  \bibfield  {author} {\bibinfo {author} {\bibfnamefont {L.}~\bibnamefont
  {Zhao}}, \bibinfo {author} {\bibfnamefont {Z.}~\bibnamefont {Iftikhar}},
  \bibinfo {author} {\bibfnamefont {T.~F.}\ \bibnamefont {Larson}}, \bibinfo
  {author} {\bibfnamefont {E.~G.}\ \bibnamefont {Arnault}}, \bibinfo {author}
  {\bibfnamefont {K.}~\bibnamefont {Watanabe}}, \bibinfo {author}
  {\bibfnamefont {T.}~\bibnamefont {Taniguchi}}, \bibinfo {author}
  {\bibfnamefont {F.}~\bibnamefont {Amet}},\ and\ \bibinfo {author}
  {\bibfnamefont {G.}~\bibnamefont {Finkelstein}},\ }\bibfield  {title}
  {\bibinfo {title} {{Loss and decoherence at the quantum Hall - superconductor
  interface}},\ }\href@noop {} {\bibfield  {journal} {\bibinfo  {journal}
  {arXiv:2210.04842}\ } (\bibinfo {year} {2022})}\BibitemShut {NoStop}%
\bibitem [{\citenamefont {Ma}\ and\ \citenamefont {Zyuzin}(1993)}]{Ma1993}%
  \BibitemOpen
  \bibfield  {author} {\bibinfo {author} {\bibfnamefont {M.}~\bibnamefont
  {Ma}}\ and\ \bibinfo {author} {\bibfnamefont {A.~Y.}\ \bibnamefont
  {Zyuzin}},\ }\bibfield  {title} {\bibinfo {title} {{Josephson effect in the
  quantum hall regime}},\ }\href {https://doi.org/10.1209/0295-5075/21/9/011}
  {\bibfield  {journal} {\bibinfo  {journal} {Europhysics Letters}\ }\textbf
  {\bibinfo {volume} {21}},\ \bibinfo {pages} {941} (\bibinfo {year}
  {1993})}\BibitemShut {NoStop}%
\bibitem [{\citenamefont {Zyuzin}(1994)}]{Zyuzin94}%
  \BibitemOpen
  \bibfield  {author} {\bibinfo {author} {\bibfnamefont {A.~Y.}\ \bibnamefont
  {Zyuzin}},\ }\bibfield  {title} {\bibinfo {title}
  {Superconductor--normal-metal--superconductor junction in a strong magnetic
  field},\ }\href {https://doi.org/10.1103/PhysRevB.50.323} {\bibfield
  {journal} {\bibinfo  {journal} {Phys. Rev. B}\ }\textbf {\bibinfo {volume}
  {50}},\ \bibinfo {pages} {323} (\bibinfo {year} {1994})}\BibitemShut
  {NoStop}%
\bibitem [{\citenamefont {Stone}\ and\ \citenamefont {Lin}(2011)}]{Stone2011}%
  \BibitemOpen
  \bibfield  {author} {\bibinfo {author} {\bibfnamefont {M.}~\bibnamefont
  {Stone}}\ and\ \bibinfo {author} {\bibfnamefont {Y.}~\bibnamefont {Lin}},\
  }\bibfield  {title} {\bibinfo {title} {{Josephson currents in quantum Hall
  devices}},\ }\href {https://doi.org/10.1103/PhysRevB.83.224501} {\bibfield
  {journal} {\bibinfo  {journal} {Physical Review B}\ }\textbf {\bibinfo
  {volume} {83}},\ \bibinfo {pages} {224501} (\bibinfo {year}
  {2011})}\BibitemShut {NoStop}%
\bibitem [{\citenamefont {{Van Ostaay}}\ \emph {et~al.}(2011)\citenamefont
  {{Van Ostaay}}, \citenamefont {Akhmerov},\ and\ \citenamefont
  {Beenakker}}]{VanOstaay2011}%
  \BibitemOpen
  \bibfield  {author} {\bibinfo {author} {\bibfnamefont {J.~A.~M.}\
  \bibnamefont {{Van Ostaay}}}, \bibinfo {author} {\bibfnamefont {A.~R.}\
  \bibnamefont {Akhmerov}},\ and\ \bibinfo {author} {\bibfnamefont {C.~W.~J.}\
  \bibnamefont {Beenakker}},\ }\bibfield  {title} {\bibinfo {title}
  {{Spin-triplet supercurrent carried by quantum Hall edge states through a
  Josephson junction}},\ }\href {https://doi.org/10.1103/PhysRevB.83.195441}
  {\bibfield  {journal} {\bibinfo  {journal} {Physical Review B}\ }\textbf
  {\bibinfo {volume} {83}},\ \bibinfo {pages} {195441} (\bibinfo {year}
  {2011})}\BibitemShut {NoStop}%
\bibitem [{\citenamefont {Alavirad}\ \emph {et~al.}(2018)\citenamefont
  {Alavirad}, \citenamefont {Lee}, \citenamefont {Lin},\ and\ \citenamefont
  {Sau}}]{Alavirad2018}%
  \BibitemOpen
  \bibfield  {author} {\bibinfo {author} {\bibfnamefont {Y.}~\bibnamefont
  {Alavirad}}, \bibinfo {author} {\bibfnamefont {J.}~\bibnamefont {Lee}},
  \bibinfo {author} {\bibfnamefont {Z.-X.}\ \bibnamefont {Lin}},\ and\ \bibinfo
  {author} {\bibfnamefont {J.~D.}\ \bibnamefont {Sau}},\ }\bibfield  {title}
  {\bibinfo {title} {{Chiral supercurrent through a quantum Hall weak link}},\
  }\href {https://doi.org/10.1103/PhysRevB.98.214504} {\bibfield  {journal}
  {\bibinfo  {journal} {Physical Review B}\ }\textbf {\bibinfo {volume} {98}},\
  \bibinfo {pages} {214504} (\bibinfo {year} {2018})}\BibitemShut {NoStop}%
\bibitem [{\citenamefont {Manesco}\ \emph
  {et~al.}(2022{\natexlab{a}})\citenamefont {Manesco}, \citenamefont
  {Fl{\'o}r}, \citenamefont {Liu},\ and\ \citenamefont {Akhmerov}}]{Manesco22}%
  \BibitemOpen
  \bibfield  {author} {\bibinfo {author} {\bibfnamefont {A.~L.~R.}\
  \bibnamefont {Manesco}}, \bibinfo {author} {\bibfnamefont {I.~M.}\
  \bibnamefont {Fl{\'o}r}}, \bibinfo {author} {\bibfnamefont {C.-X.}\
  \bibnamefont {Liu}},\ and\ \bibinfo {author} {\bibfnamefont {A.~R.}\
  \bibnamefont {Akhmerov}},\ }\bibfield  {title} {\bibinfo {title} {{Mechanisms
  of Andreev reflection in quantum Hall graphene}},\ }\href
  {https://doi.org/10.21468/SciPostPhysCore.5.3.045} {\bibfield  {journal}
  {\bibinfo  {journal} {SciPost Phys. Core}\ }\textbf {\bibinfo {volume} {5}},\
  \bibinfo {pages} {045} (\bibinfo {year} {2022}{\natexlab{a}})}\BibitemShut
  {NoStop}%
\bibitem [{\citenamefont {Kurilovich}\ \emph {et~al.}(2022)\citenamefont
  {Kurilovich}, \citenamefont {Raines},\ and\ \citenamefont
  {Glazman}}]{Kurilovich22}%
  \BibitemOpen
  \bibfield  {author} {\bibinfo {author} {\bibfnamefont {V.~D.}\ \bibnamefont
  {Kurilovich}}, \bibinfo {author} {\bibfnamefont {Z.~M.}\ \bibnamefont
  {Raines}},\ and\ \bibinfo {author} {\bibfnamefont {L.~I.}\ \bibnamefont
  {Glazman}},\ }\bibfield  {title} {\bibinfo {title} {{Disorder in Andreev
  reflection of a quantum Hall edge}},\ }\href@noop {} {\bibfield  {journal}
  {\bibinfo  {journal} {arXiv:2201.00273}\ } (\bibinfo {year}
  {2022})}\BibitemShut {NoStop}%
\bibitem [{\citenamefont {Tang}\ \emph {et~al.}(2022)\citenamefont {Tang},
  \citenamefont {Knapp},\ and\ \citenamefont {Alicea}}]{Tang22}%
  \BibitemOpen
  \bibfield  {author} {\bibinfo {author} {\bibfnamefont {Y.}~\bibnamefont
  {Tang}}, \bibinfo {author} {\bibfnamefont {C.}~\bibnamefont {Knapp}},\ and\
  \bibinfo {author} {\bibfnamefont {J.}~\bibnamefont {Alicea}},\ }\bibfield
  {title} {\bibinfo {title} {{Vortex-enabled Andreev processes in quantum
  Hall--superconductor hybrids}},\ }\href
  {https://doi.org/10.1103/PhysRevB.106.245411} {\bibfield  {journal} {\bibinfo
   {journal} {Phys. Rev. B}\ }\textbf {\bibinfo {volume} {106}},\ \bibinfo
  {pages} {245411} (\bibinfo {year} {2022})}\BibitemShut {NoStop}%
\bibitem [{\citenamefont {Qi}\ \emph {et~al.}(2010)\citenamefont {Qi},
  \citenamefont {Hughes},\ and\ \citenamefont {Zhang}}]{Qi2010}%
  \BibitemOpen
  \bibfield  {author} {\bibinfo {author} {\bibfnamefont {X.~L.}\ \bibnamefont
  {Qi}}, \bibinfo {author} {\bibfnamefont {T.~L.}\ \bibnamefont {Hughes}},\
  and\ \bibinfo {author} {\bibfnamefont {S.~C.}\ \bibnamefont {Zhang}},\
  }\bibfield  {title} {\bibinfo {title} {{Chiral topological superconductor
  from the quantum Hall state}},\ }\href
  {https://doi.org/10.1103/PhysRevB.82.184516} {\bibfield  {journal} {\bibinfo
  {journal} {Physical Review B}\ }\textbf {\bibinfo {volume} {82}},\ \bibinfo
  {pages} {184516} (\bibinfo {year} {2010})}\BibitemShut {NoStop}%
\bibitem [{\citenamefont {Clarke}\ \emph {et~al.}(2012)\citenamefont {Clarke},
  \citenamefont {Alicea},\ and\ \citenamefont {Shtengel}}]{Clarke2012}%
  \BibitemOpen
  \bibfield  {author} {\bibinfo {author} {\bibfnamefont {D.~J.}\ \bibnamefont
  {Clarke}}, \bibinfo {author} {\bibfnamefont {J.}~\bibnamefont {Alicea}},\
  and\ \bibinfo {author} {\bibfnamefont {K.}~\bibnamefont {Shtengel}},\
  }\bibfield  {title} {\bibinfo {title} {{Exotic non-Abelian anyons from
  conventional fractional quantum Hall states}},\ }\href
  {https://doi.org/10.1038/ncomms2340} {\bibfield  {journal} {\bibinfo
  {journal} {Nature Communications}\ }\textbf {\bibinfo {volume} {4}},\
  \bibinfo {pages} {1348} (\bibinfo {year} {2012})}\BibitemShut {NoStop}%
\bibitem [{\citenamefont {Lindner}\ \emph {et~al.}(2012)\citenamefont
  {Lindner}, \citenamefont {Berg}, \citenamefont {Refael},\ and\ \citenamefont
  {Stern}}]{Lindner2012}%
  \BibitemOpen
  \bibfield  {author} {\bibinfo {author} {\bibfnamefont {N.~H.}\ \bibnamefont
  {Lindner}}, \bibinfo {author} {\bibfnamefont {E.}~\bibnamefont {Berg}},
  \bibinfo {author} {\bibfnamefont {G.}~\bibnamefont {Refael}},\ and\ \bibinfo
  {author} {\bibfnamefont {A.}~\bibnamefont {Stern}},\ }\bibfield  {title}
  {\bibinfo {title} {{Fractionalizing majorana fermions: Non-abelian statistics
  on the edges of abelian quantum hall states}},\ }\href
  {https://doi.org/10.1103/PhysRevX.2.041002} {\bibfield  {journal} {\bibinfo
  {journal} {Physical Review X}\ }\textbf {\bibinfo {volume} {2}},\ \bibinfo
  {pages} {041002} (\bibinfo {year} {2012})}\BibitemShut {NoStop}%
\bibitem [{\citenamefont {Vaezi}(2013)}]{Vaezi13}%
  \BibitemOpen
  \bibfield  {author} {\bibinfo {author} {\bibfnamefont {A.}~\bibnamefont
  {Vaezi}},\ }\bibfield  {title} {\bibinfo {title} {{Fractional topological
  superconductor with fractionalized Majorana fermions}},\ }\href
  {https://doi.org/10.1103/PhysRevB.87.035132} {\bibfield  {journal} {\bibinfo
  {journal} {Phys. Rev. B}\ }\textbf {\bibinfo {volume} {87}},\ \bibinfo
  {pages} {035132} (\bibinfo {year} {2013})}\BibitemShut {NoStop}%
\bibitem [{\citenamefont {Clarke}\ \emph {et~al.}(2014)\citenamefont {Clarke},
  \citenamefont {Alicea},\ and\ \citenamefont {Shtengel}}]{Clarke2014}%
  \BibitemOpen
  \bibfield  {author} {\bibinfo {author} {\bibfnamefont {D.~J.}\ \bibnamefont
  {Clarke}}, \bibinfo {author} {\bibfnamefont {J.}~\bibnamefont {Alicea}},\
  and\ \bibinfo {author} {\bibfnamefont {K.}~\bibnamefont {Shtengel}},\
  }\bibfield  {title} {\bibinfo {title} {{Exotic circuit elements from
  zero-modes in hybrid superconductor--quantum-Hall systems}},\ }\href
  {https://doi.org/10.1038/nphys3114} {\bibfield  {journal} {\bibinfo
  {journal} {Nature Physics}\ }\textbf {\bibinfo {volume} {10}},\ \bibinfo
  {pages} {877} (\bibinfo {year} {2014})}\BibitemShut {NoStop}%
\bibitem [{\citenamefont {Mong}\ \emph {et~al.}(2014)\citenamefont {Mong},
  \citenamefont {Clarke}, \citenamefont {Alicea}, \citenamefont {Lindner},
  \citenamefont {Fendley}, \citenamefont {Nayak}, \citenamefont {Oreg},
  \citenamefont {Stern}, \citenamefont {Berg}, \citenamefont {Shtengel},\ and\
  \citenamefont {Fisher}}]{Mong2014}%
  \BibitemOpen
  \bibfield  {author} {\bibinfo {author} {\bibfnamefont {R.~S.~K.}\
  \bibnamefont {Mong}}, \bibinfo {author} {\bibfnamefont {D.~J.}\ \bibnamefont
  {Clarke}}, \bibinfo {author} {\bibfnamefont {J.}~\bibnamefont {Alicea}},
  \bibinfo {author} {\bibfnamefont {N.~H.}\ \bibnamefont {Lindner}}, \bibinfo
  {author} {\bibfnamefont {P.}~\bibnamefont {Fendley}}, \bibinfo {author}
  {\bibfnamefont {C.}~\bibnamefont {Nayak}}, \bibinfo {author} {\bibfnamefont
  {Y.}~\bibnamefont {Oreg}}, \bibinfo {author} {\bibfnamefont {A.}~\bibnamefont
  {Stern}}, \bibinfo {author} {\bibfnamefont {E.}~\bibnamefont {Berg}},
  \bibinfo {author} {\bibfnamefont {K.}~\bibnamefont {Shtengel}},\ and\
  \bibinfo {author} {\bibfnamefont {M.~P.~A.}\ \bibnamefont {Fisher}},\
  }\bibfield  {title} {\bibinfo {title} {{Universal topological quantum
  computation from a superconductor-abelian quantum hall heterostructure}},\
  }\href {https://doi.org/10.1103/PhysRevX.4.011036} {\bibfield  {journal}
  {\bibinfo  {journal} {Physical Review X}\ }\textbf {\bibinfo {volume} {4}},\
  \bibinfo {pages} {011036} (\bibinfo {year} {2014})}\BibitemShut {NoStop}%
\bibitem [{\citenamefont {Beenakker}(2014)}]{Beenakker14}%
  \BibitemOpen
  \bibfield  {author} {\bibinfo {author} {\bibfnamefont {C.~W.~J.}\
  \bibnamefont {Beenakker}},\ }\bibfield  {title} {\bibinfo {title}
  {{Annihilation of Colliding Bogoliubov Quasiparticles Reveals their Majorana
  Nature}},\ }\href {https://doi.org/10.1103/PhysRevLett.112.070604} {\bibfield
   {journal} {\bibinfo  {journal} {Phys. Rev. Lett.}\ }\textbf {\bibinfo
  {volume} {112}},\ \bibinfo {pages} {070604} (\bibinfo {year}
  {2014})}\BibitemShut {NoStop}%
\bibitem [{\citenamefont {San-Jose}\ \emph {et~al.}(2015)\citenamefont
  {San-Jose}, \citenamefont {Lado}, \citenamefont {Aguado}, \citenamefont
  {Guinea},\ and\ \citenamefont {Fern\'andez-Rossier}}]{SanJose15}%
  \BibitemOpen
  \bibfield  {author} {\bibinfo {author} {\bibfnamefont {P.}~\bibnamefont
  {San-Jose}}, \bibinfo {author} {\bibfnamefont {J.~L.}\ \bibnamefont {Lado}},
  \bibinfo {author} {\bibfnamefont {R.}~\bibnamefont {Aguado}}, \bibinfo
  {author} {\bibfnamefont {F.}~\bibnamefont {Guinea}},\ and\ \bibinfo {author}
  {\bibfnamefont {J.}~\bibnamefont {Fern\'andez-Rossier}},\ }\bibfield  {title}
  {\bibinfo {title} {{Majorana Zero Modes in Graphene}},\ }\href
  {https://doi.org/10.1103/PhysRevX.5.041042} {\bibfield  {journal} {\bibinfo
  {journal} {Phys. Rev. X}\ }\textbf {\bibinfo {volume} {5}},\ \bibinfo {pages}
  {041042} (\bibinfo {year} {2015})}\BibitemShut {NoStop}%
\bibitem [{\citenamefont {Finocchiaro}\ \emph {et~al.}(2018)\citenamefont
  {Finocchiaro}, \citenamefont {Guinea},\ and\ \citenamefont
  {San-Jose}}]{Finocchiaro18}%
  \BibitemOpen
  \bibfield  {author} {\bibinfo {author} {\bibfnamefont {F.}~\bibnamefont
  {Finocchiaro}}, \bibinfo {author} {\bibfnamefont {F.}~\bibnamefont
  {Guinea}},\ and\ \bibinfo {author} {\bibfnamefont {P.}~\bibnamefont
  {San-Jose}},\ }\bibfield  {title} {\bibinfo {title} {Topological
  $\ensuremath{\pi}$ junctions from crossed andreev reflection in the quantum
  hall regime},\ }\href {https://doi.org/10.1103/PhysRevLett.120.116801}
  {\bibfield  {journal} {\bibinfo  {journal} {Phys. Rev. Lett.}\ }\textbf
  {\bibinfo {volume} {120}},\ \bibinfo {pages} {116801} (\bibinfo {year}
  {2018})}\BibitemShut {NoStop}%
\bibitem [{\citenamefont {Snizhko}\ \emph {et~al.}(2018)\citenamefont
  {Snizhko}, \citenamefont {Egger},\ and\ \citenamefont {Gefen}}]{Snizhko18}%
  \BibitemOpen
  \bibfield  {author} {\bibinfo {author} {\bibfnamefont {K.}~\bibnamefont
  {Snizhko}}, \bibinfo {author} {\bibfnamefont {R.}~\bibnamefont {Egger}},\
  and\ \bibinfo {author} {\bibfnamefont {Y.}~\bibnamefont {Gefen}},\ }\bibfield
   {title} {\bibinfo {title} {{Measurement and control of a Coulomb-blockaded
  parafermion box}},\ }\href {https://doi.org/10.1103/PhysRevB.97.081405}
  {\bibfield  {journal} {\bibinfo  {journal} {Phys. Rev. B}\ }\textbf {\bibinfo
  {volume} {97}},\ \bibinfo {pages} {081405} (\bibinfo {year}
  {2018})}\BibitemShut {NoStop}%
\bibitem [{\citenamefont {Nielsen}\ \emph {et~al.}(2022)\citenamefont
  {Nielsen}, \citenamefont {Flensberg}, \citenamefont {Egger},\ and\
  \citenamefont {Burrello}}]{Nielsen22}%
  \BibitemOpen
  \bibfield  {author} {\bibinfo {author} {\bibfnamefont {I.~E.}\ \bibnamefont
  {Nielsen}}, \bibinfo {author} {\bibfnamefont {K.}~\bibnamefont {Flensberg}},
  \bibinfo {author} {\bibfnamefont {R.}~\bibnamefont {Egger}},\ and\ \bibinfo
  {author} {\bibfnamefont {M.}~\bibnamefont {Burrello}},\ }\bibfield  {title}
  {\bibinfo {title} {Readout of parafermionic states by transport
  measurements},\ }\href {https://doi.org/10.1103/PhysRevLett.129.037703}
  {\bibfield  {journal} {\bibinfo  {journal} {Phys. Rev. Lett.}\ }\textbf
  {\bibinfo {volume} {129}},\ \bibinfo {pages} {037703} (\bibinfo {year}
  {2022})}\BibitemShut {NoStop}%
\bibitem [{\citenamefont {Galambos}\ \emph {et~al.}(2022)\citenamefont
  {Galambos}, \citenamefont {Ronetti}, \citenamefont {Het\'enyi}, \citenamefont
  {Loss},\ and\ \citenamefont {Klinovaja}}]{Galambos22}%
  \BibitemOpen
  \bibfield  {author} {\bibinfo {author} {\bibfnamefont {T.~H.}\ \bibnamefont
  {Galambos}}, \bibinfo {author} {\bibfnamefont {F.}~\bibnamefont {Ronetti}},
  \bibinfo {author} {\bibfnamefont {B.}~\bibnamefont {Het\'enyi}}, \bibinfo
  {author} {\bibfnamefont {D.}~\bibnamefont {Loss}},\ and\ \bibinfo {author}
  {\bibfnamefont {J.}~\bibnamefont {Klinovaja}},\ }\bibfield  {title} {\bibinfo
  {title} {{Crossed Andreev reflection in spin-polarized chiral edge states due
  to the Meissner effect}},\ }\href
  {https://doi.org/10.1103/PhysRevB.106.075410} {\bibfield  {journal} {\bibinfo
   {journal} {Phys. Rev. B}\ }\textbf {\bibinfo {volume} {106}},\ \bibinfo
  {pages} {075410} (\bibinfo {year} {2022})}\BibitemShut {NoStop}%
\bibitem [{\citenamefont {Nayak}\ \emph {et~al.}(2008)\citenamefont {Nayak},
  \citenamefont {Simon}, \citenamefont {Stern}, \citenamefont {Freedman},\ and\
  \citenamefont {{Das Sarma}}}]{Nayak2008}%
  \BibitemOpen
  \bibfield  {author} {\bibinfo {author} {\bibfnamefont {C.}~\bibnamefont
  {Nayak}}, \bibinfo {author} {\bibfnamefont {S.~H.}\ \bibnamefont {Simon}},
  \bibinfo {author} {\bibfnamefont {A.}~\bibnamefont {Stern}}, \bibinfo
  {author} {\bibfnamefont {M.}~\bibnamefont {Freedman}},\ and\ \bibinfo
  {author} {\bibfnamefont {S.}~\bibnamefont {{Das Sarma}}},\ }\bibfield
  {title} {\bibinfo {title} {{Non-Abelian anyons and topological quantum
  computation}},\ }\href {https://doi.org/10.1103/RevModPhys.80.1083}
  {\bibfield  {journal} {\bibinfo  {journal} {Reviews of Modern Physics}\
  }\textbf {\bibinfo {volume} {80}},\ \bibinfo {pages} {1083} (\bibinfo {year}
  {2008})}\BibitemShut {NoStop}%
\bibitem [{\citenamefont {Takagaki}(1998)}]{Takagaki98}%
  \BibitemOpen
  \bibfield  {author} {\bibinfo {author} {\bibfnamefont {Y.}~\bibnamefont
  {Takagaki}},\ }\bibfield  {title} {\bibinfo {title} {Transport properties of
  semiconductor-superconductor junctions in quantizing magnetic fields},\
  }\href {https://doi.org/10.1103/PhysRevB.57.4009} {\bibfield  {journal}
  {\bibinfo  {journal} {Phys. Rev. B}\ }\textbf {\bibinfo {volume} {57}},\
  \bibinfo {pages} {4009} (\bibinfo {year} {1998})}\BibitemShut {NoStop}%
\bibitem [{\citenamefont {Hoppe}\ \emph {et~al.}(2000)\citenamefont {Hoppe},
  \citenamefont {Z{\"{u}}licke},\ and\ \citenamefont
  {Sch{\"{o}}n}}]{Hoppe2000}%
  \BibitemOpen
  \bibfield  {author} {\bibinfo {author} {\bibfnamefont {H.}~\bibnamefont
  {Hoppe}}, \bibinfo {author} {\bibfnamefont {U.}~\bibnamefont
  {Z{\"{u}}licke}},\ and\ \bibinfo {author} {\bibfnamefont {G.}~\bibnamefont
  {Sch{\"{o}}n}},\ }\bibfield  {title} {\bibinfo {title} {{Andreev reflection
  in strong magnetic fields}},\ }\href
  {https://doi.org/10.1103/PhysRevLett.84.1804} {\bibfield  {journal} {\bibinfo
   {journal} {Physical Review Letters}\ }\textbf {\bibinfo {volume} {84}},\
  \bibinfo {pages} {1804} (\bibinfo {year} {2000})}\BibitemShut {NoStop}%
\bibitem [{\citenamefont {Wang}\ \emph {et~al.}(2013)\citenamefont {Wang},
  \citenamefont {Meric}, \citenamefont {Huang}, \citenamefont {Gao},
  \citenamefont {Gao}, \citenamefont {Tran}, \citenamefont {Taniguchi},
  \citenamefont {Watanabe}, \citenamefont {Campos}, \citenamefont {Muller},
  \citenamefont {Guo}, \citenamefont {Kim}, \citenamefont {Hone}, \citenamefont
  {Shepard},\ and\ \citenamefont {Dean}}]{Wang13}%
  \BibitemOpen
  \bibfield  {author} {\bibinfo {author} {\bibfnamefont {L.}~\bibnamefont
  {Wang}}, \bibinfo {author} {\bibfnamefont {I.}~\bibnamefont {Meric}},
  \bibinfo {author} {\bibfnamefont {P.~Y.}\ \bibnamefont {Huang}}, \bibinfo
  {author} {\bibfnamefont {Q.}~\bibnamefont {Gao}}, \bibinfo {author}
  {\bibfnamefont {Y.}~\bibnamefont {Gao}}, \bibinfo {author} {\bibfnamefont
  {H.}~\bibnamefont {Tran}}, \bibinfo {author} {\bibfnamefont {T.}~\bibnamefont
  {Taniguchi}}, \bibinfo {author} {\bibfnamefont {K.}~\bibnamefont {Watanabe}},
  \bibinfo {author} {\bibfnamefont {L.~M.}\ \bibnamefont {Campos}}, \bibinfo
  {author} {\bibfnamefont {D.~A.}\ \bibnamefont {Muller}}, \bibinfo {author}
  {\bibfnamefont {J.}~\bibnamefont {Guo}}, \bibinfo {author} {\bibfnamefont
  {P.}~\bibnamefont {Kim}}, \bibinfo {author} {\bibfnamefont {J.}~\bibnamefont
  {Hone}}, \bibinfo {author} {\bibfnamefont {K.~L.}\ \bibnamefont {Shepard}},\
  and\ \bibinfo {author} {\bibfnamefont {C.~R.}\ \bibnamefont {Dean}},\
  }\bibfield  {title} {\bibinfo {title} {{One-Dimensional Electrical Contact to
  a Two-Dimensional Material}},\ }\href
  {https://doi.org/10.1126/science.1244358} {\bibfield  {journal} {\bibinfo
  {journal} {Science}\ }\textbf {\bibinfo {volume} {342}},\ \bibinfo {pages}
  {614} (\bibinfo {year} {2013})}\BibitemShut {NoStop}%
\bibitem [{\citenamefont {Abanin}\ and\ \citenamefont
  {Levitov}(2008)}]{Abanin2008}%
  \BibitemOpen
  \bibfield  {author} {\bibinfo {author} {\bibfnamefont {D.~A.}\ \bibnamefont
  {Abanin}}\ and\ \bibinfo {author} {\bibfnamefont {L.~S.}\ \bibnamefont
  {Levitov}},\ }\bibfield  {title} {\bibinfo {title} {{Conformal invariance and
  shape-dependent conductance of graphene samples}},\ }\href
  {https://doi.org/10.1103/PhysRevB.78.035416} {\bibfield  {journal} {\bibinfo
  {journal} {Physical Review B}\ }\textbf {\bibinfo {volume} {78}},\ \bibinfo
  {pages} {035416} (\bibinfo {year} {2008})}\BibitemShut {NoStop}%
\bibitem [{\citenamefont {Williams}\ \emph {et~al.}(2009)\citenamefont
  {Williams}, \citenamefont {Abanin}, \citenamefont {Dicarlo}, \citenamefont
  {Levitov},\ and\ \citenamefont {Marcus}}]{Williams2009}%
  \BibitemOpen
  \bibfield  {author} {\bibinfo {author} {\bibfnamefont {J.~R.}\ \bibnamefont
  {Williams}}, \bibinfo {author} {\bibfnamefont {D.~A.}\ \bibnamefont
  {Abanin}}, \bibinfo {author} {\bibfnamefont {L.}~\bibnamefont {Dicarlo}},
  \bibinfo {author} {\bibfnamefont {L.~S.}\ \bibnamefont {Levitov}},\ and\
  \bibinfo {author} {\bibfnamefont {C.~M.}\ \bibnamefont {Marcus}},\ }\bibfield
   {title} {\bibinfo {title} {{Quantum Hall conductance of two-terminal
  graphene devices}},\ }\href {https://doi.org/10.1103/PhysRevB.80.045408}
  {\bibfield  {journal} {\bibinfo  {journal} {Physical Review B}\ }\textbf
  {\bibinfo {volume} {80}},\ \bibinfo {pages} {045408} (\bibinfo {year}
  {2009})}\BibitemShut {NoStop}%
\bibitem [{\citenamefont {Coissard}\ \emph {et~al.}(2022)\citenamefont
  {Coissard}, \citenamefont {Grushin}, \citenamefont {Repellin}, \citenamefont
  {Veyrat}, \citenamefont {Watanabe}, \citenamefont {Taniguchi}, \citenamefont
  {Gay}, \citenamefont {Courtois}, \citenamefont {Sellier},\ and\ \citenamefont
  {Sac{\'e}p{\'e}}}]{Coissard22b}%
  \BibitemOpen
  \bibfield  {author} {\bibinfo {author} {\bibfnamefont {A.}~\bibnamefont
  {Coissard}}, \bibinfo {author} {\bibfnamefont {A.~G.}\ \bibnamefont
  {Grushin}}, \bibinfo {author} {\bibfnamefont {C.}~\bibnamefont {Repellin}},
  \bibinfo {author} {\bibfnamefont {L.}~\bibnamefont {Veyrat}}, \bibinfo
  {author} {\bibfnamefont {K.}~\bibnamefont {Watanabe}}, \bibinfo {author}
  {\bibfnamefont {T.}~\bibnamefont {Taniguchi}}, \bibinfo {author}
  {\bibfnamefont {F.}~\bibnamefont {Gay}}, \bibinfo {author} {\bibfnamefont
  {H.}~\bibnamefont {Courtois}}, \bibinfo {author} {\bibfnamefont
  {H.}~\bibnamefont {Sellier}},\ and\ \bibinfo {author} {\bibfnamefont
  {B.}~\bibnamefont {Sac{\'e}p{\'e}}},\ }\bibfield  {title} {\bibinfo {title}
  {{Absence of edge reconstruction for quantum Hall edge channels in graphene
  devices}},\ }\href@noop {} {\bibfield  {journal} {\bibinfo  {journal}
  {arXiv:2210.08152}\ } (\bibinfo {year} {2022})}\BibitemShut {NoStop}%
\bibitem [{\citenamefont {D{\'e}prez}\ \emph {et~al.}(2021)\citenamefont
  {D{\'e}prez}, \citenamefont {Veyrat}, \citenamefont {Vignaud}, \citenamefont
  {Nayak}, \citenamefont {Watanabe}, \citenamefont {Taniguchi}, \citenamefont
  {Gay}, \citenamefont {Sellier},\ and\ \citenamefont
  {Sac{\'e}p{\'e}}}]{Deprez21}%
  \BibitemOpen
  \bibfield  {author} {\bibinfo {author} {\bibfnamefont {C.}~\bibnamefont
  {D{\'e}prez}}, \bibinfo {author} {\bibfnamefont {L.}~\bibnamefont {Veyrat}},
  \bibinfo {author} {\bibfnamefont {H.}~\bibnamefont {Vignaud}}, \bibinfo
  {author} {\bibfnamefont {G.}~\bibnamefont {Nayak}}, \bibinfo {author}
  {\bibfnamefont {K.}~\bibnamefont {Watanabe}}, \bibinfo {author}
  {\bibfnamefont {T.}~\bibnamefont {Taniguchi}}, \bibinfo {author}
  {\bibfnamefont {F.}~\bibnamefont {Gay}}, \bibinfo {author} {\bibfnamefont
  {H.}~\bibnamefont {Sellier}},\ and\ \bibinfo {author} {\bibfnamefont
  {B.}~\bibnamefont {Sac{\'e}p{\'e}}},\ }\bibfield  {title} {\bibinfo {title}
  {{A tunable Fabry--P{\'e}rot quantum Hall interferometer in graphene}},\
  }\href {https://doi.org/10.1038/s41565-021-00847-x} {\bibfield  {journal}
  {\bibinfo  {journal} {Nature Nanotechnology}\ }\textbf {\bibinfo {volume}
  {16}},\ \bibinfo {pages} {555} (\bibinfo {year} {2021})}\BibitemShut
  {NoStop}%
\bibitem [{\citenamefont {Ronen}\ \emph {et~al.}(2021)\citenamefont {Ronen},
  \citenamefont {Werkmeister}, \citenamefont {Haie~Najafabadi}, \citenamefont
  {Pierce}, \citenamefont {Anderson}, \citenamefont {Shin}, \citenamefont
  {Lee}, \citenamefont {Lee}, \citenamefont {Johnson}, \citenamefont
  {Watanabe}, \citenamefont {Taniguchi}, \citenamefont {Yacoby},\ and\
  \citenamefont {Kim}}]{Ronen2021}%
  \BibitemOpen
  \bibfield  {author} {\bibinfo {author} {\bibfnamefont {Y.}~\bibnamefont
  {Ronen}}, \bibinfo {author} {\bibfnamefont {T.}~\bibnamefont {Werkmeister}},
  \bibinfo {author} {\bibfnamefont {D.}~\bibnamefont {Haie~Najafabadi}},
  \bibinfo {author} {\bibfnamefont {A.~T.}\ \bibnamefont {Pierce}}, \bibinfo
  {author} {\bibfnamefont {L.~E.}\ \bibnamefont {Anderson}}, \bibinfo {author}
  {\bibfnamefont {Y.~J.}\ \bibnamefont {Shin}}, \bibinfo {author}
  {\bibfnamefont {S.~Y.}\ \bibnamefont {Lee}}, \bibinfo {author} {\bibfnamefont
  {Y.~H.}\ \bibnamefont {Lee}}, \bibinfo {author} {\bibfnamefont
  {B.}~\bibnamefont {Johnson}}, \bibinfo {author} {\bibfnamefont
  {K.}~\bibnamefont {Watanabe}}, \bibinfo {author} {\bibfnamefont
  {T.}~\bibnamefont {Taniguchi}}, \bibinfo {author} {\bibfnamefont
  {A.}~\bibnamefont {Yacoby}},\ and\ \bibinfo {author} {\bibfnamefont
  {P.}~\bibnamefont {Kim}},\ }\bibfield  {title} {\bibinfo {title}
  {{Aharonov--Bohm effect in graphene-based Fabry--P{\'e}rot quantum Hall
  interferometers}},\ }\href@noop {} {\bibfield  {journal} {\bibinfo  {journal}
  {Nature Nanotechnology}\ }\textbf {\bibinfo {volume} {16}},\ \bibinfo {pages}
  {563} (\bibinfo {year} {2021})}\BibitemShut {NoStop}%
\bibitem [{\citenamefont {Kurilovich}\ and\ \citenamefont
  {Glazman}(2022)}]{Kurilovich2022b}%
  \BibitemOpen
  \bibfield  {author} {\bibinfo {author} {\bibfnamefont {V.~D.}\ \bibnamefont
  {Kurilovich}}\ and\ \bibinfo {author} {\bibfnamefont {L.~I.}\ \bibnamefont
  {Glazman}},\ }\bibfield  {title} {\bibinfo {title} {{Criticality in the
  crossed Andreev reflection of a quantum Hall edge}},\ }\href@noop {}
  {\bibfield  {journal} {\bibinfo  {journal} {arXiv:2209.12932}\ } (\bibinfo
  {year} {2022})}\BibitemShut {NoStop}%
\bibitem [{\citenamefont {Chtchelkatchev}\ and\ \citenamefont
  {Burmistrov}(2007)}]{Chtchelkatchev2007}%
  \BibitemOpen
  \bibfield  {author} {\bibinfo {author} {\bibfnamefont {N.~M.}\ \bibnamefont
  {Chtchelkatchev}}\ and\ \bibinfo {author} {\bibfnamefont {I.~S.}\
  \bibnamefont {Burmistrov}},\ }\bibfield  {title} {\bibinfo {title}
  {{Conductance oscillations with magnetic field of a two-dimensional electron
  gas-superconductor junction}},\ }\href
  {https://doi.org/10.1103/PhysRevB.75.214510} {\bibfield  {journal} {\bibinfo
  {journal} {Physical Review B}\ }\textbf {\bibinfo {volume} {75}},\ \bibinfo
  {pages} {214510} (\bibinfo {year} {2007})}\BibitemShut {NoStop}%
\bibitem [{\citenamefont {Batov}\ \emph {et~al.}(2009)\citenamefont {Batov},
  \citenamefont {Schaepers}, \citenamefont {Chtchelkatchev}, \citenamefont
  {Golubov}, \citenamefont {Hardtdegen},\ and\ \citenamefont
  {Ustinov}}]{Batov09}%
  \BibitemOpen
  \bibfield  {author} {\bibinfo {author} {\bibfnamefont {I.}~\bibnamefont
  {Batov}}, \bibinfo {author} {\bibfnamefont {T.}~\bibnamefont {Schaepers}},
  \bibinfo {author} {\bibfnamefont {N.}~\bibnamefont {Chtchelkatchev}},
  \bibinfo {author} {\bibfnamefont {A.}~\bibnamefont {Golubov}}, \bibinfo
  {author} {\bibfnamefont {H.}~\bibnamefont {Hardtdegen}},\ and\ \bibinfo
  {author} {\bibfnamefont {A.}~\bibnamefont {Ustinov}},\ }\bibfield  {title}
  {\bibinfo {title} {{Electronic transport in mesoscopic superconductor/2D
  electron gas junctions in strong magnetic field}},\ }\href@noop {} {\bibfield
   {journal} {\bibinfo  {journal} {Bulletin of the Russian Academy of Sciences:
  Physics}\ }\textbf {\bibinfo {volume} {73}},\ \bibinfo {pages} {880}
  (\bibinfo {year} {2009})}\BibitemShut {NoStop}%
\bibitem [{\citenamefont {Blonder}\ \emph {et~al.}(1982)\citenamefont
  {Blonder}, \citenamefont {Tinkham},\ and\ \citenamefont
  {Klapwijk}}]{Blonder82}%
  \BibitemOpen
  \bibfield  {author} {\bibinfo {author} {\bibfnamefont {G.~E.}\ \bibnamefont
  {Blonder}}, \bibinfo {author} {\bibfnamefont {M.}~\bibnamefont {Tinkham}},\
  and\ \bibinfo {author} {\bibfnamefont {T.~M.}\ \bibnamefont {Klapwijk}},\
  }\bibfield  {title} {\bibinfo {title} {{Transition from metallic to tunneling
  regimes in superconducting microconstrictions: Excess current, charge
  imbalance, and supercurrent conversion}},\ }\href
  {https://doi.org/10.1103/PhysRevB.25.4515} {\bibfield  {journal} {\bibinfo
  {journal} {Phys. Rev. B}\ }\textbf {\bibinfo {volume} {25}},\ \bibinfo
  {pages} {4515} (\bibinfo {year} {1982})}\BibitemShut {NoStop}%
\bibitem [{\citenamefont {Manesco}\ \emph
  {et~al.}(2022{\natexlab{b}})\citenamefont {Manesco}, \citenamefont
  {Fl{\'o}r}, \citenamefont {Liu},\ and\ \citenamefont
  {Akhmerov}}]{Manesco2021}%
  \BibitemOpen
  \bibfield  {author} {\bibinfo {author} {\bibfnamefont {A.}~\bibnamefont
  {Manesco}}, \bibinfo {author} {\bibfnamefont {I.~M.}\ \bibnamefont
  {Fl{\'o}r}}, \bibinfo {author} {\bibfnamefont {C.-X.}\ \bibnamefont {Liu}},\
  and\ \bibinfo {author} {\bibfnamefont {A.}~\bibnamefont {Akhmerov}},\
  }\bibfield  {title} {\bibinfo {title} {{Mechanisms of Andreev reflection in
  quantum Hall graphene}},\ }\href@noop {} {\bibfield  {journal} {\bibinfo
  {journal} {SciPost Physics Core}\ }\textbf {\bibinfo {volume} {5}},\ \bibinfo
  {pages} {045} (\bibinfo {year} {2022}{\natexlab{b}})}\BibitemShut {NoStop}%
\bibitem [{\citenamefont {Halperin}\ \emph {et~al.}(2011)\citenamefont
  {Halperin}, \citenamefont {Stern}, \citenamefont {Neder},\ and\ \citenamefont
  {Rosenow}}]{halperin11}%
  \BibitemOpen
  \bibfield  {author} {\bibinfo {author} {\bibfnamefont {B.~I.}\ \bibnamefont
  {Halperin}}, \bibinfo {author} {\bibfnamefont {A.}~\bibnamefont {Stern}},
  \bibinfo {author} {\bibfnamefont {I.}~\bibnamefont {Neder}},\ and\ \bibinfo
  {author} {\bibfnamefont {B.}~\bibnamefont {Rosenow}},\ }\bibfield  {title}
  {\bibinfo {title} {{Theory of the Fabry-P{\'e}rot quantum Hall
  interferometer}},\ }\href@noop {} {\bibfield  {journal} {\bibinfo  {journal}
  {Physical Review B}\ }\textbf {\bibinfo {volume} {83}},\ \bibinfo {pages}
  {155440} (\bibinfo {year} {2011})}\BibitemShut {NoStop}%
\bibitem [{\citenamefont {Teyssandier}\ and\ \citenamefont
  {Pr{\^e}le}(2010)}]{teyssandier10}%
  \BibitemOpen
  \bibfield  {author} {\bibinfo {author} {\bibfnamefont {F.}~\bibnamefont
  {Teyssandier}}\ and\ \bibinfo {author} {\bibfnamefont {D.}~\bibnamefont
  {Pr{\^e}le}},\ }\bibfield  {title} {\bibinfo {title} {Commercially available
  capacitors at cryogenic temperatures},\ }\href@noop {} {\bibfield  {journal}
  {\bibinfo  {journal} {Ninth International Workshop on Low Temperature
  Electronics-WOLTE9, https://hal.science/hal-00623399}\ } (\bibinfo {year}
  {2010})}\BibitemShut {NoStop}%
\bibitem [{\citenamefont {Z{\"u}licke}\ \emph {et~al.}(2001)\citenamefont
  {Z{\"u}licke}, \citenamefont {Hoppe},\ and\ \citenamefont
  {Sch{\"o}n}}]{Zulicke01}%
  \BibitemOpen
  \bibfield  {author} {\bibinfo {author} {\bibfnamefont {U.}~\bibnamefont
  {Z{\"u}licke}}, \bibinfo {author} {\bibfnamefont {H.}~\bibnamefont {Hoppe}},\
  and\ \bibinfo {author} {\bibfnamefont {G.}~\bibnamefont {Sch{\"o}n}},\
  }\bibfield  {title} {\bibinfo {title} {Andreev reflection at
  superconductor-semiconductor interfaces in high magnetic fields},\
  }\href@noop {} {\bibfield  {journal} {\bibinfo  {journal} {Physica B:
  Condensed Matter}\ }\textbf {\bibinfo {volume} {298}},\ \bibinfo {pages}
  {453} (\bibinfo {year} {2001})}\BibitemShut {NoStop}%
\bibitem [{\citenamefont {Kim}\ \emph {et~al.}(2018)\citenamefont {Kim},
  \citenamefont {Gay}, \citenamefont {Del~Maestro}, \citenamefont
  {Sac{\'e}p{\'e}},\ and\ \citenamefont {Rogachev}}]{Kim2018}%
  \BibitemOpen
  \bibfield  {author} {\bibinfo {author} {\bibfnamefont {H.}~\bibnamefont
  {Kim}}, \bibinfo {author} {\bibfnamefont {F.}~\bibnamefont {Gay}}, \bibinfo
  {author} {\bibfnamefont {A.}~\bibnamefont {Del~Maestro}}, \bibinfo {author}
  {\bibfnamefont {B.}~\bibnamefont {Sac{\'e}p{\'e}}},\ and\ \bibinfo {author}
  {\bibfnamefont {A.}~\bibnamefont {Rogachev}},\ }\bibfield  {title} {\bibinfo
  {title} {Pair-breaking quantum phase transition in superconducting
  nanowires},\ }\href@noop {} {\bibfield  {journal} {\bibinfo  {journal}
  {Nature Physics}\ }\textbf {\bibinfo {volume} {14}},\ \bibinfo {pages} {912}
  (\bibinfo {year} {2018})}\BibitemShut {NoStop}%
\bibitem [{\citenamefont {Mazin}\ \emph {et~al.}(1995)\citenamefont {Mazin},
  \citenamefont {Golubov},\ and\ \citenamefont {Zaikin}}]{Mazin95}%
  \BibitemOpen
  \bibfield  {author} {\bibinfo {author} {\bibfnamefont {I.~I.}\ \bibnamefont
  {Mazin}}, \bibinfo {author} {\bibfnamefont {A.~A.}\ \bibnamefont {Golubov}},\
  and\ \bibinfo {author} {\bibfnamefont {A.~D.}\ \bibnamefont {Zaikin}},\
  }\bibfield  {title} {\bibinfo {title} {{``Chain Scenario'' for Josephson
  Tunneling with $\ensuremath{\pi}$ Shift in
  YB${\mathrm{a}}_{2}$C${\mathrm{u}}_{3}$${\mathrm{O}}_{7}$}},\ }\href
  {https://doi.org/10.1103/PhysRevLett.75.2574} {\bibfield  {journal} {\bibinfo
   {journal} {Phys. Rev. Lett.}\ }\textbf {\bibinfo {volume} {75}},\ \bibinfo
  {pages} {2574} (\bibinfo {year} {1995})}\BibitemShut {NoStop}%
\bibitem [{\citenamefont {Calado}\ \emph {et~al.}(2015)\citenamefont {Calado},
  \citenamefont {Goswami}, \citenamefont {Nanda}, \citenamefont {Diez},
  \citenamefont {Akhmerov}, \citenamefont {Watanabe}, \citenamefont
  {Taniguchi}, \citenamefont {Klapwijk},\ and\ \citenamefont
  {Vandersypen}}]{Calado2015a}%
  \BibitemOpen
  \bibfield  {author} {\bibinfo {author} {\bibfnamefont {V.~E.}\ \bibnamefont
  {Calado}}, \bibinfo {author} {\bibfnamefont {S.}~\bibnamefont {Goswami}},
  \bibinfo {author} {\bibfnamefont {G.}~\bibnamefont {Nanda}}, \bibinfo
  {author} {\bibfnamefont {M.}~\bibnamefont {Diez}}, \bibinfo {author}
  {\bibfnamefont {A.~R.}\ \bibnamefont {Akhmerov}}, \bibinfo {author}
  {\bibfnamefont {K.}~\bibnamefont {Watanabe}}, \bibinfo {author}
  {\bibfnamefont {T.}~\bibnamefont {Taniguchi}}, \bibinfo {author}
  {\bibfnamefont {T.~M.}\ \bibnamefont {Klapwijk}},\ and\ \bibinfo {author}
  {\bibfnamefont {L.~M.~K.}\ \bibnamefont {Vandersypen}},\ }\bibfield  {title}
  {\bibinfo {title} {{Ballistic Josephson junctions in edge-contacted
  graphene}},\ }\href {https://doi.org/10.1038/nnano.2015.156} {\bibfield
  {journal} {\bibinfo  {journal} {Nature Nanotechnology}\ }\textbf {\bibinfo
  {volume} {10}},\ \bibinfo {pages} {761} (\bibinfo {year} {2015})}\BibitemShut
  {NoStop}%
\bibitem [{\citenamefont {Allen}\ \emph {et~al.}(2017)\citenamefont {Allen},
  \citenamefont {Shtanko}, \citenamefont {Fulga}, \citenamefont {Wang},
  \citenamefont {Nurgaliev}, \citenamefont {Watanabe}, \citenamefont
  {Taniguchi}, \citenamefont {Akhmerov}, \citenamefont {Jarillo-Herrero},
  \citenamefont {Levitov},\ and\ \citenamefont {Yacoby}}]{Allen2017}%
  \BibitemOpen
  \bibfield  {author} {\bibinfo {author} {\bibfnamefont {M.~T.}\ \bibnamefont
  {Allen}}, \bibinfo {author} {\bibfnamefont {O.}~\bibnamefont {Shtanko}},
  \bibinfo {author} {\bibfnamefont {I.~C.}\ \bibnamefont {Fulga}}, \bibinfo
  {author} {\bibfnamefont {J.~I.}\ \bibnamefont {Wang}}, \bibinfo {author}
  {\bibfnamefont {D.}~\bibnamefont {Nurgaliev}}, \bibinfo {author}
  {\bibfnamefont {K.}~\bibnamefont {Watanabe}}, \bibinfo {author}
  {\bibfnamefont {T.}~\bibnamefont {Taniguchi}}, \bibinfo {author}
  {\bibfnamefont {A.~R.}\ \bibnamefont {Akhmerov}}, \bibinfo {author}
  {\bibfnamefont {P.}~\bibnamefont {Jarillo-Herrero}}, \bibinfo {author}
  {\bibfnamefont {L.~S.}\ \bibnamefont {Levitov}},\ and\ \bibinfo {author}
  {\bibfnamefont {A.}~\bibnamefont {Yacoby}},\ }\bibfield  {title} {\bibinfo
  {title} {{Observation of Electron Coherence and Fabry-P\'{e}rot Standing
  Waves at a Graphene Edge}},\ }\href
  {https://doi.org/10.1021/acs.nanolett.7b03156} {\bibfield  {journal}
  {\bibinfo  {journal} {Nano Letters}\ }\textbf {\bibinfo {volume} {17}},\
  \bibinfo {pages} {7380} (\bibinfo {year} {2017})}\BibitemShut {NoStop}%
\bibitem [{\citenamefont {Octavio}\ \emph {et~al.}(1983)\citenamefont
  {Octavio}, \citenamefont {Tinkham}, \citenamefont {Blonder},\ and\
  \citenamefont {Klapwijk}}]{Octavio1983}%
  \BibitemOpen
  \bibfield  {author} {\bibinfo {author} {\bibfnamefont {M.}~\bibnamefont
  {Octavio}}, \bibinfo {author} {\bibfnamefont {M.}~\bibnamefont {Tinkham}},
  \bibinfo {author} {\bibfnamefont {G.~E.}\ \bibnamefont {Blonder}},\ and\
  \bibinfo {author} {\bibfnamefont {T.~M.}\ \bibnamefont {Klapwijk}},\
  }\bibfield  {title} {\bibinfo {title} {{Subharmonic energy-gap structure in
  superconducting constrictions}},\ }\href
  {https://doi.org/10.1103/PhysRevB.27.6739} {\bibfield  {journal} {\bibinfo
  {journal} {Phys. Rev. B}\ }\textbf {\bibinfo {volume} {27}},\ \bibinfo
  {pages} {6739} (\bibinfo {year} {1983})}\BibitemShut {NoStop}%
\bibitem [{\citenamefont {B\"uttiker}(1988)}]{Buttiker88}%
  \BibitemOpen
  \bibfield  {author} {\bibinfo {author} {\bibfnamefont {M.}~\bibnamefont
  {B\"uttiker}},\ }\bibfield  {title} {\bibinfo {title} {{Absence of
  backscattering in the quantum Hall effect in multiprobe conductors}},\ }\href
  {https://doi.org/10.1103/PhysRevB.38.9375} {\bibfield  {journal} {\bibinfo
  {journal} {Phys. Rev. B}\ }\textbf {\bibinfo {volume} {38}},\ \bibinfo
  {pages} {9375} (\bibinfo {year} {1988})}\BibitemShut {NoStop}%
\bibitem [{\citenamefont {Hong}\ \emph {et~al.}(2020)\citenamefont {Hong},
  \citenamefont {Lee}, \citenamefont {Lee}, \citenamefont {Lee}, \citenamefont
  {Ma}, \citenamefont {Kim}, \citenamefont {Yoon}, \citenamefont {Ihm},
  \citenamefont {Kim}, \citenamefont {Shin}, \citenamefont {Kim}, \citenamefont
  {Jeon}, \citenamefont {Jeon}, \citenamefont {Kim}, \citenamefont {Lee},
  \citenamefont {Lee}, \citenamefont {Antidormi}, \citenamefont {Roche},
  \citenamefont {Chhowalla}, \citenamefont {Shin},\ and\ \citenamefont
  {Shin}}]{Hong2020}%
  \BibitemOpen
  \bibfield  {author} {\bibinfo {author} {\bibfnamefont {S.}~\bibnamefont
  {Hong}}, \bibinfo {author} {\bibfnamefont {C.-S.}\ \bibnamefont {Lee}},
  \bibinfo {author} {\bibfnamefont {M.-H.}\ \bibnamefont {Lee}}, \bibinfo
  {author} {\bibfnamefont {Y.}~\bibnamefont {Lee}}, \bibinfo {author}
  {\bibfnamefont {K.~Y.}\ \bibnamefont {Ma}}, \bibinfo {author} {\bibfnamefont
  {G.}~\bibnamefont {Kim}}, \bibinfo {author} {\bibfnamefont {S.~I.}\
  \bibnamefont {Yoon}}, \bibinfo {author} {\bibfnamefont {K.}~\bibnamefont
  {Ihm}}, \bibinfo {author} {\bibfnamefont {K.-J.}\ \bibnamefont {Kim}},
  \bibinfo {author} {\bibfnamefont {T.~J.}\ \bibnamefont {Shin}}, \bibinfo
  {author} {\bibfnamefont {S.~W.}\ \bibnamefont {Kim}}, \bibinfo {author}
  {\bibfnamefont {E.-c.}\ \bibnamefont {Jeon}}, \bibinfo {author}
  {\bibfnamefont {H.}~\bibnamefont {Jeon}}, \bibinfo {author} {\bibfnamefont
  {J.-Y.}\ \bibnamefont {Kim}}, \bibinfo {author} {\bibfnamefont {H.-I.}\
  \bibnamefont {Lee}}, \bibinfo {author} {\bibfnamefont {Z.}~\bibnamefont
  {Lee}}, \bibinfo {author} {\bibfnamefont {A.}~\bibnamefont {Antidormi}},
  \bibinfo {author} {\bibfnamefont {S.}~\bibnamefont {Roche}}, \bibinfo
  {author} {\bibfnamefont {M.}~\bibnamefont {Chhowalla}}, \bibinfo {author}
  {\bibfnamefont {H.-J.}\ \bibnamefont {Shin}},\ and\ \bibinfo {author}
  {\bibfnamefont {H.~S.}\ \bibnamefont {Shin}},\ }\bibfield  {title} {\bibinfo
  {title} {Ultralow-dielectric-constant amorphous boron nitride},\ }\href@noop
  {} {\bibfield  {journal} {\bibinfo  {journal} {Nature}\ }\textbf {\bibinfo
  {volume} {582}},\ \bibinfo {pages} {511} (\bibinfo {year}
  {2020})}\BibitemShut {NoStop}%
\bibitem [{\citenamefont {Flensberg}\ \emph {et~al.}(1988)\citenamefont
  {Flensberg}, \citenamefont {Hansen},\ and\ \citenamefont
  {Octavio}}]{Flensberg1988}%
  \BibitemOpen
  \bibfield  {author} {\bibinfo {author} {\bibfnamefont {K.}~\bibnamefont
  {Flensberg}}, \bibinfo {author} {\bibfnamefont {J.~B.}\ \bibnamefont
  {Hansen}},\ and\ \bibinfo {author} {\bibfnamefont {M.}~\bibnamefont
  {Octavio}},\ }\bibfield  {title} {\bibinfo {title} {{Subharmonic energy-gap
  structure in superconducting weak links}},\ }\href
  {https://doi.org/10.1103/PhysRevB.38.8707} {\bibfield  {journal} {\bibinfo
  {journal} {Phys. Rev. B}\ }\textbf {\bibinfo {volume} {38}},\ \bibinfo
  {pages} {8707} (\bibinfo {year} {1988})}\BibitemShut {NoStop}%
\bibitem [{\citenamefont {Niebler}\ \emph {et~al.}(2009)\citenamefont
  {Niebler}, \citenamefont {Cuniberti},\ and\ \citenamefont
  {Novotn{\'{y}}}}]{Niebler2009}%
  \BibitemOpen
  \bibfield  {author} {\bibinfo {author} {\bibfnamefont {G.}~\bibnamefont
  {Niebler}}, \bibinfo {author} {\bibfnamefont {G.}~\bibnamefont {Cuniberti}},\
  and\ \bibinfo {author} {\bibfnamefont {T.}~\bibnamefont {Novotn{\'{y}}}},\
  }\bibfield  {title} {\bibinfo {title} {{Analytical calculation of the excess
  current in the Octavio-Tinkham-Blonder-Klapwijk theory}},\ }\bibfield
  {journal} {\bibinfo  {journal} {Superconductor Science and Technology}\
  }\textbf {\bibinfo {volume} {22}},\ \href
  {https://doi.org/10.1088/0953-2048/22/8/085016}
  {10.1088/0953-2048/22/8/085016} (\bibinfo {year} {2009})\BibitemShut
  {NoStop}%
\bibitem [{\citenamefont {Chiodi}\ \emph {et~al.}(2012)\citenamefont {Chiodi},
  \citenamefont {Ferrier}, \citenamefont {Gu\'eron}, \citenamefont {Cuevas},
  \citenamefont {Montambaux}, \citenamefont {Fortuna}, \citenamefont
  {Kasumov},\ and\ \citenamefont {Bouchiat}}]{Chiodi12}%
  \BibitemOpen
  \bibfield  {author} {\bibinfo {author} {\bibfnamefont {F.}~\bibnamefont
  {Chiodi}}, \bibinfo {author} {\bibfnamefont {M.}~\bibnamefont {Ferrier}},
  \bibinfo {author} {\bibfnamefont {S.}~\bibnamefont {Gu\'eron}}, \bibinfo
  {author} {\bibfnamefont {J.~C.}\ \bibnamefont {Cuevas}}, \bibinfo {author}
  {\bibfnamefont {G.}~\bibnamefont {Montambaux}}, \bibinfo {author}
  {\bibfnamefont {F.}~\bibnamefont {Fortuna}}, \bibinfo {author} {\bibfnamefont
  {A.}~\bibnamefont {Kasumov}},\ and\ \bibinfo {author} {\bibfnamefont
  {H.}~\bibnamefont {Bouchiat}},\ }\bibfield  {title} {\bibinfo {title}
  {{Geometry-related magnetic interference patterns in long $SNS$ Josephson
  junctions}},\ }\href {https://doi.org/10.1103/PhysRevB.86.064510} {\bibfield
  {journal} {\bibinfo  {journal} {Phys. Rev. B}\ }\textbf {\bibinfo {volume}
  {86}},\ \bibinfo {pages} {064510} (\bibinfo {year} {2012})}\BibitemShut
  {NoStop}%
\bibitem [{\citenamefont {Tinkham}(1996)}]{Tinkham2014}%
  \BibitemOpen
  \bibfield  {author} {\bibinfo {author} {\bibfnamefont {M.}~\bibnamefont
  {Tinkham}},\ }\href@noop {} {\emph {\bibinfo {title} {Introduction to
  superconductivity}}}\ (\bibinfo  {publisher} {Dover, Mineola},\ \bibinfo
  {year} {1996})\BibitemShut {NoStop}%
\end{thebibliography}%

~\vfill
\begin{table*}[!ht]
\centering
\begin{tabular}{|c||c|c|c|c|c|c|c|}
\hline
Device & $\vphantom{M^{0^0}}$	 $L$ (nm) & $W$ (nm) & $A \times 10^3 \ (\upmu m^2)$ & $C_{\rm{g}} \ (\rm{mF}/\rm{m}^2)$\\
\hline
$\vphantom{M^{0^0}}$ HV88-B & 140 & 178 & $25 \pm 8$ & 1.27  \\
\hline
$\vphantom{M^{0^0}}$ HV88-C & 170 & 210  & $36 \pm 9$ & 1.39 \\
\hline
$\vphantom{M^{0^0}}$ HV88-D & 200 & 247 & $49 \pm 11$ & 1.43\\
\hline
$\vphantom{M^{0^0}}$ HV88-E & 240 & 288  & $69 \pm 13$ & 1.14 \\
\hline
$\vphantom{M^{0^0}}$ HV88-F & 270 & 334  & $90 \pm 14$ & 1.55 \\
\hline
$\vphantom{M^{0^0}}$ HV88-G & 107 & 2332 & $250 \pm 50$ & 2.12 \\
\hline
$\vphantom{M^{0^0}}$ HV88-H & 202 & 2434 & $492 \pm 56$ & 1.93 \\
\hline
$\vphantom{M^{0^0}}$ DP24-C & 170 & 125  & $21 \pm 8$ & 1.12 \\
\hline
$\vphantom{M^{0^0}}$ DP24-D & 200 & 125  & $25 \pm 8$ & 1.89 \\
\hline
\end{tabular}
\caption{\textbf{Device parameters.} $L$ is the length separating the junction electrodes and $W$ the width of the junction (see schematics in Fig. \ref{ExtDataFig1}a). $A$ is the geometric area of the graphene (see Methods for uncertainty calculation). $C_{g}$ is the capacitance per unit area deduced from the slope of the $\nu = 2$ quantum Hall plateau center (see SI section II.C). }
\label{TabI}
\end{table*}

\newpage

~\vfill

\newpage

~\vfill

\newpage

~\vfill

\newpage

~\vfill

\clearpage

\clearpage


\clearpage
\onecolumngrid
\setcounter{figure}{0}
\setcounter{section}{0}
\renewcommand{\thefigure}{S\arabic{figure}}

\newpage

~\vspace{5em}
\part*{ \centering Supplementary Information}
\bigskip
\vspace{2em}

\section{Critical temperature and upper critical field of the M\MakeLowercase{o}G\MakeLowercase{e}  electrodes}
\label{secHc}

\begin{figure}[ht!]
	\centering
		\includegraphics{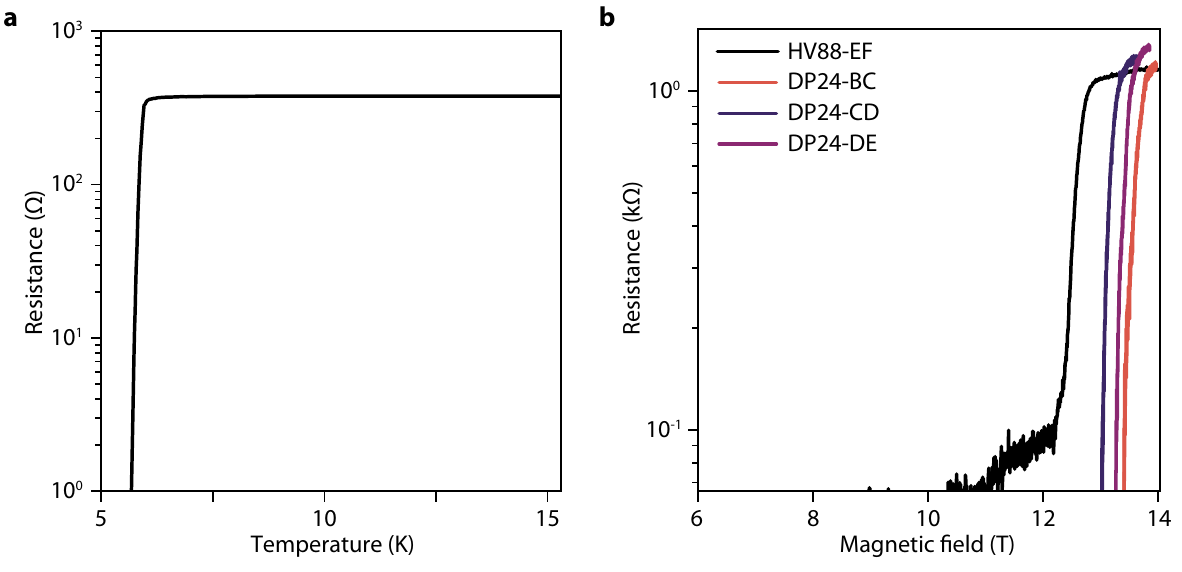}
	\caption{\textbf{MoGe superconducting transition.} \textbf{a}, Four-terminal resistance of a MoGe thin film as a function of temperature. \textbf{b}, Two-terminal resistance of superconducting electrodes on sample HV88 and DP24, a constant wiring resistance was subtracted. The legend indicates the measured MoGe electrodes (e.g. HV88-EF corresponds to the common electrode shared by junctions HV88-E and HV88-F).}
	\label{figHc}
\end{figure}

We present in this section the superconducting properties of the MoGe electrodes. 
We measured a superconducting transition temperature of $5.9$ K on a separate bare MoGe film (see Fig. \ref{figHc}a.) deposited in the same conditions as the electrodes of the samples.
Transitions to normal state with magnetic field for some MoGe electrodes of both samples HV88 and DP24 are shown in Fig. \ref{figHc}b. The resistance was measured in a two-terminal configuration with a lock-in amplifier technique, at a temperature of 0.05 K. A constant resistance of 1.06 $\rm{k}\Omega$ was subtracted to account for the wiring resistance in series with the electrodes. We measured a critical field of about 12.5 T for HV88 and 13.5 T for DP24.

\section{Device characterization}

Here, we present the characterization of all devices, including their normal state properties at 6 K, the quantum Hall regime, and the Josephson junctions properties at zero and low magnetic field. 

\subsection{Graphene nanoribbon devices}
\label{FP}

\begin{figure}[ht!]
	\centering
		\includegraphics[width=160mm]{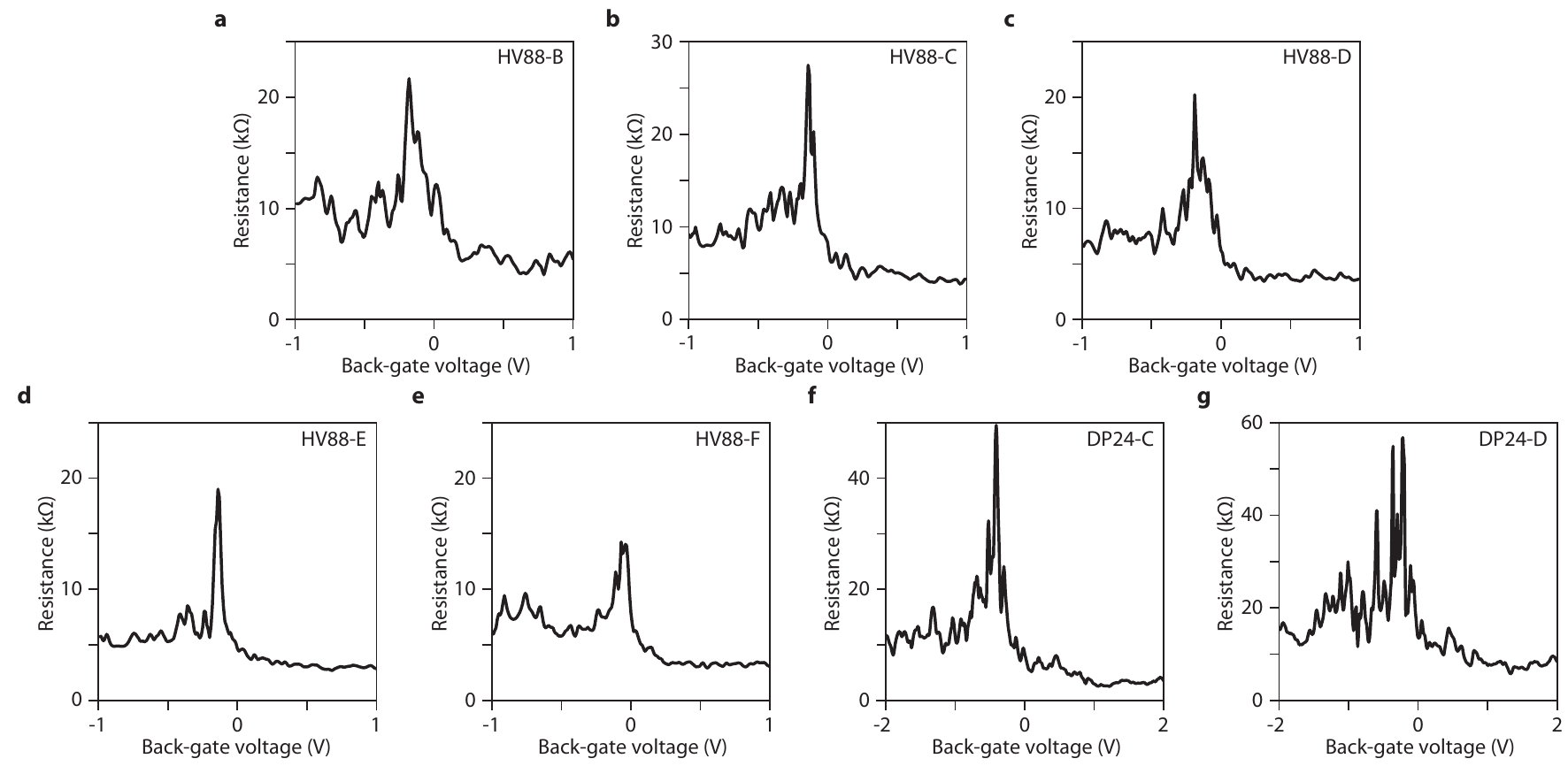}
	\caption{\textbf{Normal state resistance of the devices.} Pseudo four-terminal resistance of the devices as a function of back-gate voltage, measured at $B=0$ T.}
	\label{figFabryPerot}
\end{figure}

Figure \ref{figFabryPerot} displays the resistance versus back-gate voltage measured at a temperature of 6 K for HV88 and 4 K for DP24. The high value of resistance results mainly from the contact resistance, which is high due to the very narrow width of the contacts. The charge neutrality points of the devices are located at low negative back-gate voltage, indicating a residual electron-type doping. The higher value of the resistance for negative back-gate voltage is characteristic of charge transfer from the contacts, which induces an electron doping and thus a bipolar npn regime with reduced transparency. This bipolar npn regime is consistent with the resistance fluctuations seen in all junctions, which we ascribe to the Fabry-P\'{e}rot interferences between the pn interfaces routinely observed in high mobility devices~\cite{Calado2015a,Shalom2015,Allen2017}. The electron doping from the MoGe electrodes also explains why the quantum Hall effect is not well defined on the hole side (negative back-gate voltage), as shown in section \ref{sec:LandauFan}.

\subsection{Quantum Hall effect in the graphene nanoribbons}
\label{sec:LandauFan}

\begin{figure}[ht!]
	\centering
		\includegraphics[width=160mm]{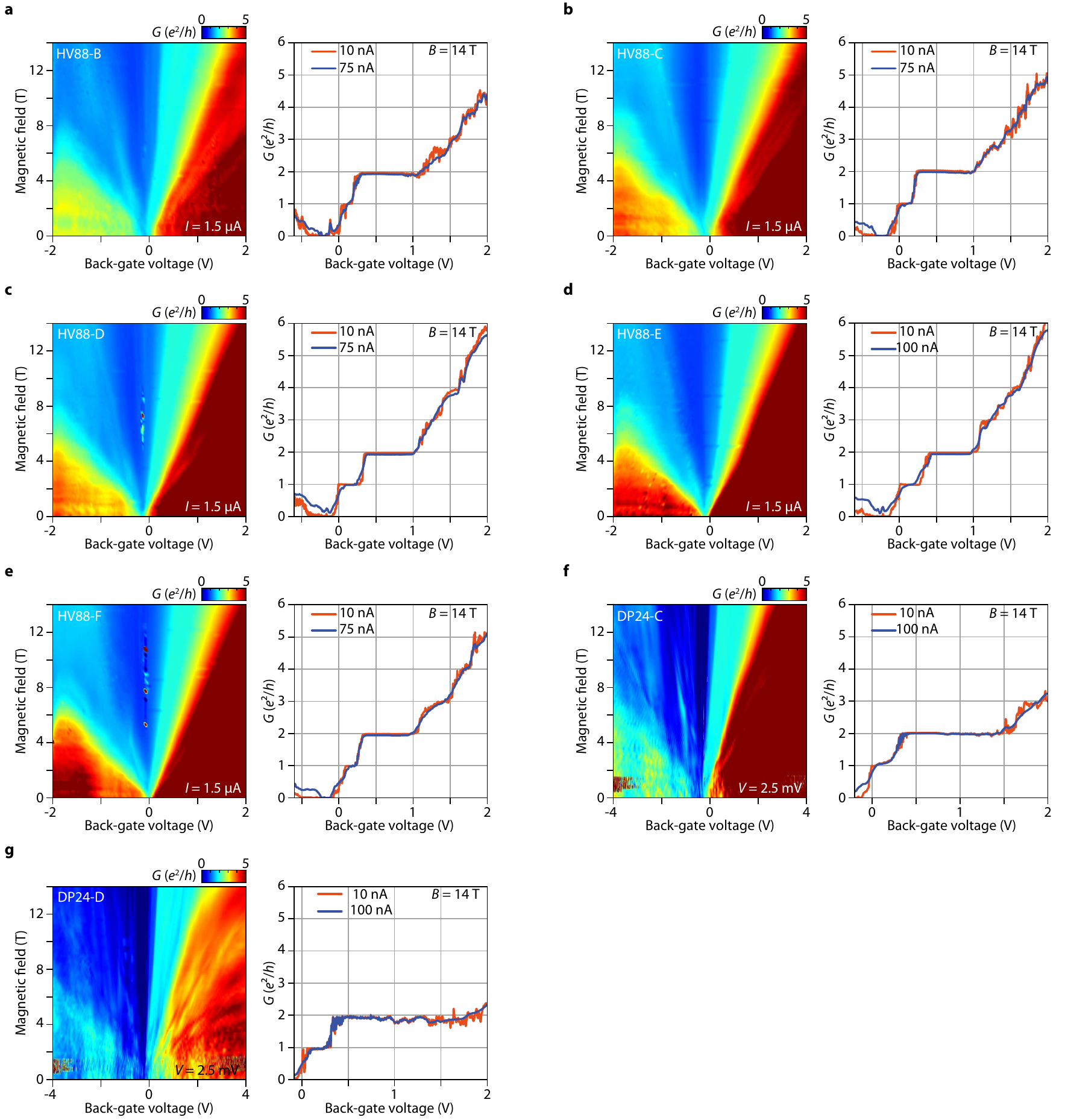}
	\caption{\textbf{Quantum Hall effect characterization.} Measurements on sample HV88 junction B to F in \textbf{a-e} and on sample DP24 junctions C and D in \textbf{f-g}. The left panel is the junction differential resistance measured with a current of $1.5 \ \mu \rm{A}$ at a temperature of 10 mK (except for DP24-C and D measured with a bias voltage of 2.5 mV at 4 K) as a function of back-gate voltage and magnetic field. The colormaps display the usual fan observed in the quantum Hall regime. The right panel corresponds to separate measurements with smaller current excitations at 14 T and 10 mK. Several broken-symmetry states emerge when measured with a current of 10 nA, while a current of 75 nA is sufficient to significantly reduce these plateaus widths.}
	\label{figFan}
\end{figure}

The quantum Hall effect was characterized with a dc current bias of $I = 1.5 \ \mu \rm{A}$ for HV88 at 10 mK and with a dc voltage bias of $V = 2.5 \ \rm{mV}$ for DP24 at 4 K, to prevent deviation from quantization due to sub-gap features~\cite{Octavio1983}. The resistance maps of the different junctions as a function of back-gate voltage and magnetic field are shown in Fig.~\ref{figFan}. As discussed in the previous section, the hole side is badly quantized due to the npn bipolar regime. On the electron side, the $\nu=2$ quantum Hall plateau is well quantized from low magnetic field (from $\sim 1.5$ T depending on the device for narrow junctions). The visibility of other QH plateaus  with smaller energy gap, as for instance the $\nu=1$ state, requires that the critical current for their QH breakdown is larger than the non-linearities induced by the Andreev reflections at finite bias (typically up to $2\Delta /e$). As a result, the $\nu = 1$ state emerges as a well quantized plateau from $B\sim 6$ T. It is clearly visible in the line-cuts taken at $B=14$ T in Fig.~\ref{figFan}. Plateaus for filling factors $\nu>2$ meet the same issues. Their absence can furthermore be accounted for by a lack of equilibration between the edge channel of the zeroth Landau level and the edge channels of the higher Landau levels in our mesoscopic devices~\cite{Buttiker88}.

The line-cuts of Fig. \ref{figFan} measured with a small current excitation $I = 10 \ \rm{nA}$ at 14 T, that is, above the upper critical field of the electrodes are not altered by superconducting sub-gap features nor quantum Hall breakdown. We can see in Fig. \ref{figFan}e and d the emergence of the $\nu=3$ and 4 broken-symmetry states.

\subsection{Back-gate capacitance estimation}
\label{sec:CapaEstimation}

\begin{figure}[ht!]
	\centering
		\includegraphics[width=0.8\textwidth]{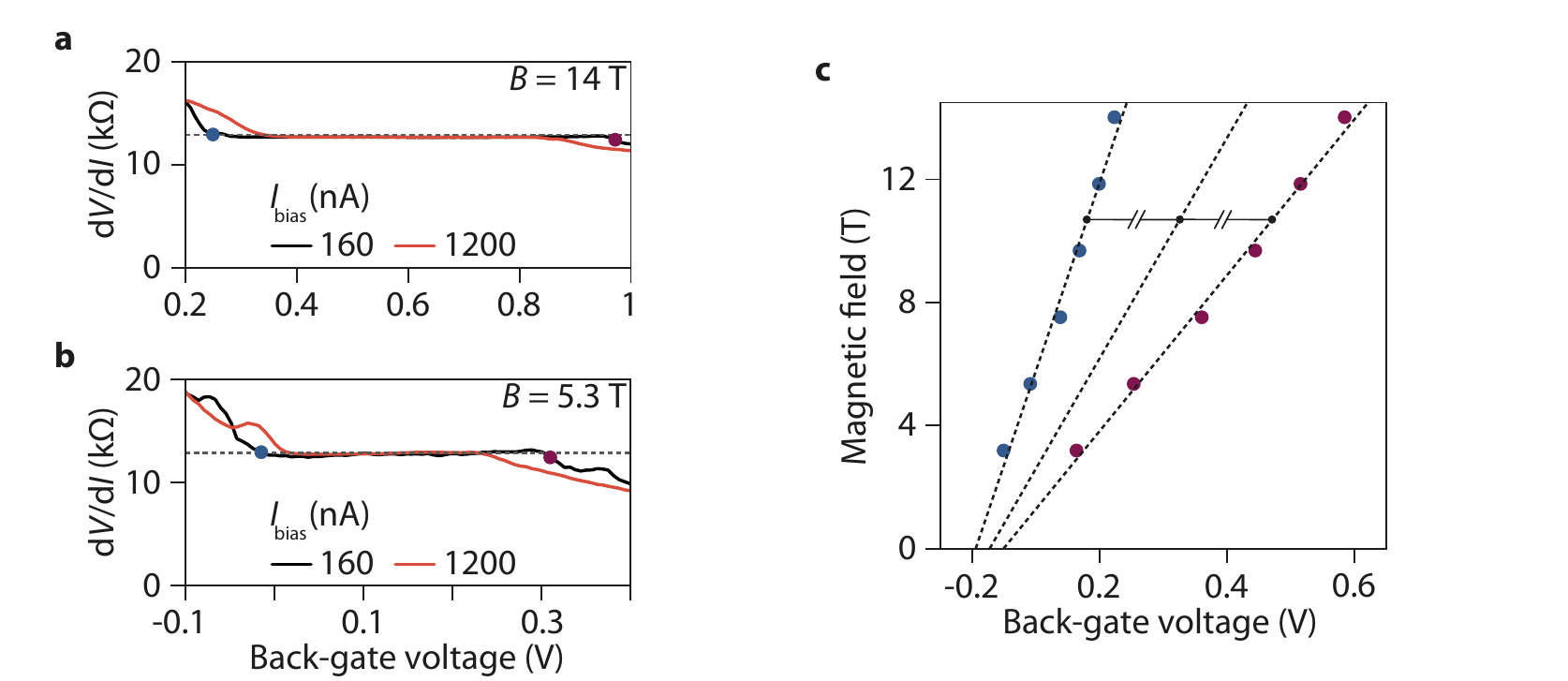}
	\caption{\textbf{Filling factor 2 plateau characterization in sample HV88-C.} \textbf{a-b}, Differential resistance at 14 T and 5.3 T at two different dc bias currents (160 nA and 1.2 $\mu \rm{A}$), blue (purple) dots delimit the plateau edge defined by a deviation of 2\% from the plateau resistance. \textbf{c}, Detected plateau edge and plateau center evolution with magnetic field.}
	\label{figGateEstimate}
\end{figure}

The quantum Hall data of Fig. \ref{figFan} are used to quantify the back-gate capacitance. While a two-plate capacitor model gives $C = 1.4 \ \rm{mF}/\rm{m}^2$ for sample HV88 and $C = 0.89 \ \rm{mF}/\rm{m}^2$ for sample DP24 using a dielectric constant of $\varepsilon_{\rm{BN}} = 3.3$ for boron nitride~\cite{Hong2020}, in very small structure the capacitance can be modified due to screening by the electrodes. We thus estimated the back-gate capacitance through the dispersion of the filling factor 2 plateau as shown in Fig. \ref{figGateEstimate}. The middle of the plateau is defined as $\nu=2.0$. The curve at 160 nA is chosen as a reference to detect the plateau edges as it shows limited deviation from quantization compared to both lower and higher currents. These deviations originate from the non-linearities of the superconducting interface at low current, and from the quantum Hall breakdown at high current. The resulting capacitances that we used to calculate the filling factor are reported in the Extended Data Table I.\\

\subsection{Contact resistance}
\label{sec:ContactResistance}

If one assume ballistic transport within the junction, we can estimate the graphene resistance in the pure quantum limit \cite{Shalom2015}: 
$R_Q = h/e^{2} g_s g_v N$, where $g_s = g_v = 2$ account for the spin and valley degeneracies, respectively, and N is the number of transverse modes propagating in the graphene $N = k_F W / \pi$. $k_F =\sqrt{\pi n}$ and the electron density is defined as $n = C(V_g - V_{CNP})/e$.

The difference between the measured resistance $R$ and $R_Q$ provides the contact resistance $R_c = (R - R_Q)/2$. At $V_g = 1$ V, we obtain $460 \pm 21$ $\Omega.\mu m$ averaged over 13 devices for the MoGe contacts. This corresponds to an average transmission probability $Tr = R_Q / (R_Q + R_c) \sim 0.23$. This transmission probability is half that obtained from the IV curves in the superconducting state (see section II.E). Note that a similar, though less pronounced, difference was observed in \cite{Shalom2015}.

\subsection{Supercurrent characterization at $B=0$ T}

\begin{figure}[ht!]
	\centering
		\includegraphics[width=0.8\textwidth]{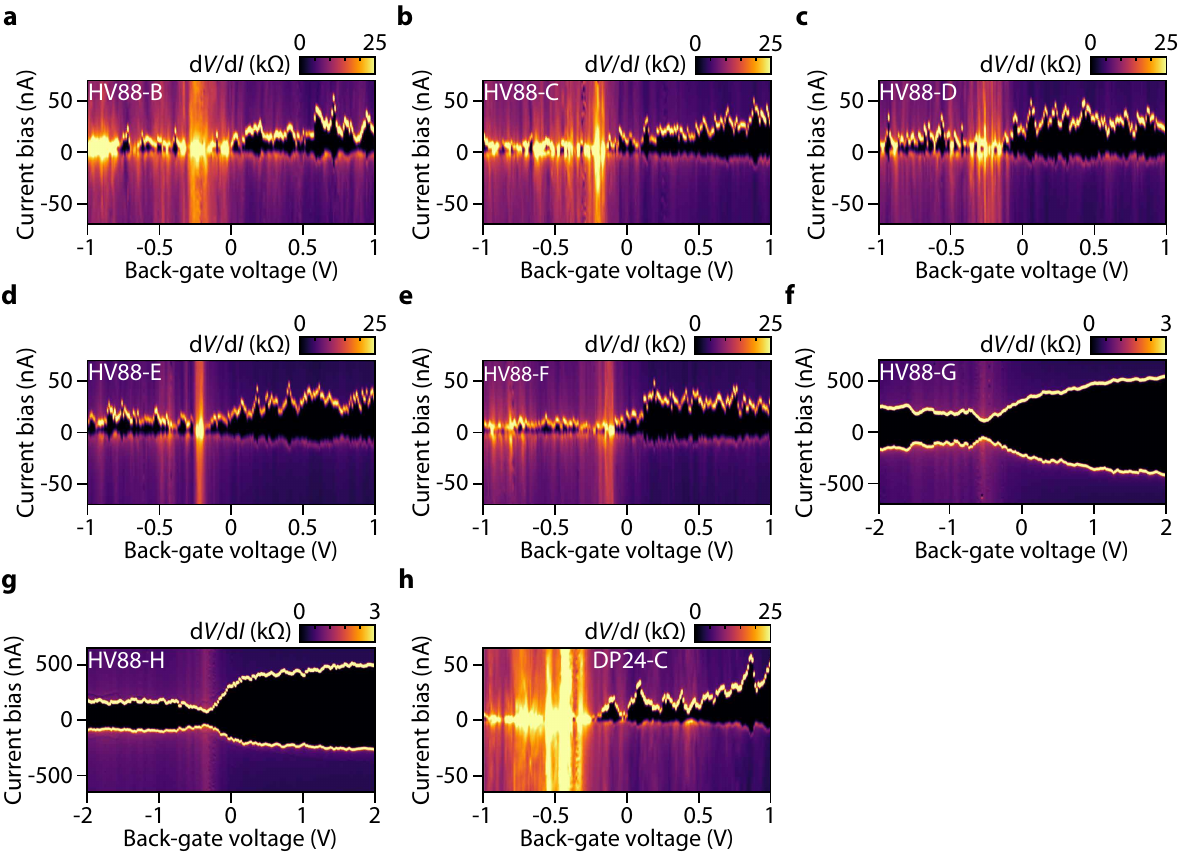}
	\caption{\textbf{Supercurrent at zero field and 10 mK.} Differential resistance versus back-gate voltage and dc current bias for sample HV88, junctions B to H in \textbf{a-g}, and for junction DP24-C in \textbf{h}.}
	\label{figSCgateSweep}
\end{figure}

Figure \ref{figSCgateSweep} displays the differential resistance as a function of dc current bias and back-gate voltage at zero magnetic field for all devices. 
The decrease of resistance with back-gate voltage leads to an increase of switching current as expected for a constant $I_cR_{\rm{n}}$ product, where $R_{\rm{n}}$ is the normal state resistance. We obtained $I_cR_{\rm{n}}$ ranging from $\sim 87 \,\mu$V to  $\sim 243 \,\mu$V.  

\begin{figure}[ht!]
	\centering
		\includegraphics[width=0.9\textwidth]{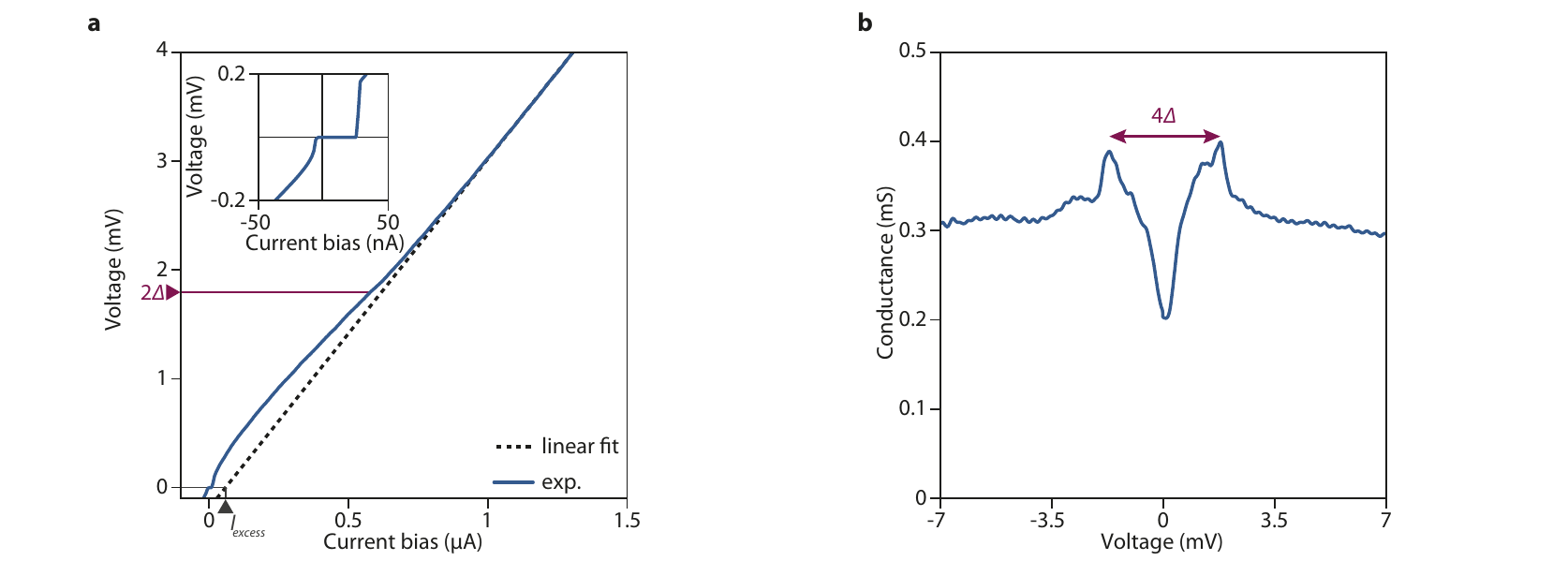}
	\caption{\textbf{Excess current and superconducting gap of the MoGe electrode.}
	\textbf{a}, $I-V$ characteristic of the junction HV88-F at $B = 0 \ \rm{T}$ and $V_{\rm{g}} = 1 \ \rm{V}$. The dashed line is a linear fit of the high current bias part. The intercept of this fit with the x-axis gives the excess current $I_{\rm{excess}}$ (see grey arrow). Inset: Higher resolution $I-V$ measurement at low current bias showing more clearly the supercurrent. \textbf{b}, Numerical derivative $dI/dV$ of \textbf{a} versus measured voltage. The voltages of the two conductance maxima are associated with $\pm 2 \Delta /e$ according to Ref. \cite{Octavio1983}.}
	\label{figTransparency}
\end{figure}

We estimated the contact transparency from $I-V$ characteristics following Ref. \cite{Octavio1983,Flensberg1988,Niebler2009}. The normal state resistance $R_{\rm{n}}$ is estimated from a linear fit at high current bias (see Fig.\ref{figTransparency}), and the excess current $I_{\rm{exc}}$ is defined as the intercept of this fit at zero voltage. The conductance plotted as a function of measured voltage exhibits two maxima, symmetric with respect to zero voltage, that we ascribe to twice the superconducting gap $2 \Delta$. Eq. (25) in Ref. \cite{Niebler2009} is used to deduce the Blonder-Tinkham-Klapwijk parameter $Z$, which leads to the transparency defined as $\mathcal{T}=1/(1+Z^2)$. The transparency of our junctions are reported in Table \ref{Transparency} for 7 junctions, along with $R_{\rm{n}}$, $I_{\rm{exc}}$, and $\Delta$. 

\begin{table*}[!ht]
\centering
\begin{tabular}{|c|c|c|c|c|c|c|c|c|}
\hline
$\vphantom{M^{0^0}}$ Junction & $V_{\rm{CNP}}$ (V) & $V_{\rm{CNP}}^{\nu = 2}$ (V) & $V_{\rm{g}} \ (\rm{V})$ & $\Delta \ (\mu \rm{eV})$ & $ I_{\rm{exc}}$ (nA) & $ R_{\rm{n}}$ ($k\Omega$) & $T$ & $R_{\rm{n}}I_{\rm{c}}$  ($\mu \rm{V}$) \\
\hline
$\vphantom{M^{0^0}}$ HV88-C & -0.13 & -0.14 & 1 & 852 & 49 & 4.5 & 0.49 & 150 \\
\hline
$\vphantom{M^{0^0}}$ HV88-D & -0.19 & -0.15 & 1 & 865 & 46 & 3.8 & 0.47 & 91 \\
\hline
$\vphantom{M^{0^0}}$ HV88-E & -0.14 & -0.25 & 1 & 876 & 86 & 3.0 & 0.5 & 96 \\
\hline
$\vphantom{M^{0^0}}$ HV88-F & -0.04 & -0.08 & 1 & 897 & 59 & 3.2 & 0.47 & 87 \\
\hline
$\vphantom{M^{0^0}}$ HV88-G  & -0.52 & -0.24 & 2 & 762 & 552 & 0.34 & 0.48 & 146 \\
\hline
$\vphantom{M^{0^0}}$ HV88-H & -0.34 & -0.23 & 2 &  773 & 434 & 0.34 & 0.47 & 137 \\
\hline
$\vphantom{M^{0^0}}$ DP24-C & -0.4 & -0.12 & 1 & 754 & 83 & 4.2 & 0.56 & 243 \\
\hline
\end{tabular}
\caption{Junction characteristics at $B = 0 \, \rm{T}$. Charge neutrality point positions $V_{\rm{CNP}}$ were extracted from the Dirac peak positions in plots of the resistance as a function of the gate-voltage (see Figure \ref{figFabryPerot}). We also provide the charge neutrality point positions $V_{\rm{CNP}}^{\nu = 2}$ estimated from the fits to the $\nu = 2$ plateau center used for capacitance estimation (see section \ref{sec:CapaEstimation}). Note that filling factor evaluation was done using $V_{\rm{CNP}}^{\nu = 2}$ for the Dirac point gate voltage value. The back-gate voltage value at which the gap $\Delta$ of the superconducting electrodes, the junctions excess current $ I_{\rm{exc}}$, the normal state resistance $ R_{\rm{n}}$, the graphene-superconducting electrode interface transparency $T$ as well as the $R_{\rm{n}}I_{\rm{c}}$ products were evaluated is specified in the $V_{\rm{g}}$ column.}
\label{Transparency}
\end{table*}

\subsection{Fraunhofer patterns}

\begin{figure}[ht!]
	\centering
		\includegraphics[width=120mm]{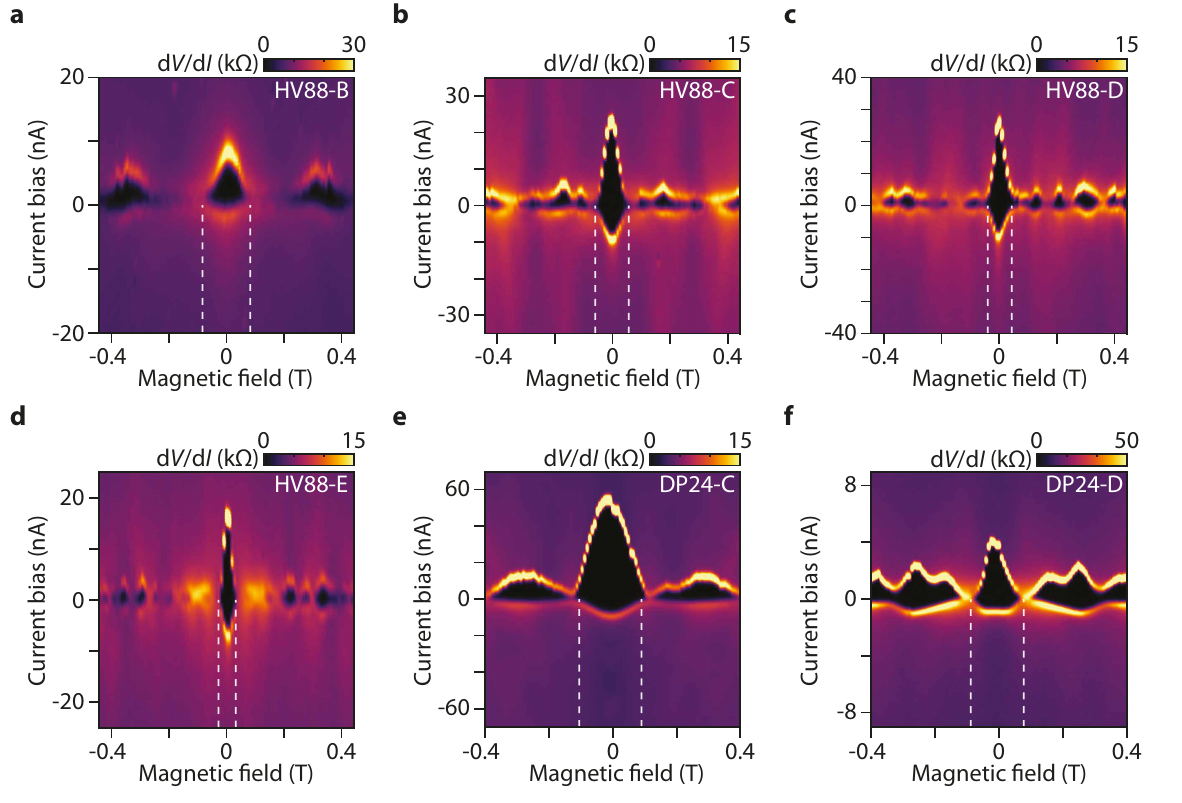}
	\caption{\textbf{Quantum interference of the critical current.} Measurements on sample HV88 junction B to E in \textbf{a-d} and on sample DP24 junctions C and D in \textbf{e-f}. \textbf{a-f}, Differential resistance versus magnetic field and dc current bias. The supercurrent delimiting zero (in black) and finite resistance state (in yellow and purple) shows oscillatory behavior resembling Fraunhofer pattern. The white dashed lines indicate the expected periodicity for a superconducting flux quantum $\phi_0 = h/2e$ through the graphene area.}
	\label{figFraunhofer}
\end{figure}

In this section we present the quantum interference (Fraunhofer) patterns of the Josephson junctions at low magnetic field. Figure \ref{figFraunhofer} displays the differential resistance as a function of current bias and magnetic field for 6 devices. The evolutions of the switching current do not follow the usual Fraunhofer pattern with a sinus cardinal decay. This is not clearly understood but we cannot exclude vortices trapped in the superconducting electrodes due to the relatively large field needed for reaching a flux quantum through the small graphene area of the junctions (e.g. $B = 83$ mT for device HV88-B), as well as deviation from the Fraunhofer pattern due to the square geometry of the graphene \cite{Chiodi12}. Still, we can extract a flux period from the central lobs of these quantum interference patterns. The magnetic field at which the switching current vanishes for the central lob corresponds theoretically to $\phi_0=h/2e$ \cite{Tinkham2014}. We thus added white dashed lines in Fig. \ref{figFraunhofer} that indicates the equivalent magnetic field for one flux quantum through the graphene area. We obtain a remarkably good agreement for all devices. Interestingly, Fraunhofer patterns nearly matching the graphene area have been observed in similar devices with square geometry ($W\sim L$) in \cite{Calado2015a}.

\section{Additional Shapiro maps}

\begin{figure}[ht!]
	\centering
		\includegraphics[width=0.7\textwidth]{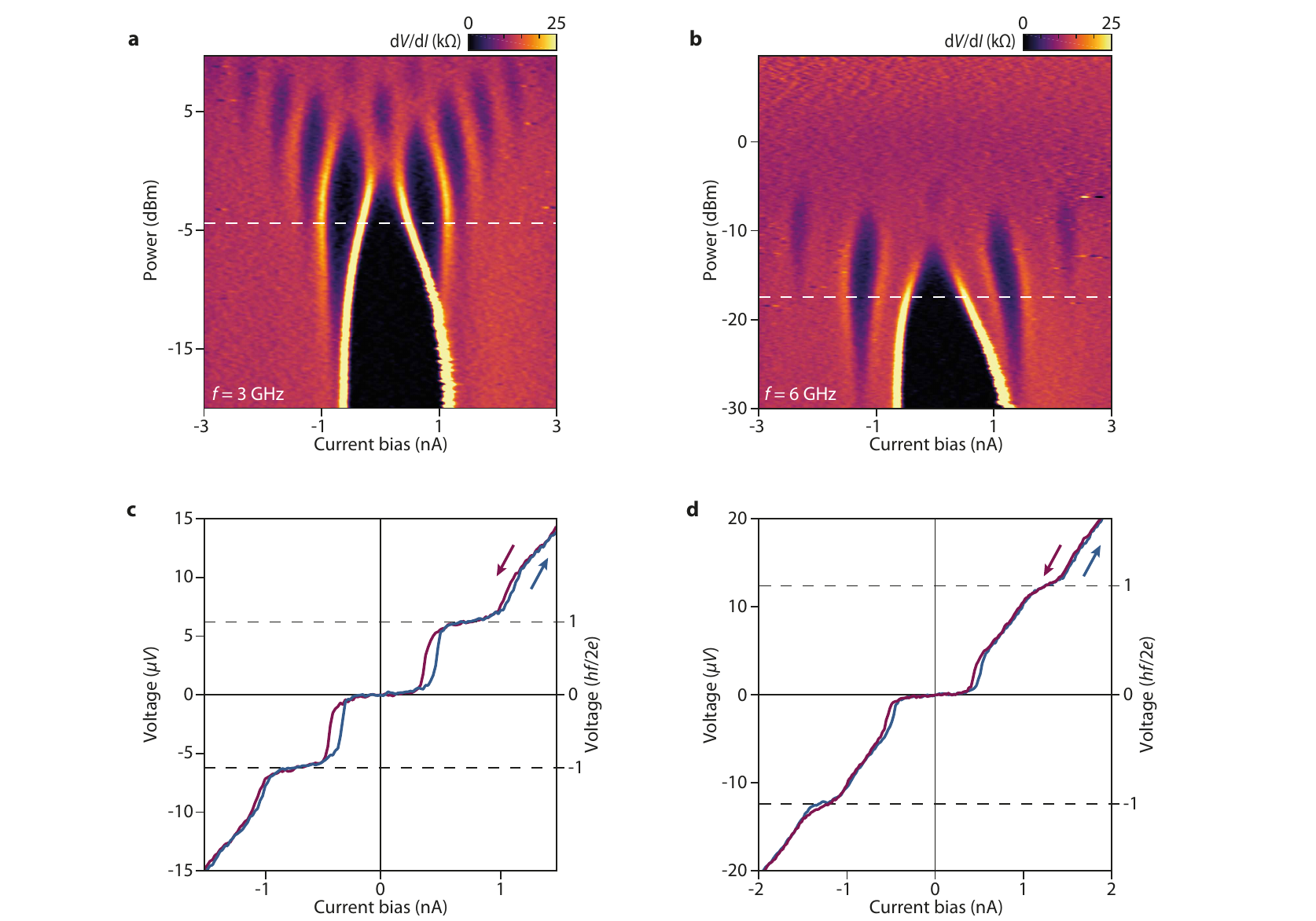}
	\caption{\textbf{Additional Shapiro maps.} \textbf{a-b}, Differential resistance as a function of current bias (sweep up) and microwave power at 3 GHz (\textbf{a}) and 6 GHz (\textbf{b}), measured in junction HV88-C with the same parameters of magnetic field and back-gate voltage as in Fig. 1e. \textbf{c-d}, $I-V$ curves at a fixed microwave power indicated by the white lines in \textbf{a} and \textbf{b}, showing successive Shapiro voltage steps of height $hf/2e$. The blue $I-V$ is sweep up and purple is sweep down.}
	\label{figShapiro}
\end{figure}

We show in this section additional Shapiro maps, complementing the data of Fig. 1e. Figure \ref{figShapiro} displays the Shapiro maps of the differential resistance along with $I-V$ characteristics measured with the same parameters of magnetic field and back-gate voltage as in Fig. 1e but with microwave frequencies of 3 GHz, in Fig. \ref{figShapiro}a and c, and at 6 GHz in Fig. \ref{figShapiro}b and d.

\section{$2\phi_0$-periodic oscillations}

\subsection{Constant back-gate voltage versus constant filling factor}

\begin{figure}[ht!]
	\centering
		\includegraphics[width=120mm]{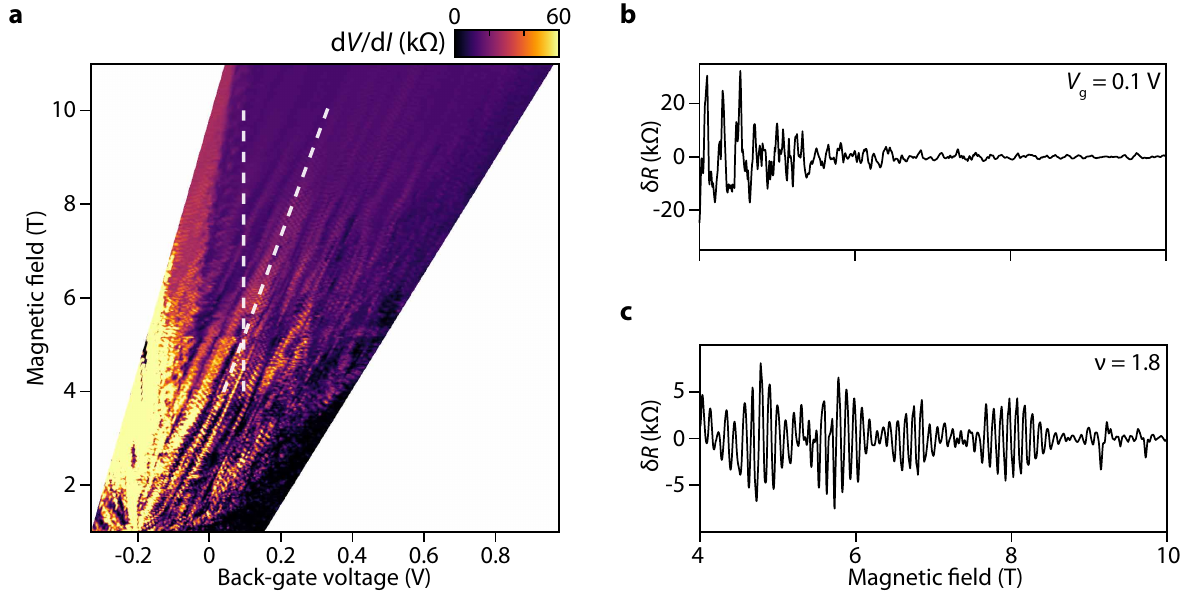}
	\caption{\textbf{Absence of oscillation with back-gate voltage.} \textbf{a}, Differential resistance map as a function of back-gate voltage and magnetic field at zero current bias in sample HV88-D. \textbf{b}, Resistance versus magnetic field at a fixed back-gate voltage of $V_{\rm{g}} = 0.1 \ \rm{V}$ (taken along the vertical dashed line in (\textbf{a})) after background subtraction. \textbf{c}, Resistance versus magnetic field at a fixed filling factor $\nu = 1.8$ (taken along the diagonal dashed line in (\textbf{a})) after background subtraction.}
	\label{figCstBackgate}
\end{figure}

In this section we illustrate the relevance of studying the flux periodicity of the supercurrent and resistance oscillations at constant filling factor compared to constant back-gate voltage. Figure \ref{figCstBackgate} displays the zero bias differential resistance of sample HV88-D as a function of back-gate voltage and magnetic field. The measurement is restricted to the $\nu=2$ QH plateau to limit the the data acquisition to a reasonable duration. Fringes similar to that of Fig. 2d are visible in this resistance map. However, extracting a line-cut at constant back-gate voltage (along the vertical dashed line at $V_{\rm g} = 0.1$ V) does not exhibit any oscillatory behavior, as shown in Fig. \ref{figCstBackgate}b. On the other hand, a line-cut at constant filling factor (along the diagonal dashed line at $\nu = 1.8$) shown in Fig. \ref{figCstBackgate}c clearly unveils the $2\phi_0$-periodic oscillations. 

\subsection{Switching current and resistance oscillations on additional devices}

\begin{figure}[ht!]
	\centering
		\includegraphics[width=120mm]{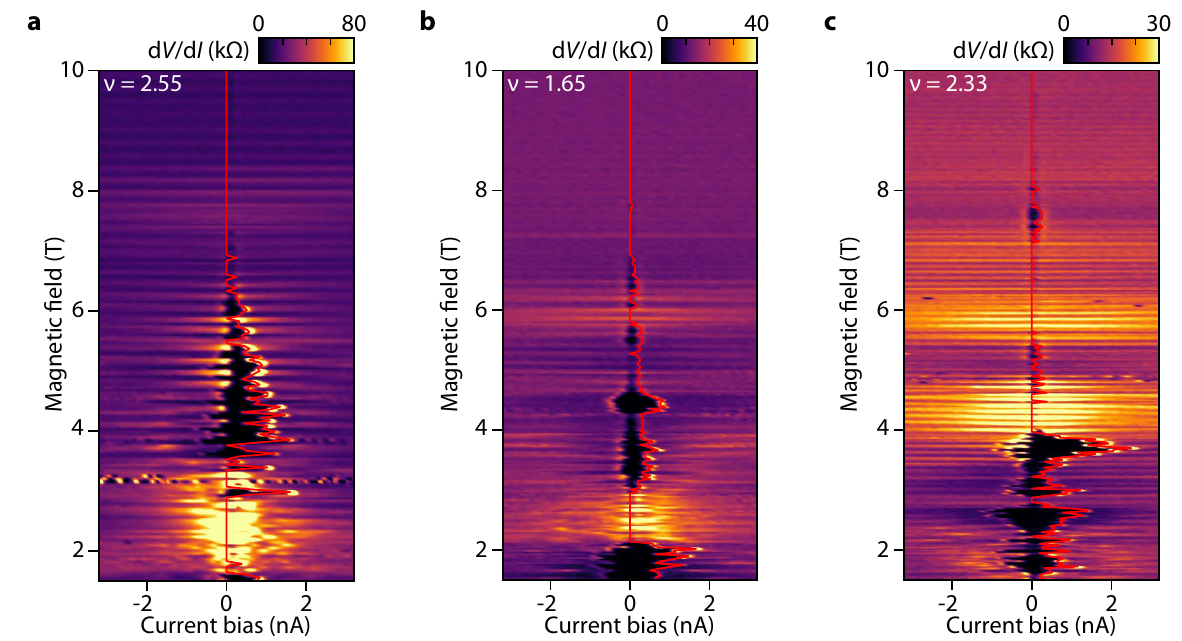}
	\caption{\textbf{Critical current oscillation with magnetic field in sample HV88.} \textbf{a}, Junction B. \textbf{b}, Junction C. \textbf{c}, Junction D. Differential resistance as a function of current bias and magnetic field at constant filling factor. The extracted switching current is superimposed as a red line.}
	\label{figIsVsB}
\end{figure}

We show in Fig. \ref{figIsVsB} additional measurements of the $2\phi_0$-periodic oscillations of the supercurrent for the junctions HV88-B, HV88-C and HV88-D. The differential resistance map is plotted as a function of dc current bias and magnetic field, at constant filling factor. The red traces superimposed on each Figures show the detected switching current. Note that such a detection was used for the switching current map in Fig 2c for HV88-B. 

\begin{figure}[ht!]
	\centering
		\includegraphics[width=120 mm]{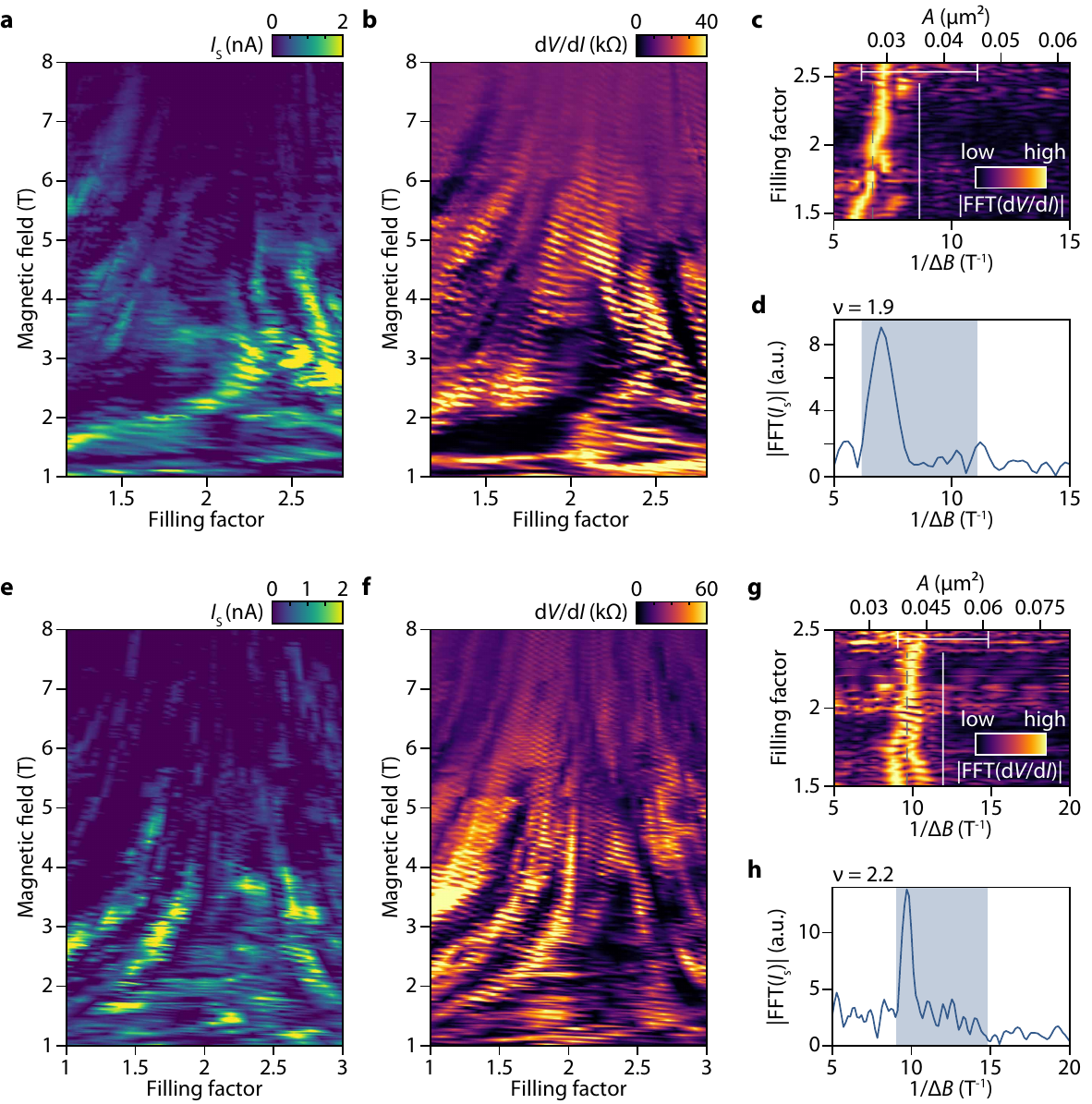}
	\caption{\textbf{$2 \phi_0$ oscillations of critical current and resistance in junctions C and D of sample HV88.} \textbf{a-d}, Junction HV88-C. \textbf{e-h}, Junction HV88-D. 
	\textbf{a}, \textbf{e}, Critical current and \textbf{b}, \textbf{f}, differential resistance at zero bias are periodically modulated with both magnetic field and filling factor. 
	\textbf{c}, \textbf{g}, The normalized Fourier transform of the resistance at finite current (5 nA) shows a small increase of the oscillation frequency with filling factor. 
	In \textbf{c} and \textbf{g}, the vertical lines correspond to the graphene area (right line) while the dashed lines correspond to the graphene area corrected as $(L - 2l_{B})$x$(L - 2l_{B})$ (left line). The magnetic length $l_{B}$ is taken at $4.75$ T (resp. $5$ T) in (\textbf{c}) (resp. \textbf{g}). The Fourier transform shown in (\textbf{c}) was obtained in a magnetic field window ranging from $2.5$ T to $7$ T. The Fourier transform shown in (\textbf{g}) was obtained in a magnetic field window ranging from $3$ T to $7$ T.  The error bar corresponds to the uncertainty on the graphene area detailed in Methods. 
	\textbf{d}, \textbf{h}, The Fourier transform of the critical current shows a peak at a frequency compatible with the expected one. The colored rectangle is centered at the frequency corresponding to $2 \phi_0$ oscillations for the graphene area, its width is given by the uncertainty on this area. The Fourier transform signal in (\textbf{d}) (resp. (\textbf{h})) was obtained in a magnetic field window going from $2$ T to $7$ T (resp. from $1$ to $8$ T).}
	\label{figIsVsBVg-CD}
\end{figure}

In Fig. \ref{figIsVsBVg-CD} we show the switching current map and the zero bias resistance as a function of filling factor and magnetic field for the junctions HV88-C and HV88-D. Fig. \ref{figIsVsBVg-CD}c and g display the Fourier transform of the resistance map as a function of frequency $1/\Delta B$ and filling factor for HV88-C and HV88-D, respectively. Fig. \ref{figIsVsBVg-CD}d and h display the Fourier transform of the critical current oscillations as a function of frequency $1/\Delta B$ for HV88-C and HV88-D, respectively. The frequencies of these oscillations are in excellent agreement with the graphene area corrected by the magnetic length, therefore confirming the results presented in the main text.

\begin{figure}[ht!]
	\centering
		\includegraphics[width=120 mm]{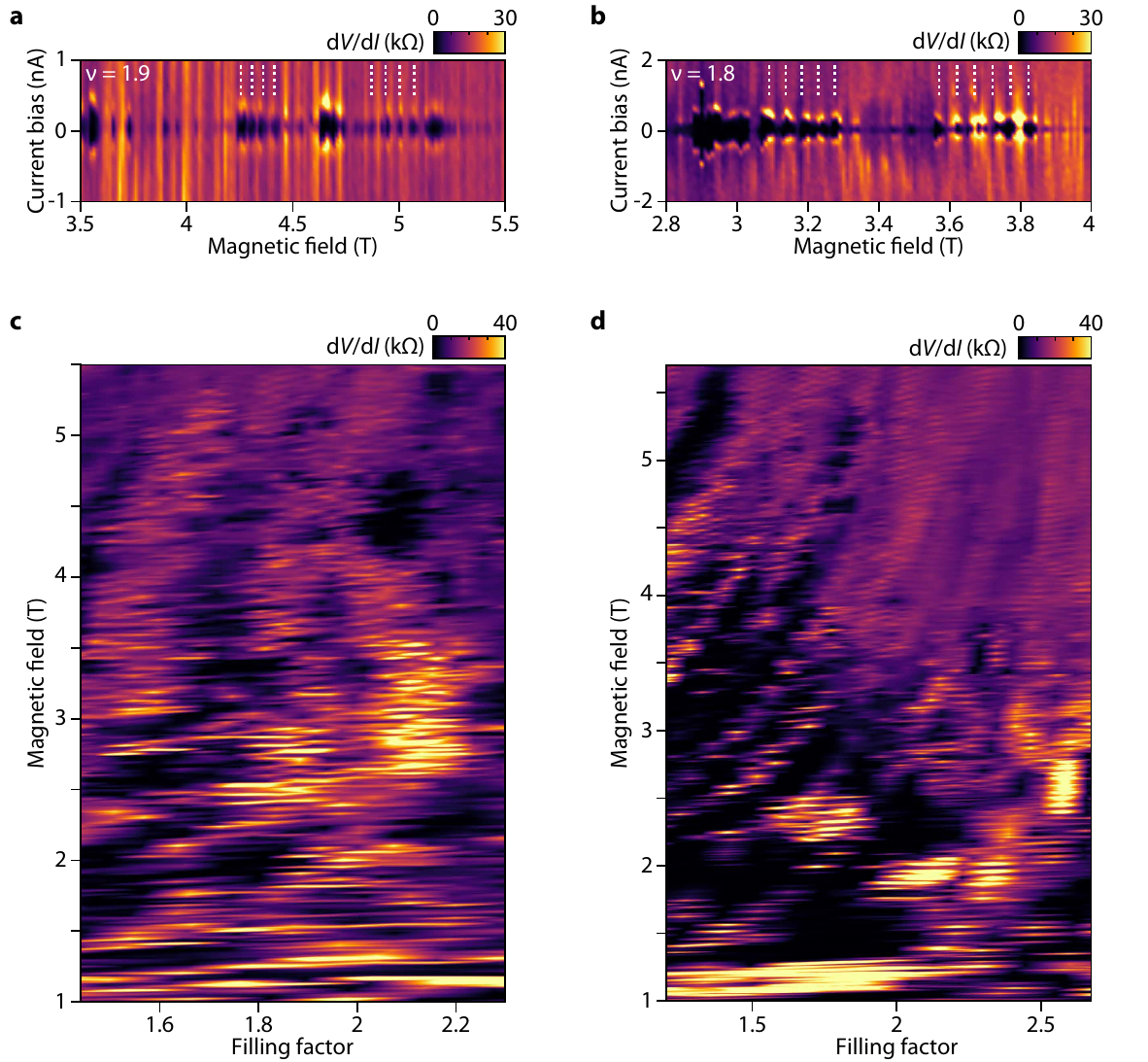}
	\caption{\textbf{$2 \phi_0$ oscillations of critical current and resistance in junctions E and F of sample HV88.} \textbf{a, c}, Junction HV88-E. \textbf{b, d}, Junction HV88-F. \textbf{a-b}, Critical current and differential resistance are periodically modulated with magnetic field. In \textbf{a}, the left (resp. right) white comb indicates the $2 \phi_0$ magnetic field periodicity corresponding to $113 \%$ (resp. $89\%$) of the graphene area. In \textbf{b}, the left (resp. right) white comb indicates the $2 \phi_0$ magnetic field periodicity corresponding to $99 \%$ (resp. $90\%$) of the graphene area. \textbf{c-d}, The differential resistance at 0 nA shows oscillations with filling factor and magnetic field.}
	\label{figIsVsBVg-EF}
\end{figure}

\begin{figure}[ht!]
	\centering
		\includegraphics[width=120 mm]{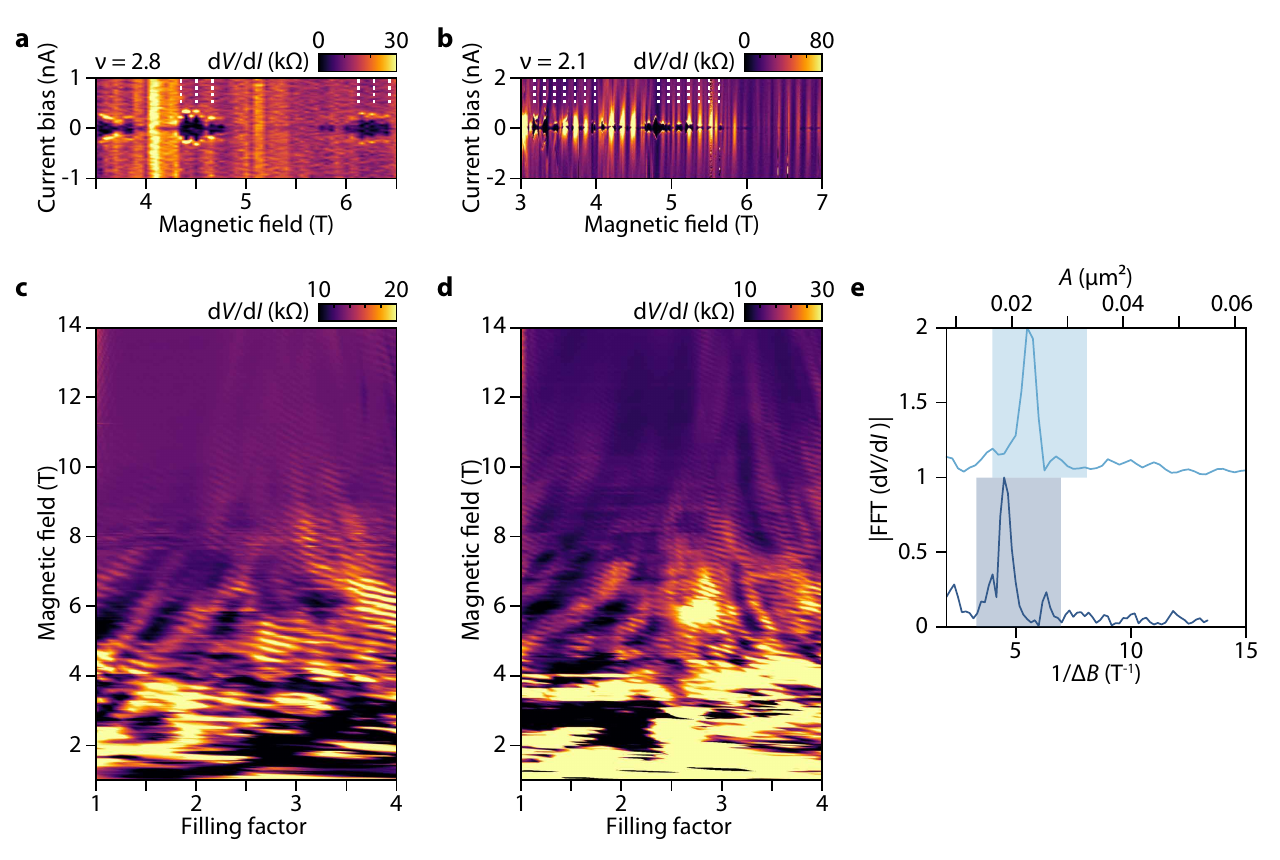}
	\caption{\textbf{$2 \phi_0$ oscillations of critical current and resistance in junctions C and D of sample DP24.} \textbf{a-b}, Differential resistance vs current bias and magnetic field at a constant filling factor for junction DP24-C in \textbf{a} and for junction DP24-D in \textbf{b}. $2 \phi_0$ oscillations of resistance and critical current are visible, the dashed lines indicating the theoretical period within the device area uncertainty, that is $125\%$ of the device DP24-C graphene area in (\textbf{a}) and $123\%$ of the device DP24-D graphene area in (\textbf{b}). \textbf{c-d}, Differential resistance mapping as a function of filling factor and magnetic field for junction DP24-C in (\textbf{c}) and for junction DP24-D in (\textbf{d}) with a current bias of 1 nA. \textbf{e}, Fourier transform of the resistance oscillations at a fixed filling factor showing that the oscillation period correspond to a flux quantum of $2 \phi_0$. Bottom curve DP24-C (obtained at $\nu = 3 $), top curve DP24-D (obtained at $\nu = 1.44$).}
	\label{figHseDP24}
\end{figure}

The data of the two longest and widest nanoribbons, HV88-E, HV88-F, are shown in Fig. \ref{figIsVsBVg-EF}. Despite a smaller magnetic field period, we resolve the supercurrent and resistance oscillations on the QH plateau. Differential resistance maps as a function of magnetic field and current bias at fixed filling factors in Fig. \ref{figIsVsBVg-EF}a and b clearly show the $2\phi_0$-periodic oscillations of the supercurrent and of the resistance, as indicated by the white dashed lines. Similar results have been obtained on junctions DP24-C and DP24-D, as shown in Fig. \ref{figHseDP24}.

\section{Additional wide junction}

\begin{figure}[ht!]
	\centering
		\includegraphics[width=120 mm]{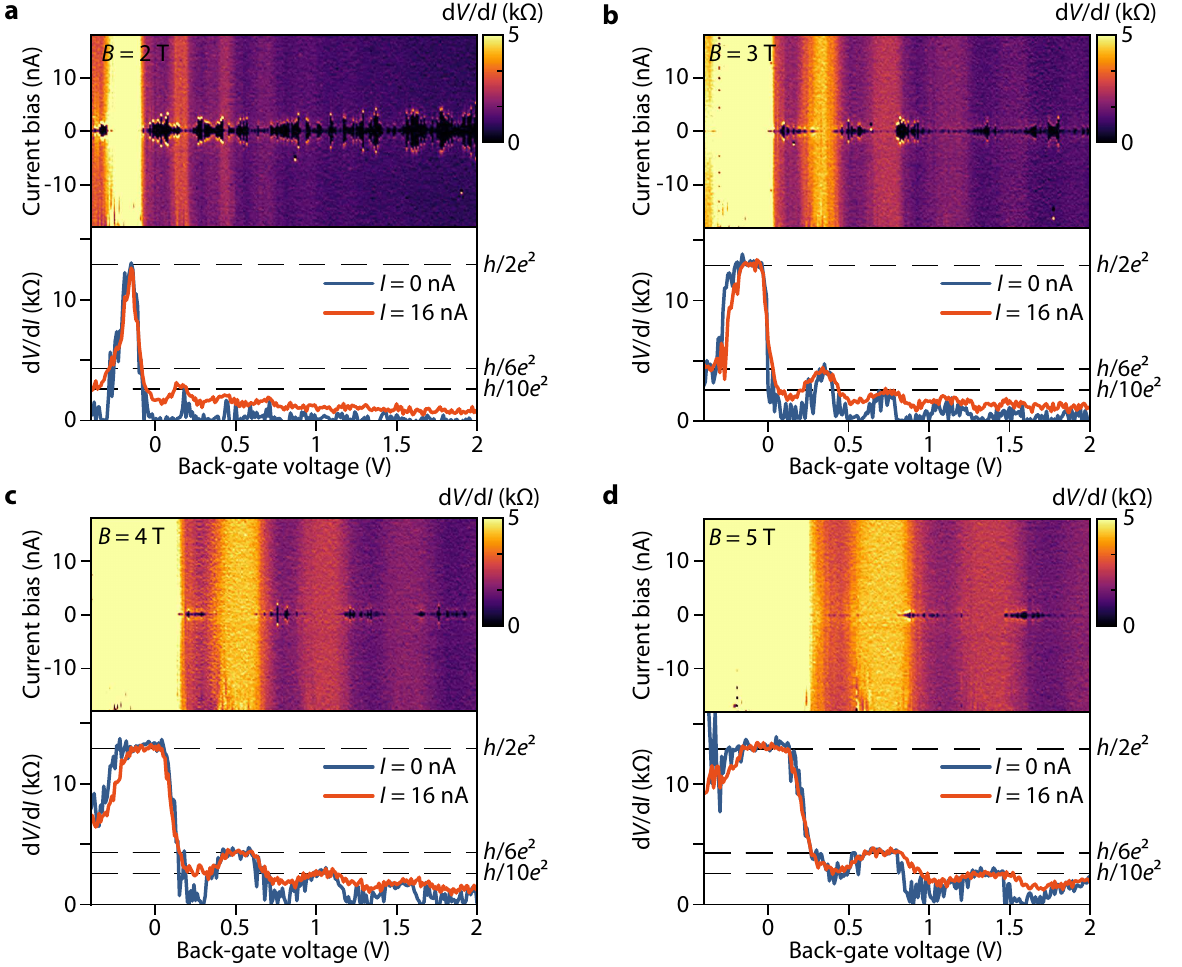}
\caption{\textbf{Additional wide graphene Josephson junction.} 
 \textbf{a}-\textbf{d}, Top panels show differential resistance maps as a function of back-gate voltage and dc current bias for device HV88-H ($W = 2.4\,\mu$m, $L=202$ nm) at $B = 2$ T, $3$ T, $4$ T and $5$ T. Bottom panels are linecuts of the differential resistance at dc current biases of 0 nA and 16 nA, which show the emergence of supercurrent pockets (zero resistance reached by the blue curve), and the corresponding resistive state (orange curve). The magnetic field is indicated in each top panel. Supercurrent is visible only when the resistance is not quantized, that is, at filling factor of QH plateau that are not developed, or in between plateaus. When QH plateau emerges, as for instance the $h/2e^2$ plateau for $B\geq3$ T or the $h/6e^2$ plateau for $B\geq3$ T, the supercurrent vanishes. Note that the oscillatory behavior of the resistance (orange curve) is characteristic of QH devices in two-terminal configuration with $L\ll W$~\cite{Abanin2008, Williams2009}.} 
	\label{figJunctionG}
\end{figure}

We present in this section the data for a second wide junction (HV88-H) in complement to Extended Data Fig. 4. This junction has nearly the same width ($W=2.4\, \mu$m) but is twice longer than junction HV88-G. Figure \ref{figJunctionG} displays the differential resistance as a function of back-gate voltage and dc current bias at different magnetic field values, along with the differential resistance measured at 0 and 26 nA to show the superconducting pocket and the resistive state. As in Extended Data Fig. 4, supercurrent is conspicuously absent on the $h/2e^2$ QH plateau. For low magnetic field there is a finite supercurrent in the $h/6e^2$ QH plateau. However, this supercurrent vanishes when the resistance becomes quantized upon increasing magnetic field, indicating that this junction cannot maintain supercurrent in the edge transport regime. This is consistent with the suppression of the Thouless energy by the long superconducting interfaces, which in turn suppress the supercurrent \cite{Alavirad2018}.

\clearpage



\end{document}